\documentclass[12pt,a4paper]{article}
\pdfoutput=1

\usepackage{color}
\usepackage{amssymb,amsmath,bm,bbm}
\usepackage{epsf}
\usepackage{epsfig}
\usepackage{afterpage}
\usepackage{longtable}
\usepackage[dvipsnames]{xcolor}
\usepackage[linktoc=page,bookmarks=false,colorlinks=false,linkbordercolor=RoyalBlue,citebordercolor=ForestGreen,urlbordercolor=CornflowerBlue]{hyperref}
\usepackage{latexsym,mathrsfs,dsfont}
\usepackage[normalem]{ulem} 
\usepackage[compress]{cite}
\usepackage{graphicx}
\usepackage{url}
\usepackage{paralist}
\usepackage{bbold}
\usepackage{slashed}
\usepackage{multirow}

\usepackage[hypcap]{caption, subcaption}

\usepackage{booktabs}
\usepackage{listings}
\usepackage{siunitx} 
\usepackage{textcomp}

\allowdisplaybreaks[1]

\numberwithin{equation}{section}

\usepackage{fancyhdr}
\advance \headheight by 3.0truept       

\def \refeq#1{(\ref{#1})}
\def \refsec#1{Section~\ref{#1}}

\def \refapp#1{Appendix~\ref{#1}}
\def \reffig#1{Figure~\ref{#1}}
\def \reftab#1{Table~\ref{#1}}

\definecolor{darkgreen}{rgb}{0.0,0.6,0.0}

\newcommand{\eps}{\varepsilon}

\def \eps{\varepsilon}

\newcommand{\kepe}{\kappa_{\varepsilon^\prime}}

\def \cLdB1{{{\cal L}_{\Delta B = 1}^{\rm EW}}} 
\def \BR{{\cal B}}                               

\def \One{\leavevmode\hbox{\small1\kern-3.6pt\normalsize1}} 

\newcommand{\muNP}{{\mu_{\Lambda}}}
\newcommand{\muEW}{{\mu_{\rm ew}}}

\def\epe{\varepsilon'/\varepsilon}
\def\kpn{K^+\rightarrow\pi^+\nu\bar\nu}
\def\klpn{K_{L}\rightarrow\pi^0\nu\bar\nu}
\def\ksm{K_S\to\mu\bar\mu}
\def\klm{K_L\to\mu\bar\mu}
\def\klpll{K_L\to\pi^0\ell\bar\ell}
\def\klpee{K_L\to\pi^0 e\bar{e}}
\def\klpmm{K_L\to\pi^0\mu\bar\mu}

\newcommand{\be}{\begin{equation}}
\newcommand{\ee}{\end{equation}}
\newcommand{\tev}{\, {\rm TeV}}
\newcommand{\gev}{\, {\rm GeV}}

\newcommand{\GSM}{{\mathrm{G_{SM}}}}

\newcommand{\SUthreeC}{{\mathrm{SU(3)_c}}}
\newcommand{\SUtwoL}{{\mathrm{SU(2)_L}}}
\newcommand{\UoneY}{{\mathrm{U(1)_Y}}}

\newcommand{\wc}[3][{}]{[{\cal C}_{#2}^{#1}]_{#3}}
\newcommand{\dotwc}[3][{}]{[\dot{\cal C}_{#2}^{#1}]_{#3}}

\newcommand{\Wc}[2][{}]{{\cal C}_{#2}^{#1}}

\newcommand{\dotWc}[2][{}]{\dot{\cal C}_{#2}^{#1}}

\newcommand{\Op}[2][{}]{{\cal O}_{#2}^{#1}}
\newcommand{\op}[3][{}]{[{\cal O}_{#2}^{#1}]_{#3}}

\newcommand \oL[1]{{\overline{#1}}}
\newcommand \wTil[1]{{\widetilde{#1}}}

\begin{document}


\vspace{-14mm}
\begin{flushright}
  TUM-HEP-1114/17
\end{flushright}

\vspace{4mm}

\begin{center}
{\Large\bf\boldmath
  Leptoquarks meet $\epe$ and rare Kaon processes}
\\[8mm]
{\large\bf Christoph Bobeth and  Andrzej~J.~Buras}
\\[0.5cm]
{\small 
  TUM Institute for Advanced Study, Lichtenbergstr.~2a, D-85748 Garching, Germany \\
  Physik Department, TU M\"unchen, James-Franck-Stra{\ss}e, D-85748 Garching, Germany \\
  Excellence Cluster Universe, Technische Universit\"at M\"unchen,
  Boltzmannstr. 2, D-85748 Garching, Germany

}
\end{center}

\vspace{4mm}

\begin{abstract}
\noindent

We analyse for the first time the CP violating ratio $\epe$ in $K\to \pi\pi$
decays in leptoquark (LQ) models. Assuming a mass gap to the electroweak (EW)
scale, the main mechanism for LQs to contribute to $\epe$ is EW gauge-mixing of
semi-leptonic into non-leptonic operators, which we treat in the Standard Model
effective theory (SMEFT). We perform also the one-loop decoupling for scalar
LQs, finding that in all models with both left-handed and right-handed LQ
couplings box-diagrams generate numerically strongly enhanced EW-penguin
operators $Q_{8,8'}$ already at the LQ scale.  We then investigate correlations
of $\epe$ with rare Kaon processes ($\klpn$, $\kpn$, $\klpll$, $\ksm$,
$\Delta M_K$ and $\eps_K$) and find that even imposing only a moderate
enhancement of $(\epe)_{\rm NP} = 5 \times 10^{-4}$ to explain the current
anomaly hinted by the Dual QCD approach and RBC-UKQCD lattice QCD calculations
leads to conflicts with experimental upper bounds on rare Kaon processes. They
exclude all LQ models with only a single coupling as an explanation of the
$\epe$ anomaly and put strong-to-serious constraints on parameter spaces of the
remaining models. Future results on $\kpn$ from the NA62 collaboration, $\klpn$
from the KOTO experiment and $\ksm$ from LHCb will even stronger exhibit the
difficulty of LQ models in explaining the measured $\epe$, in case the $\epe$
anomaly will be confirmed by improved lattice QCD calculations. Hopefully also
improved measurements of $\klpll$ decays will one day help in this context.

\end{abstract}

\setcounter{page}{0}
\thispagestyle{empty}
\newpage

\tableofcontents

\newpage

%
%
%

\section{Introduction}

Leptoquarks (LQs) are very special new particles as they couple directly quarks
to leptons and consequently carry both baryon and lepton number, $B$ and $L$.
Moreover, they are strongly interacting, carry fractional electric charges but
in contrast to quarks they are bosons: either scalars or vectors
\cite{Buchmuller:1986zs, Davies:1990sc, Davidson:1993qk, Dorsner:2016wpm}.
Because of these rather special properties of LQs the phenomenological
implications of models containing them are markedly different than the ones
where new scalars and gauge bosons couple directly only leptons to leptons and
quarks to quarks.

Indeed quite generally the pattern of flavour violations within LQ models 
is as follows.
\begin{itemize}
\item The semi-leptonic and leptonic decays of mesons are privileged as in these
  models they can naturally appear already at tree-level. This applies in
  particular to many rare decays which are loop suppressed within the
  SM. Therefore, the presence of departures from SM expectations for such decays
  can naturally be explained in some LQ models.
\item On the other hand all non-leptonic decays of mesons and also purely
  leptonic processes are loop-suppressed within LQ models.
\item In consequence, large LQ effects in semi-leptonic and leptonic decays of
  mesons do not necessarily imply large modifications of SM predictions for
  non-leptonic observables for which the SM, with few notable exceptions
  discussed below, offers a good description of the data. On the other hand in
  view of a very strong suppression of FCNCs in purely leptonic processes within
  the SM, still large LQ effects in these processes can be found in spite of
  their loop suppression providing thereby strong constraints on LQ models.
\item Moreover, it should be emphasized that in LQ models in which the only new
  particles are LQs, the fermions exchanged in the loops are leptons in
  non-leptonic meson decays but SM quarks in the case of purely leptonic
  processes. Thus these processes are sensitive to the sum over all LQ couplings
  of the particles in the loop.
\end{itemize}

Most flavour analyses of LQs in the literature concentrated on semi-leptonic
decays of mesons, purely leptonic processes and $B^0_{s,d}-\bar B^0_{s,d}$
mixing.  We refer to selected recent papers \cite{Hiller:2017bzc,
  Dorsner:2017ufx, Crivellin:2017zlb, Buttazzo:2017ixm, Calibbi:2017qbu,
  DiLuzio:2017vat} for further references to a very rich literature.  In this
context, the loop-suppression of LQ contributions to non-leptonic transitions is
certainly useful for those non-leptonic observables for which the SM predictions
agree well with data as strong constraints on LQ parameters from them can be
avoided making the explanation of $B$ physics anomalies easier.

In the present paper we will address the ratio $\epe$ and its correlation with
rare Kaon decays, which to our knowledge has never been studied in LQ
models. This is motivated by the recent results on $\epe$ from lattice QCD
\cite{Blum:2015ywa, Bai:2015nea} and Dual QCD large $N$ approach
\cite{Buras:2015xba, Buras:2016fys} that have shown the emerging anomaly in
$\epe$ \cite{Buras:2015yba} with its value in the SM being significantly below
the experimental world average from NA48 \cite{Batley:2002gn} and KTeV
\cite{AlaviHarati:2002ye, Abouzaid:2010ny} collaborations. This finding has been
confirmed in \cite{Kitahara:2016nld}.  We will assume a mass gap between the new
physics scale of the order of the LQ mass $\muNP \sim M_{\rm LQ}$ and the
electroweak (EW) scale $\muEW$, which is conveniently done in the framework of
Standard Model effective theory (SMEFT). In our analysis of LQ models we will
assume nonvanishing contributions to $\epe$ to explain the anomaly, and find
that it would automatically imply too large deviations in $\klpn$, $\kpn$,
$\ksm$ and $\klpll$ decays.  We will also point out the important role of the
$K_L-K_S$ mass difference, $\Delta M_K$, even if CP-conserving, and also of
$\varepsilon_K$ for such analyses.

While our paper will deal in details with $\epe$, rare Kaon decays, $\Delta M_K$
and $\varepsilon_K$ in LQ models, we will present first general formulae in the
framework of SMEFT for the interplay of semi-leptonic operators, who's Wilson
coefficients receive tree-level contributions in these models, with non-leptonic
operators that are generated through renormalisation group (RG) effects at
one-loop level. This should allow in the future to analyse systematically $\epe$
in other models with a similar pattern of flavour violation.

It should also be emphasized that the $\epe$ anomaly is a challenge for those
analyses of $B$-physics anomalies in which all NP couplings have been chosen to
be real and those to the first generation set to zero. It should also be
realised that the anomalies $R_D$ and $R_{D^*}$ being very significant can be in
LQ models explained through a tree-level exchange, while the $\epe$ anomaly,
being even larger, if the bound on $\epe$ in \cite{Buras:2015xba, Buras:2016fys}
is assumed, can only be addressed in these models at one-loop level.  This shows
in a different manner that the hinted $\epe$ anomaly is a big challenge for LQ
models.

The outline of our paper is as follows. In Section~\ref{sec:2} we will summarise
briefly the present status of $\epe$ and we will list the relevant operators
contributing to it. We will also briefly discuss $\klpn$, $\kpn$, $\ksm$,
$\klpll$, $\Delta M_K$ and $\varepsilon_K$. In \refsec{sec:SMEFT} we will first
list all semi-leptonic and non-leptonic operators in the SMEFT that are relevant
for our paper and provide for the most important ones the results of decoupling
of LQs. Subsequently we will list RG equations that are responsible for the
generation of the Wilson coefficients of four-quark operators in SMEFT from
semi-leptonic four-fermion operators. This will allow us to calculate the Wilson
coefficients of standard low-energy operators of \refsec{sec:2} in terms the
Wilson coefficients of SMEFT semi-leptonic operators.  We further provide the
one-loop decoupling of scalar LQs in \refsec{sec:SMEFT}.  Having these equations
we will in \refsec{sec:4} analyse $\epe$ in all LQ models taking constraints
from the processes listed above into account. Our main results are given in
numerous figures that demonstrate strong correlations of LQ contributions in
$\epe$ and rare Kaon processes in these models, and showing in part strong
conflicts with existing experimental bounds. In \refsec{sec:5} we summarise the
results of our analysis, also high-lighted in \reftab{tab:LQ-models}, and
provide further remarks.  Useful details on LQ models and $\epe$ are collected
in four appendices.

%
%
%

\section{\boldmath 
  Preliminaries on $\epe$ and rare Kaon decays}
\label{sec:2}

After a short recollection of $\epe$ itself, we provide in this section basic
information on rare decays $K\to \pi\nu\bar\nu$, $K_L \to \pi \ell\bar\ell$ and
$\ksm$ that turn out to be strongly related to $\epe$ in LQ scenarios.
Subsequently we continue with $\Delta M_K$ and $\eps_K$, which place also strong
constraints on LQ models in the presence of $\epe$ anomaly.

%
%

\subsection{\boldmath $\epe$}

In our work, we consider only the operators that are part of the SM effective
Hamiltonian given in \refeq{eq:Heffeprime} and their chirality-flipped
analogues, with the following operators

{\bf QCD--Penguins:}
\begin{equation}
  \label{eq:QCD-peng-op}
\begin{aligned}
  Q_3 & = (\bar s d)_{V-A}\!\!\sum_{q=u,d,s,c,b}(\bar qq)_{V-A} , &   
  Q_4 & = (\bar s_{\alpha} d_{\beta})_{V-A}\!\!\sum_{q=u,d,s,c,b}(\bar q_{\beta} q_{\alpha})_{V-A} , 
\\
  Q_5 & = (\bar s d)_{V-A}\!\!\sum_{q=u,d,s,c,b}(\bar qq)_{V+A} , &
  Q_6 & = (\bar s_{\alpha} d_{\beta})_{V-A}\!\!\sum_{q=u,d,s,c,b}
      (\bar q_{\beta} q_{\alpha})_{V+A} ,
\end{aligned}
\end{equation}

{\bf Electroweak Penguins:}
\begin{equation}
  \label{eq:QED-peng-op}
\begin{aligned}
  Q_7 & = \frac{3}{2}\,(\bar s d)_{V-A}\!\!\sum_{q=u,d,s,c,b} e_q\,(\bar qq)_{V+A} , &
  Q_8 & = \frac{3}{2}\,(\bar s_{\alpha} d_{\beta})_{V-A}\!\!\sum_{q=u,d,s,c,b}
      e_q\,(\bar q_{\beta} q_{\alpha})_{V+A} ,
\\
  Q_9 & = \frac{3}{2}\,(\bar s d)_{V-A}\!\!\sum_{q=u,d,s,c,b}e_q\,(\bar q q)_{V-A} , &
  Q_{10} & =\frac{3}{2}\, (\bar s_{\alpha} d_{\beta})_{V-A}\!\!\sum_{q=u,d,s,c,b}e_q\,
       (\bar q_{\beta}q_{\alpha})_{V-A} .
\end{aligned}
\end{equation}
Here, $\alpha,\beta$ denote colour indices and $e_q$ the electric quark
charges reflecting the electroweak origin of $Q_7,\ldots,Q_{10}$. Finally,
$(\bar sd)_{V\pm A}\equiv \bar s_\alpha\gamma_\mu(1\pm\gamma_5) d_\alpha$. 

The dominant contributions to $\epe$ at the low-energy scale come from $Q_6$ and
$Q_8$ operators for which the matrix elements are known from RBC-UKQCD
collaboration \cite{Bai:2015nea, Blum:2015ywa}. Using these results and
including isospin breaking corrections one finds \cite{Buras:2015yba}
\begin{align}
  \label{eq:epe-LBGJJ}
  (\epe)_\text{SM} & = (1.9 \pm 4.5) \times 10^{-4} ,
\end{align}
with a similar result obtained in \cite{Blum:2015ywa,Kitahara:2016nld}. The
lattice results for hadronic matrix elements of $Q_6$ and $Q_8$ are supported by
the calculations in the Dual QCD approach \cite{Buras:2015xba, Buras:2016fys}
from which one finds the upper bound
\begin{align}
  \label{eq:epe-BoundBGJJ}
  (\epe)_\text{SM} & \le (6.0\pm 2.4) \times 10^{-4} .
\end{align}

All these results are far below the world average from NA48 \cite{Batley:2002gn}
and KTeV \cite{AlaviHarati:2002ye, Abouzaid:2010ny} collaborations
\begin{align}
  \label{eq:epe-EXP}
  (\epe)_\text{exp}=(16.6\pm 2.3)\times 10^{-4} 
\end{align}
hinting for the presence of new physics in $\epe$ so that we can write
\cite{Buras:2015jaq}
\begin{align}
  \label{eq:epe-SM+NP}
  \frac{\varepsilon'}{\varepsilon} &
  = \left(\frac{\varepsilon'}{\varepsilon}\right)_{\rm SM}
  + \left(\frac{\varepsilon'}{\varepsilon}\right)_{\rm NP} , & 
  \left(\frac{\varepsilon'}{\varepsilon}\right)_{\rm NP} &
  \equiv \kepe \times 10^{-3} , &   
  0.5 & \le \kepe \le 1.5 .
\end{align}

It should be emphasized that present lattice results in
\cite{Bai:2015nea,Blum:2015ywa} are not accurate enough to claim the presence of
new physics in $\epe$ with high confidence. It is rather the bound from the Dual
QCD approach \cite{Buras:2015xba, Buras:2016fys} in \refeq{eq:epe-BoundBGJJ}
that gives us the strongest motivation for this analysis.

As stressed in \cite{Buras:2015jaq} the only NP scenarios that have a chance to
provide such a large upward shift in $\epe$ are those which can modify
significantly the Wilson coefficients of {$(V\mp A)\otimes(V\pm A)$ operators
$Q_{6,6'}$ and $Q_{8,8'}$ at the low scale $\mu \approx m_c$. Yet it is often
sufficient that one of the operators $Q_{5,5',7,7'}$ is generated at the
electroweak or higher energy scale. Then RG evolution to the low scale $\mu$ at
which hadronic matrix elements are evaluated generate subsequently contributions
of $Q_{6,6',8,8'}$ to $\epe$, respectively. In \refsec{sec:SMEFT} we will find
that LQ models can generate $Q_{7,5'}$ at the intermediate electroweak scale
$\muEW$ via EW gauge-mixing of semi-leptonic into non-leptonic operators, and
that one-loop box-diagram contributions can generate $Q_{6,6',8,8'}$ at the very
LQ scale $\muNP$ in LQ models with left-handed and right-handed LQ couplings.
Useful phenomenological expressions for $\epe$, derived in \cite{Bobeth:2016llm}
and extended here, are collected in \refapp{app:d->dqq}. As can be seen in
\refeq{eq:epe-seminum}, $\epe$ depends on the imaginary part of the Wilson
coefficients and hence on the imaginary parts of the underlying fundamental
couplings.

Several examples of NP scenarios that are able to provide sufficient upward
shift in $\epe$ are presented in \cite{Buras:2015jaq}. These include in
particular tree-level $Z^\prime$ exchanges with explicit realisation in 331
models \cite{Buras:2015kwd, Buras:2016dxz} or models with tree-level $Z$
exchanges \cite{Bobeth:2017xry, Endo:2016tnu} with explicit realisation in
models with mixing of heavy vector-like fermions with ordinary fermions
\cite{Bobeth:2016llm} and Littlest Higgs model with T-parity
\cite{Blanke:2015wba}. Also simplified $Z^\prime$ scenarios \cite{Buras:2015yca}
are of help here. But the interest in studying LQ models in this context is
their ability in the explanations of $B$-physics anomalies \cite{Hiller:2017bzc,
  Dorsner:2017ufx, Crivellin:2017zlb, Buttazzo:2017ixm, Calibbi:2017qbu,
  DiLuzio:2017vat}, which some of the scenarios listed above are not able to do.

%

\subsection{\boldmath 
  $\kpn$, $\klpn$,  $\klpll$ and $\ksm$
  \label{sec:DF1-semileptonic-EFT}
}

As will be explained in \refsec{sec:SMEFT} in more detail, EW gauge mixing of
semi-leptonic operators into non-leptonic ones in SMEFT gives rise to
correlations between $\epe$ and observables in semi-leptonic Kaon
decays. Especially the branching fractions of $\klpn$, $\klpll$ and $\ksm$ are
highly sensitive to imaginary parts of Wilson coefficients, with additional
constraints from $\kpn$.  But as pointed out in \cite{Buras:2015jaq} also
$\Delta M_K$, even if CP-conserving, depends sensitively on these imaginary
parts.

The rare decays in question are described by the general $\Delta F = 1$
Hamiltonian of the semi-leptonic FCNC transition of down-type quarks into
leptons and neutrinos below $\muEW$
\begin{align}
  \label{eq:DF1-eff-H}
  {\cal H}_{d\to d(\ell \ell,\nu\nu)} & =
  -\frac{4 G_F}{\sqrt{2}} \lambda^{ji}_t \frac{\alpha_e}{4 \pi}
   \sum_k C_k^{baji} Q_k^{baji} + \mbox{h.c.} \, 
\end{align}
with $a,b$ being lepton indices, $i,j$ down-quark indices and
\begin{align}
  \lambda_u^{ji} & \equiv V^\ast_{uj} V^{}_{ui} , &
  u = \{u, c, t\} .
\end{align}
There are eight semi-leptonic operators relevant for $d_i \ell_a \to d_j \ell_b$
when considering UV completions that give rise to SMEFT above the electroweak
scale \cite{Alonso:2014csa}
\begin{equation}
\begin{aligned}
  Q_{9(9')}^{baji} &
  = [\bar{d}_j \gamma_\mu P_{L(R)} d_i] [\bar{\ell}_b \gamma^\mu \ell_a] ,
& \qquad
  Q_{10(10')}^{baji} &
  = [\bar{d}_j \gamma_\mu P_{L(R)} d_i] [\bar{\ell}_b \gamma^\mu \gamma_5 \ell_a] ,
\\
  Q_{S(S')}^{baji} &
  = [\bar{d}_j P_{R(L)} d_i] [\bar{\ell}_b \ell_a] ,
& \qquad
  Q_{P(P')}^{baji} &
  = [\bar{d}_j P_{R(L)} d_i] [\bar{\ell}_b \gamma_5 \ell_a] ,
\end{aligned}
\end{equation}
and two for $d_i \nu_a\to d_j \nu_b$
\begin{align}
  Q_{L(R)}^{baji} &
  = [\bar{d}_j \gamma_\mu P_{L(R)} d_i] [\bar{\nu}_b \gamma^\mu (1-\gamma_5) \nu_a] .
\end{align}
The SM contribution to these Wilson coefficients is lepton-flavour diagonal
\begin{align}
  C_k^{baji} & 
  = C_{k,{\rm SM}} \, \delta_{ba}
  + \frac{\pi}{\alpha_e} \frac{v^2}{\lambda_t^{ji}} \, C_{k,{\rm NP}}^{baji}  
\end{align}
where $v=246\gev$ and a normalisation factor has been introduced for the NP
contribution that proves convenient for later matching with SMEFT in
\refsec{sec:semi-lept-SMEFT-matching}. The non-vanishing SM contributions
\begin{align}
  C_{9,{\rm SM}} &
  = \frac{Y(x_t)}{s_W^2} - 4 Z(x_t) , &
  C_{10,{\rm SM}} &
  = - \frac{Y(x_t)}{s_W^2} , &
  C_{L,{\rm SM}} &
  = - \frac{X(x_t)}{s_W^2} ,
\end{align}
are given by the gauge-independent functions $X(x_t)$, $Y(x_t)$ and
$Z(x_t)$~\cite{Buchalla:1990qz} that depend on the ratio
$x_t \equiv m_t^2/m_W^2$ of the top-quark and $W$-boson masses. Here
$s_W \equiv \sin \theta_W$.


\subsubsection{\boldmath $K\to \pi \nu\bar\nu$
  \label{sec:Kpivv-basics}
}

The branching fractions of the $K\to \pi\nu\bar\nu$ modes involve a sum over all
lepton flavours of the neutrinos in the final state
\begin{align}
  \label{eq:Br-klpn}
  \BR(\klpn) & =
  \frac{\kappa_L }{3\,\lambda^{10}} \sum_{a,b}
  \mbox{Im}^2 \Big(\lambda^{sd}_t X_{t}^{ab} \Big) ,                         
\\
  \label{eq:Br-kpn}
  \BR(\kpn) & =
  \frac{\kappa_+ (1 + \Delta_{\rm EM})}{3\,\lambda^{10}} \sum_{a,b}
  \left[  \mbox{Im}^2 \Big(\lambda^{sd}_t X_{t}^{ab} \Big)
        + \mbox{Re}^2 \Big(\lambda^{sd}_c X_{c}^{aa}
                         + \lambda^{sd}_t X_{t}^{ab} \Big)\right] ,
\end{align}
with more details on these general formula in app. C.2 of \cite{Bobeth:2016llm}.
As mentioned above, $\BR(\klpn)$ is especially sensitive to imaginary parts of
couplings. The LQ tree-level exchange contributes to the short-distance quantity
$X_t^{ab} \equiv X(x_t) \delta_{ab} + X_{\rm LQ}^{ab}$ with the lepton-flavour
diagonal SM contribution \cite{Buchalla:1993bv, Misiak:1999yg, Buchalla:1998ba,
  Brod:2010hi}
\begin{align}
  X(x_t) &
  = 1.481 \pm 0.009
\end{align}
as extracted in \cite{Buras:2015qea} from original papers. The charm contribution
$X_c^{aa}$ of the SM in $\kpn$ is also lepton-flavour diagonal. The LQ contribution
enters the Wilson coefficients at $\muEW$ as
\begin{align}
  X_{\rm LQ}^{ab} &
  = - s_W^2 v^2 \frac{\pi}{\alpha_e} 
    \frac{\left( C_{L,{\rm NP}}^{basd} + C_{R,{\rm NP}}^{basd} \right)}{\lambda_t^{sd}} 
\end{align}
with matching conditions to SMEFT in \refeq{eq:SMEFT-matching-ddllvv}.
The SM predictions \cite{Bobeth:2016llm}
\begin{align}
  \label{eq:SM-Br-klpn}
  \BR(\klpn)_{\rm SM} & = \left( 3.2^{+1.1}_{-0.7} \right) \times 10^{-11} ,
\\
  \label{eq:SM-Br-kpn}
  \BR(\kpn)_{\rm SM}  & = \left( 8.5^{+1.0}_{-1.2} \right) \times 10^{-11}  
\end{align}
can be confronted with the current upper bound \cite{Artamonov:2008qb}
\begin{align}
  \label{eq:exp-Br-klpn}
  \BR(\klpn)_{\rm exp} & <  2.6 \times 10^{-8} ,
\end{align}
and the measurement \cite{Artamonov:2008qb}
\begin{align}
  \label{eq:exp-Br-kpn}
  \BR(\kpn)_{\rm exp} & = \left( 17.3^{+11.5}_{-10.5} \right) \times 10^{-11} .
\end{align}
The latter measurement can be converted into an upper bound on
$\BR(\klpn) < 1.45 \times 10^{-9}$, known as Grossman-Nir bound
\cite{Grossman:1997sk}, which is stronger than the current experimental upper
bound \refeq{eq:exp-Br-klpn}.

We elaborate a bit more on the specific structure of the branching fractions by
expanding the imaginary parts
\begin{equation}
\begin{aligned}
  \label{eq:Br-klpn-NP}
  \BR(\klpn) & = \BR(\klpn)_{\rm SM} 
\\ &
  + \frac{\kappa_L}{\lambda^{10}} \frac{1}{3} \left[
    2\, \mbox{Im} (\lambda_t^{sd} X_{\rm SM}) 
    \sum_{a}   \mbox{Im}  (\lambda_t^{sd}X_{\rm LQ}^{aa})
  + \sum_{a,b} \mbox{Im}^2(\lambda_t^{sd}X_{\rm LQ}^{ab}) \right].                         
\end{aligned}
\end{equation}
The interference of the SM$\times$NP in the second line can be constructive or
destructive, whereas the NP$\times$NP contribution is purely constructive to the
SM contribution. It is customary to neglect the NP$\times$NP contribution since
it is formally suppressed by $v^2/\muNP^2$ w.r.t. SM$\times$NP, but in view of
the fact that the upper experimental bound \refeq{eq:exp-Br-klpn} on
$\BR(\klpn)$ is orders above the SM prediction \refeq{eq:SM-Br-klpn}, it turns
out to be the by far dominant contribution and we keep it here.  The expression
for $\BR(\kpn)$ receives analogous additional terms of the real parts of
$X_{\rm LQ}^{ab}$ that are in principle independent parameters, but in this case
the experimental constraint \refeq{eq:exp-Br-kpn} is much more
stringent. Moreover, the SM$\times$NP real parts can interfere constructively or
destructively with the SM and imaginary LQ parts. Yet, these effects are
naturally cut off due to the presence of the constructive NP$\times$NP
contribution of the real parts themselves for large couplings. In addition,
similar to $\klpn$ the constructive NP$\times$NP contribution from imaginary
parts will play the most important role in our analysis.


\subsubsection{\boldmath $\klpll$}

Generalising the formulae in \cite{Buchalla:2003sj, Isidori:2004rb, Friot:2004yr,
Mescia:2006jd} to include NP contributions and adapting them to our notations 
we find 
\begin{align}
  \label{eq:BrKpiLL}
  \BR(\klpll) & 
  = \left( C_\text{dir}^\ell \pm C_\text{int}^\ell \left|a_s\right|
         + C_\text{mix}^\ell \left|a_s\right|^2 
         + C_\text{CPC}^\ell \right)\times 10^{-12} ,
\end{align}
where \cite{Mescia:2006jd}
\begin{equation}
  \label{eq:klpll-num-1}
\begin{aligned}
  C_\text{dir}^e   & = (4.62\pm0.24)[(\omega_{7V}^e)^2 + (\omega_{7A}^e)^2] ,  & \qquad
  C_\text{int}^e   & = (11.3\pm0.3) \, \omega_{7V}^e , &
\\
  C_\text{dir}^\mu & = (1.09\pm0.05)[(\omega_{7V}^\mu)^2 + 2.32 (\omega_{7A}^\mu)^2] , &
  C_\text{int}^\mu & = (2.63\pm0.06) \, \omega_{7V}^\mu ,
\end{aligned}
\end{equation}
and
\begin{equation}
  \label{eq:klpll-num-2}
\begin{aligned}
  C_\text{mix}^e   & = 14.5\pm0.05 , & \qquad
  C_\text{CPC}^e   & \simeq 0 , &      \qquad
  \left|a_s\right| & = 1.2 \pm0.2 ,
\\
  C_\text{mix}^\mu & = 3.36\pm0.20 , &
  C_\text{CPC}^\mu & = 5.2\pm1.6 . &
\end{aligned}
\end{equation}
The SM and NP contributions enter through
\begin{align}
  \omega_{7V}^\ell &
  = \frac{1}{2\pi}\left(P_0+ C_{9,{\rm SM}} \right) 
    \left[ \frac{{\rm Im}\lambda_t^{sd}}{1.407 \times 10^{-4}} \right]
  + \frac{1}{\alpha_e} \frac{v^2}{2} \frac{{\rm Im} \left[ 
          C_{9,{\rm NP}}^{\ell\ell sd} + C_{9',{\rm NP}}^{\ell\ell sd}
    \right]}{1.407 \times 10^{-4}},
\\
  \omega_{7A}^\ell &
  = \frac{1}{2\pi} C_{10,{\rm SM}}
    \left[ \frac{{\rm Im}\lambda_t^{sd}}{1.407 \times 10^{-4}} \right]
  + \frac{1}{\alpha_e} \frac{v^2}{2} \frac{{\rm Im} \left[ 
          C_{10,{\rm NP}}^{\ell\ell sd} + C_{10',{\rm NP}}^{\ell\ell sd}
    \right]}{1.407 \times 10^{-4}}
\end{align}
where $P_0=2.88\pm 0.06$ \cite{Buras:1994qa} and $\ell$ either $e$ or $\mu$.

The expressions for $C_{9,{\rm NP}}^{\ell\ell sd}$ and
$C_{10,{\rm NP}}^{\ell\ell sd}$ in terms of SMEFT coefficients are given
in~\refeq{eq:SMEFT-matching-ddllvv}.  NP contributions do not depend on
$\lambda^{sd}_t$ but the factor $1.407 \times 10^{-4}$ is present because it has
been used in \cite{Mescia:2006jd} to obtain the numbers in
\refeq{eq:klpll-num-1} and \refeq{eq:klpll-num-2}.

The present experimental bounds
\begin{align}
  \label{KLee}
  \BR(\klpee)_{\rm exp} & 
    < 28 \times 10^{-11} , & \text{\cite{AlaviHarati:2003mr}}
\\
  \label{KLmm}
  \BR(\klpmm)_{\rm exp} &
    < 38 \times 10^{-11} , & \text{\cite{AlaviHarati:2000hs}}
\intertext{are still by one order of magnitude larger than the
SM predictions \cite{Mescia:2006jd}}
  \label{eq:KLpee}
  \BR(\klpee)_\text{SM} &
  = 3.54^{+0.98}_{-0.85}\left(1.56^{+0.62}_{-0.49}\right) \times 10^{-11}\,,
\\
  \label{eq:KLpmm}
  \BR(\klpmm)_\text{SM} &
  = 1.41^{+0.28}_{-0.26}\left(0.95^{+0.22}_{-0.21}\right) \times 10^{-11}
\end{align}
with the values in parentheses corresponding to the ``$-$'' sign in
\refeq{eq:BrKpiLL}, that is the destructive interference between directly and
indirectly CP-violating contributions. The last discussion of the theoretical
status of this interference sign can be found in \cite{Prades:2007ud} where the
results of \cite{Isidori:2004rb, Friot:2004yr, Bruno:1992za} are critically
analysed.  From this discussion, constructive interference seems to be favoured
though more work is necessary. We will therefore use this constructive
interference in our numerical calculations. However, when the constraint
$\kepe\ge 0.5$ will be imposed, in LQ models NP contributions present in
directly CP violating contributions will by far dominate the branching ratios
and the sign in question will not matter.


\subsubsection{\boldmath $\ksm$}

The decay $\ksm$ provides another sensitive probe of imaginary parts of
short-distance couplings. Its branching fraction receives long-distance (LD) and
short-distance (SD) contributions, which are added incoherently in the total
rate \cite{Ecker:1991ru, Isidori:2003ts}. This is in contrast to the decay
$\klm$, where LD and SD amplitudes interfere and moreover $\BR(\klm)$ is
sensitive to real parts of couplings. The SD part of $\BR(\ksm)$ is given as
\begin{equation}
  \label{eq:ksm-br-SD}
\begin{aligned}
  \BR(\ksm)_{\rm SD} &
  = \tau_{K_S} \frac{G_F^2 \alpha_e^2}{8 \pi^3} m_K f_K^2 \beta_\mu m_\mu^2
\\ & \times
    \mbox{Im}^2 \left[\lambda_t^{sd}  C_{10,{\rm SM}} + 
     \frac{\pi}{\alpha_e} v^2\left(C_{10,{\rm NP}}^{\mu\mu sd} 
                                 - C_{10',{\rm NP}}^{\mu\mu sd} \right) \right] .
\end{aligned}
\end{equation}

Recently the LHCb collaboration improved the upper bound on $\ksm$
by one order of magnitude \cite{Aaij:2017tia}
\begin{align}
  \label{ksmbound}
  \BR(\ksm)_{\rm LHCb} & < 0.8\, (1.0) \times 10^{-9} &
  \mbox{at} \; 90\%\, (95\%) \; \mbox{C.L.}
\end{align}
to be compared with the SM prediction \cite{Isidori:2003ts, DAmbrosio:2017klp}
\begin{align}
  \BR(\ksm)_{\rm SM} &
  = (4.99_{\rm LD} + 0.19_{\rm SD}) \times 10^{-12} 
  = (5.2 \pm 1.5) \times 10^{-12}.
\end{align}
While this bound is still by two orders of magnitude above its SM value, it
turns out that for several LQ models, even the saturation of this bound would
barely remove the $\epe$ anomaly, provided it is due to the muonic LQ
couplings. There are good future prospects to improve this bound, LHCb expects
\cite{Dettori:2017} with 23~fb$^{-1}$ sensitivity to regions
$\BR(\ksm) \in [4,\, 200] \times 10^{-12}$, close to the SM prediction.

%

\subsection{\boldmath 
  $\Delta M_K$ and $\eps_K$ 
  \label{sec:DeltaF2-obs-basics}
}

We will next investigate whether additional constraints on LQ models come from
$\Delta S = 2$ transitions, that is $\eps_K$ and the $K_L-K_S$ mass difference
$\Delta M_K$. At first sight one would think that it is $\eps_K$ which is more
important as similar to $\epe$ it is related to CP violation, while $\Delta M_K$
is CP-conserving.  Yet as pointed out in \cite{Buras:2015jaq} when NP is
required to have significant imaginary couplings, it is $\Delta M_K$ and not
$\eps_K$ which is directly correlated with $\epe$. The point is that $\eps_K$
is governed by the imaginary part of the square of complex couplings and
consequently is governed by the product of real and imaginary couplings. As
$\epe$ sets the constraint only on imaginary couplings, the $\eps_K$ constraint
can be removed by simply choosing the couplings to be purely imaginary. Of
course it could turn out one day that some amount of NP in $\eps_K$ is required
as suggested in \cite{Blanke:2016bhf}, but even in this case choosing real
couplings to be sufficiently small one can obtain agreement with experiment.

But $\Delta M_K$ is governed by the real part of the square of complex couplings
and consequently is governed by the difference between the squares of real and
imaginary couplings. This difference cannot be zero as in the presence of large
real and imaginary couplings one would violate the $\eps_K$ constraint.
Therefore as analysed in details in certain $Z^\prime$ scenarios in
\cite{Buras:2015jaq} the necessity of large imaginary couplings required by
$\epe$ implies automatically significant {\it negative} NP contributions to
$\Delta M_K$.  In fact this happens also in LQ models.

Let us recall that the experimental value of $\Delta M_K$ is very 
precise \cite{Olive:2016xmw}
\begin{equation}
  \label{DMEXP}
  (\Delta M_K)^{\rm exp} = M(K_L) - M(K_S) 
  = 3.484(6) \times 10^{-15} \gev .
\end{equation}
Presently the contribution of the SM dynamics to $\Delta M_K$ is subject to
large theoretical uncertainties. In the SM $\Delta M_K$ is described by the real
parts of the box diagrams with charm quark and top quark exchanges, whereby the
contribution of the charm exchanges is by far dominant. Unfortunately, the
uncertainties in the short distance QCD corrections to the charm contribution
amount to roughly $\pm 40\%$ with the central value somewhat below the
experimental one \cite{Brod:2011ty}.  Moreover, there are also non-perturbative
long distance contributions that are known to amount to $20\pm 10\%$ of the
measured $\Delta M_K$ \cite{Bijnens:1990mz, Buras:2014maa} when calculated using
the large $N$ approach to QCD. In the future they should be known more precisely
from lattice QCD \cite{Bai:2014cva,Christ:2015pwa}.  For the time being the
rough picture is that box diagrams contribute $80\%$ of the measured
$\Delta M_K$ with the rest given by long distance contributions and possibly new
dynamics beyond the SM. But even if presently we do not know whether these new
dynamics will be required to enhance or suppress the SM prediction to agree with
data, it cannot be larger than roughly $40\%$ of the experimental value.

In LQ models new contributions to $\Delta M_K$ come from box diagrams with LQs
and SM lepton exchanges. As will be seen in our analysis, $\klpll$ decays with
electrons or muons in the final state put already severe constraints on most of
LQ models, such that a simultaneous compliance of this bound and an enhancement
of $\epe$ with $\kepe=0.5$ can take place only through enhanced
$\bar\tau\tau\bar s d$ couplings in models like $U_1$.  Such a coupling has a
direct impact on $\Delta M_K$ through box diagrams with LQs, $\tau$ and in some
models $\nu_\tau$ exchanges and it is of interest to see what are the
implications of the $\epe$ anomaly on $\Delta M_K$ in LQ models.

As far as models with vector LQs are concerned $\Delta M_K$ cannot be reliably
calculated without a UV completion, but just looking at the Dirac structure of
the resulting operators we can anticipate in the corresponding models strong
constraints for the case of simultaneous presence of right-handed and
left-handed couplings as this would generate left-right operators and very large
contributions to $\Delta M_K$ \cite{Buras:2015jaq, Bobeth:2017xry}.  In turn
this will have implications on box contributions to $\epe$ in these models.

The effective Hamiltonian for neutral meson mixing in the down-type quark
sector ($d_i \bar{d}_j \to \bar{d}_i d_j$ with $i\neq j$) can be written
as~\cite{Buras:2000if}
\begin{align}
  \label{eq:DF2-hamiltonian}
  {\cal H}_{\Delta F = 2}^{ji} &
  = {\cal N}_{ji} \sum_a  C_a^{ji} Q_a^{ji} + \mbox{h.c.} , &
  {\cal N}_{ji} & 
  = \frac{G_F^2}{4 \pi^2} M_W^2 \left(\lambda_t^{ji}\right)^2
\end{align}
with $ij = ds$ for kaon mixing and $ij = bd,bs$ for $B_d$ and $B_s$
mixing, respectively.
The set of operators consists out of $(5 + 3) = 8$ operators \cite{Buras:2000if},
\begin{equation}
  \label{eq:DF2-operators}
\begin{aligned}
  Q_{{\rm VLL}}^{ji} & 
  = [\bar{d}_j \gamma_\mu P_L d_i][\bar{d}_j \gamma^\mu P_L d_i] , &
\\[0.2cm]
  Q_{{\rm LR},1}^{ji} & 
  = [\bar{d}_j \gamma_\mu P_L d_i][\bar{d}_j \gamma^\mu P_R d_i] , &
  Q_{{\rm LR},2}^{ji} & 
  = [\bar{d}_j P_L d_i][\bar{d}_j P_R d_i] ,
\\[0.2cm]
  Q_{{\rm SLL},1}^{ji} & 
  = [\bar{d}_j P_L d_i][\bar{d}_j P_L d_i] , &
  Q_{{\rm SLL},2}^{ji} & 
  = -[\bar{d}_j \sigma_{\mu\nu} P_L d_i][\bar{d}_j \sigma^{\mu\nu} P_L d_i] ,
\end{aligned}
\end{equation}
which are built out of colour-singlet currents
$[\bar{d}^\alpha_j \ldots d^\alpha_i] [\bar{d}^\beta_j \ldots d^\beta_i]$, where
$\alpha,\, \beta$ denote colour indices.  The chirality-flipped sectors VRR and
SRR are obtained from interchanging $P_L \leftrightarrow P_R$ in VLL and SLL. In
the SM only
\begin{align}
  \label{eq:DF2:S0}
  C_{{\rm VLL}}^{ji}(\muEW)|_{\rm SM} & = S_0(x_t) \delta^{ji}, &
  S_0(x) & 
  = \frac{x (4 - 11 x + x^2)}{4\, (x-1)^2} + \frac{3 x^3 \ln x}{2\, (x-1)^3}   
\end{align}
is non-zero at the scale $\muEW$.

%
%
%

\section{\boldmath
  Decoupling of Leptoquarks and SMEFT
  \label{sec:SMEFT}
}

Throughout we assume that the masses of LQs $M_{\rm LQ} \sim \muNP$ are much
heavier than the electroweak scale $\muEW \ll \muNP$, that is at least $1\tev$.
Further the UV completion is assumed to be perturbative and that LQs transform
under the SM gauge group $\SUthreeC \otimes \SUtwoL \otimes \UoneY$
\cite{Buchmuller:1986zs}, see \refapp{app:LQ-Lag} for details on LQ models,
conventions and definitions. This allows for a perturbative decoupling of LQs at
$\muNP$ and the use of SMEFT between the scales $\muNP$ and~$\muEW$. In a second
matching step at $\muEW$ the heavy degrees of freedom of the SM ($W$, $Z$, $H$,
$t$) are decoupled and SMEFT is matched onto the low-energy effective theories
of electroweak interactions introduced before in \refsec{sec:2}.  In this
section we list the SMEFT operators that are necessary for our analysis of
$\epe$ and rare Kaon processes, describe the LQ decoupling at tree- and one-loop
level, collect the relevant parts of the RG evolution from $\muNP$ to $\muEW$
and provide the matching onto the low-energy EFTs at $\muEW$.

Essentially, the tree-level decoupling of LQs gives rise to semi-leptonic
four-fermion (SL-$\psi^4$) operators that govern semi-leptonic decays, which
provide the majority of phenomenological constraints on LQ models. The Wilson
coefficients of non-leptonic (NL-$\psi^4$) operators that govern $\epe$ and
other non-leptonic processes will be then generated at $\muEW$ by the RG
evolution of the semi-leptonic coefficients from $\muNP$ to $\muEW$ via
$\SUtwoL \otimes \UoneY$ (EW) gauge-mixing.  We note that the same EW
gauge-mixing generates also leptonic $\psi^4$ operators that govern purely
leptonic processes, as previously discussed in the framework of SMEFT in the
context of violation of lepton flavour universality in $B$ decays
\cite{Feruglio:2016gvd, Feruglio:2017rjo}. LQ models also generate direct
contributions to $\epe$ at $\muNP$ at one-loop level via QCD and EW penguin
diagrams as well as box-type diagrams. It turns out that due to the structure of
$\epe$ the QCD penguins are numerically less relevant as aforementioned EW
gauge-mixing effects of semi-leptonic operators. The same applies to EW penguins
which contribute to the non-logarithmic terms of EW gauge-mixing effects.
Finally box-diagram contributions are relevant in most of the models as
discussed in \refsec{sec:4}.

%
%

\subsection{\boldmath
  Semi- and Non-leptonic Operators in SMEFT
}

The SMEFT Lagrangian at dimension six
\begin{align}
  \label{BASICSMEFT}
  \mathcal L^{(6)} & = \sum_{k} \Wc[(6)]{k} \Op[(6)]{k}
\end{align}
contains operators $\Op[(6)]{k}$ classified in full generality in
\cite{Buchmuller:1985jz, Grzadkowski:2010es}. The Wilson coefficients scale as
$\Wc[(6)]{k} \sim 1/\muNP^2$ with the new physics scale $\muNP \sim M_{\rm LQ}$,
which is for our purposes of the order of the LQ
mass. Ref.~\cite{Grzadkowski:2010es} removed certain redundant operators present
in \cite{Buchmuller:1985jz} and we will use these results here. The
corresponding RG evolution at one-loop of all these operators has been
calculated in \cite{Jenkins:2013zja, Jenkins:2013wua, Alonso:2013hga} with the
evolution governed by the Higgs self-coupling $\lambda$ \cite{Jenkins:2013zja},
Yukawa couplings \cite{Jenkins:2013wua} and SM gauge interactions
\cite{Alonso:2013hga}. In the present paper only the results from
\cite{Alonso:2013hga} will be relevant and we will recall them below in
\refsec{sec:SMEFT-RGE}.

We use the following notation for Wilson coefficients and operators in the
corresponding effective theories:
\begin{equation}
\begin{aligned}
  \mbox{SMEFT:} & &             {\cal L}_{\rm SMEFT}  
  & \sim \Wc{a} \Op{a}\,,
\\
  \Delta F = 1, 2\;\; \mbox{low-energy EFTs:} 
  & & {\cal H}_{\Delta F=1,2} & \sim C_{a} Q_{a}\,.
\end{aligned}
\end{equation}
Note the use of the Lagrangian ${\cal L}$ for SMEFT, but the Hamiltonian
${\cal H}$ for the low-energy EFTs of $\Delta F=1,2$ processes.\footnote{Note
  the relative sign ${\cal L} = - {\cal H}$ that has to be kept in mind when
  deriving formulae below.}

\begin{table}[t]
\centering
\renewcommand{\arraystretch}{1.4}
\resizebox{\textwidth}{!}{
\begin{tabular}{|c||c|c|c||c|c|}
\hline
& \multicolumn{3}{c||}{semi-leptonic (SL-$\psi^4$)} 
& \multicolumn{2}{c||}{non-leptonic (NL-$\psi^4$)}
\\
\hline\hline
  \multirow{2}{*}{$(\oL{L}L)(\oL{L}L)$} 
& $\Op[(1)]{\ell q}$ & $(\bar \ell_p \gamma_\mu \ell_r)(\bar q_s \gamma^\mu q_t)$
& $S_{1, 3}$, $U_{1,3}$
& $\Op[(1)]{qq}$     & $(\bar q_p \gamma_\mu q_r)(\bar q_s \gamma^\mu q_t)$ 
\\
& $\Op[(3)]{\ell q}$ & $(\bar \ell_p \gamma_\mu \tau^I \ell_r)(\bar q_s \gamma^\mu \tau^I q_t)$
& $S_{1, 3}$, $U_{1,3}$
& $\Op[(3)]{qq}$     & $(\bar q_p \gamma_\mu \tau^I q_r)(\bar q_s \gamma^\mu \tau^I q_t)$
\\
\hline
  \multirow{4}{*}{$(\oL{L}L)(\oL{R}R)$} 
& $\Op{\ell u}$      & $(\bar \ell_p \gamma_\mu \ell_r)(\bar u_s \gamma^\mu u_t)$ 
& $R_2$, $\wTil{V}_2$
& $\Op[(1)]{qu}$     & $(\bar q_p \gamma_\mu q_r)(\bar u_s \gamma^\mu u_t)$
\\
& $\Op{\ell d}$      & $(\bar \ell_p \gamma_\mu \ell_r)(\bar d_s \gamma^\mu d_t)$
& $\wTil{R}_2$, $V_2$
& $\Op[(8)]{qu}$     & $(\bar q_p \gamma_\mu T^A q_r)(\bar u_s \gamma^\mu T^A u_t)$ 
\\
& $\Op{qe}$          & $(\bar q_p \gamma_\mu q_r)(\bar e_s \gamma^\mu e_t)$
& $R_2$, $V_2$
& $\Op[(1)]{qd}$     & $(\bar q_p \gamma_\mu q_r)(\bar d_s \gamma^\mu d_t)$
\\
& & &
& $\Op[(8)]{qd}$     & $(\bar q_p \gamma_\mu T^A q_r)(\bar d_s \gamma^\mu T^A d_t)$
\\
\hline
  \multirow{4}{*}{$(\oL{R}R)(\oL{R}R)$} 
& $\Op{eu}$          & $(\bar e_p \gamma_\mu e_r)(\bar u_s \gamma^\mu u_t)$
& $S_1$, $\wTil{U}_1$
& $\Op{uu}$          & $(\bar u_p \gamma_\mu u_r)(\bar u_s \gamma^\mu u_t)$
\\
& $\Op{ed}$          & $(\bar e_p \gamma_\mu e_r)(\bar d_s \gamma^\mu d_t)$
& $\wTil{S}_1$, $U_1$
& $\Op{dd}$          & $(\bar d_p \gamma_\mu d_r)(\bar d_s \gamma^\mu d_t)$
\\
& &
&
& $\Op[(1)]{ud}$     & $(\bar u_p \gamma_\mu u_r)(\bar d_s \gamma^\mu d_t)$
\\
& &
&
& $\Op[(8)]{ud}$     & $(\bar u_p \gamma_\mu T^A u_r)(\bar d_s \gamma^\mu T^A d_t)$
\\
\hline
  $(\oL{L}R)(\oL{R}L)$
& $\Op{\ell edq}$    & $(\bar \ell_p^j e_r)(\bar d_s q_t^j) + \mbox{h.c.}$
& $U_1$, $V_2$
& &
\\
\hline
  $(\oL{L}R)(\oL{L}R)$
& $\Op[(1)]{\ell equ}$ & $(\bar \ell_p^j e_r) \eps_{jk} (\bar q_s^k u_t)$
& $S_1, R_2$ 
& $\Op[(1)]{quqd}$     & $(\bar q_p^j u_r) \eps_{jk} (\bar q_s^k d_t)$
\\
  $+ \mbox{h.c.}$
& $\Op[(3)]{\ell equ}$ & $(\bar \ell_p^j \sigma_{\mu\nu} e_r) \eps_{jk} (\bar q_s^k \sigma^{\mu\nu} u_t)$
& $S_1, R_2$
& $\Op[(8)]{quqd}$     & $(\bar q_p^j T^A u_r) \eps_{jk} (\bar q_s^k T^A d_t)$
\\
\hline
\end{tabular}
}
\renewcommand{\arraystretch}{1.0}
\caption{\label{tab:4ferm}
  Semi- and non-leptonic $\psi^4$ operators in SMEFT. Chirality indices have
  been omitted on $\SUtwoL$ doublet fields $q_L$, $\ell_L$ and singlet fields
  $u_R$, $d_R$, $e_R$, respectively. The $\tau^I$ denote the Pauli matrices and
  $T^A = \lambda^A/2$ the $\SUthreeC$ color generators where $\lambda^A$ are the
  Gell-Mann matrices. The third column for semi-leptonic operators indicates
  which LQ models contribute in the tree-level matching, see 
  \refapp{app:tree-decoupl} for details.
}
\end{table}

There are 10 SL-$\psi^4$ and 12 NL-$\psi^4$ operators in SMEFT
\cite{Grzadkowski:2010es} collected in \reftab{tab:4ferm}. We have omitted flavour
indices on the operators but we will expose them whenever necessary. 
For example flavour indices in the most important $(\oL{L}L)(\oL{R}R)$ non-leptonic 
operators and in their Wilson coefficients are given as follows
\begin{equation}
\begin{aligned}
  \op[(1)]{qu}{prst} & = (\bar{q}_L^p \gamma_\mu q_L^r) (\bar{u}_R^s \gamma^\mu u_R^t), &
  & \qquad \wc[(1)]{qu}{prst} ,
\\
  \op[(1)]{qd}{prst} & = (\bar{q}_L^p \gamma_\mu q_L^r) (\bar{d}_R^s \gamma^\mu d_R^t), &
  & \qquad \wc[(1)]{qd}{prst} .
\end{aligned}
\end{equation}

%
%

\subsection{Tree-level LQ decoupling
  \label{sec:tree-LQ-dcpl}
}
  
The special nature of LQs leads at tree-level, see \reffig{fig:LQ-tree}, only to
SL-$\psi^4$ SMEFT operators in \reftab{tab:4ferm}. The results of a decoupling
at the EW scale $\muEW$ onto the low-energy EFT's governing $\Delta F = 1$
charged- current and FCNC decays of mesons (see
\refsec{sec:DF1-semileptonic-EFT}) are well known and allow easily to infer the
corresponding matching relations for SMEFT SL-$\psi^4$ coefficients at $\muNP$,
summarised in \refapp{app:tree-decoupl}.  The characteristic structure of these
coefficients for a process $Q_i L_a \to Q_j L_b$ are
\begin{align}
  \wc{{\rm SL-}\psi^4}{jiba}(\muNP) & 
  \propto \frac{(Y_{jb}^\chi)^\ast Y_{ia}^{\chi'}}{M_{\rm LQ}^2} , 
\intertext{where $Y^\chi$ ($\chi = L,R$) stand for the Yukawa couplings of LQs to
SM quarks $Q = q_L, u_R, d_R$ and leptons $L = \ell_L, e_R$, depending on the 
specific LQ model, see \refapp{app:LQ-Lag}. As will be shown in detail in 
\refsec{sec:SMEFT-RGE} and \refsec{sec:non-lept-match-SMEFT}, they give rise to
NL-$\psi^4$ coefficients via EW gauge-mixing at $\muEW$}
  \label{eq:NLpsi4-EW-scaling}
  \wc{{\rm NL}-\psi^4}{ji\cdot\cdot}(\muEW) &
  \propto \frac{\alpha_e}{4\pi} \ln \frac{\muNP}{\muEW}
          \frac{\Sigma_{\chi, {\rm LQ}}^{ji}}{M_{\rm LQ}^2} ,
\end{align}
leading to 1) loop-suppression, 2) a logarithmic enhancement and 3) a
characteristic sum $\Sigma_{\chi, {\rm LQ}}^{ji}$ over lepton-flavour indices of
the products of LQ Yukawa couplings with the same chirality $\chi = (L,
R)$. This latter quantity
\begin{align}
  \label{eq:def-Sigma_LQ}
  \Sigma_{\chi,{\rm LQ}}^{ji} & \equiv \sum_a (Y_{ja}^{\chi})^\ast Y_{ia}^\chi  
\end{align}
is central to our analysis since it enters $\epe$ and other processes or
contributions governed at loop-level. They could be responsible for any
potential deviation from the SM prediction for $\epe$.  Moreover, each of the
six couplings ($i\neq j$) entering $\Sigma^{ji}$ leads to correlations between
$\epe$ and other processes that depend on them, among which the most interesting
are those that depend more or less on $\Sigma^{ji}$ itself.

\begin{figure}
  \begin{subfigure}[t]{0.19\textwidth}
    \centering
    \includegraphics[width=0.8\textwidth]{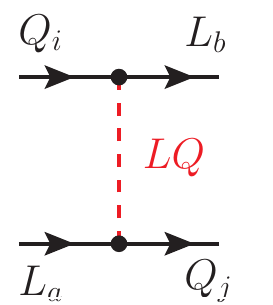}
    \caption{}
    \label{fig:LQ-tree}
  \end{subfigure}
  \begin{subfigure}[t]{0.19\textwidth}
    \centering
    \includegraphics[width=0.98\textwidth]{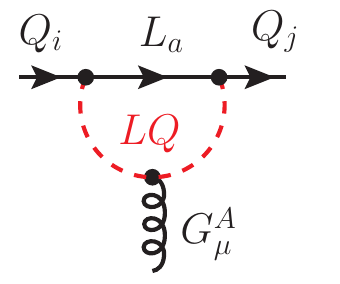}
    \caption{}
    \label{fig:LQ-QCD-peng}
  \end{subfigure}
  \begin{subfigure}[t]{0.39\textwidth}
    \centering
    \includegraphics[width=0.49\textwidth]{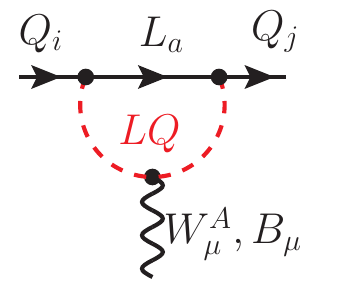}
    \includegraphics[width=0.49\textwidth]{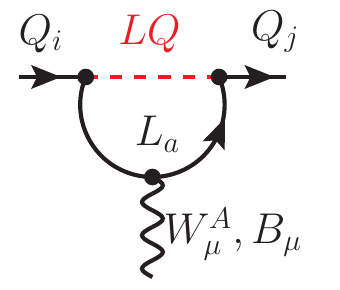}
    \caption{}
    \label{fig:LQ-EW-peng}
  \end{subfigure}
  \begin{subfigure}[t]{0.19\textwidth}
    \centering
    \includegraphics[width=0.98\textwidth]{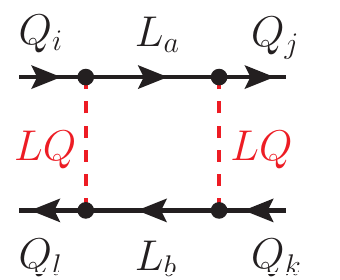}
    \caption{}
    \label{fig:LQ-box}
  \end{subfigure}
\caption{\small 
  LQ Tree-level exchange (a) gives rise to semi-leptonic operators.
  One-loop contributions to $\epe$ from QCD-penguins (b), EW-penguins (c)
  and box-type diagrams (d). Here $Q = \{q_L, u_R, d_R\}$ and 
  $L = \{\ell_L, e_R\}$ are SM quark and lepton flavour eigenstates.
  The $\SUthreeC \otimes \SUtwoL \otimes \UoneY$ gauge bosons are 
  denoted as $G_\mu^A$, $W_\mu^A$ and $B_\mu$, respectively.  
}
\end{figure}

%
%

\subsection{One-loop LQ decoupling
  \label{sec:1loop-LQ-dcpl}
}

The NL-$\psi^4$ coefficients receive direct contributions at~$\muNP$ from the LQ
decoupling first at one-loop. Loop-corrections constitute a principal problem in
massive vector LQ models when no full UV completion is specified.  In the lack
of a UV completion, simple cut-off regularisation might be used
\cite{Barbieri:2015yvd}, introducing an additional dependence on the cut-off
scale.  On the other hand this issue is of no concern in scalar LQ models, for
which we will calculate these contributions in this section. There are QCD- and
EW-penguin diagrams, \reffig{fig:LQ-QCD-peng} and \reffig{fig:LQ-EW-peng}, as
well as box-type diagrams \reffig{fig:LQ-box}.


\subsubsection{QCD penguins
  \label{sec:1loop-LQ-dcpl-QCD}
}

QCD-penguin diagrams with LQs can contribute to the three operators \footnote{The
subindex ``4'' is reminiscent to the QCD-penguin operator $P_4$ in the basis
of \cite{Chetyrkin:1996vx}.}
\begin{equation}
  \label{eq:QCD-peng-SMEFT-op}
\begin{aligned}
  \,[{\cal P}_4^{(q)}]_{ji} & 
  = [\bar{q}^j_L \gamma_\mu T^A q^i_L] \sum_k [\bar{Q}^k \gamma^\mu T^A Q^k] , &
\\
  [{\cal P}_4^{(d)}]_{ji} & 
  = [\bar{d}^j_R \gamma_\mu T^A d^i_R] \sum_k [\bar{Q}^k \gamma^\mu T^A Q^k] , &
  [{\cal P}_4^{(u)}]_{ji} & 
  = [\bar{u}^j_R \gamma_\mu T^A u^i_R] \sum_k [\bar{Q}^k \gamma^\mu T^A Q^k] , &
\end{aligned}
\end{equation}
depending on the LQ model. These operators are meant to be normalised as
in \refeq{BASICSMEFT}. The sum over flavour-diagonal quark currents
\begin{align}
  \sum_k [\bar{Q}^k \gamma^\mu T^A Q^k] & \equiv 
  \sum_k \left([\bar{q}^k_L \gamma^\mu T^A q^k_L]
              + [\bar{d}^k_R \gamma^\mu T^A d^k_R] 
              + [\bar{u}^k_R \gamma^\mu T^A u^k_R] \right)
\end{align}
arises from the quark-flavour universal gluon coupling and the matching might be
performed exploiting the generic formula given in \cite{Bobeth:1999ww} for the
$b\to s \, gluon$ off-shell vertex, see appendix in \cite{Bobeth:1999mk} for
more details. We refrain from a projection onto the operators in
\reftab{tab:4ferm} because we will neglect the RG evolution from $\muNP$ to
$\muEW$ for these operators \footnote{We denote them by ${\cal P}_i$ because
  they are linear combinations of the non-redundant set ${\cal O}_j$ given in
  \cite{Grzadkowski:2010es}.}, which is a loop correction due to
self-mixing. This will simplify the matching of SMEFT on low-energy EFT at
$\muEW$. For FCNC down-type transitions one has
\begin{align}
   \label{eq:LQ-QCD-peng-matching}
   \wc[(a)]{4}{ji} (\muEW) &
   = \frac{\alpha_s}{4\pi} r_{4,a}^{\rm LQ} 
   \frac{\Sigma^{ji}_{\chi,{\rm LQ}}}{M_{\rm LQ}^2} &
   (a & = q,d)
\end{align} 
with the constants
\begin{align}
  r_{4,a}^{S_1}        & = \frac{\delta_{aq}}{18} , & 
  r_{4,a}^{\wTil{S}_1} & = \frac{\delta_{ad}}{18} , & 
  r_{4,a}^{R_2}        & = \frac{\delta_{aq}}{18} , &
  r_{4,a}^{\wTil{R}_2} & = \frac{\delta_{ad}}{9} , & 
  r_{4,a}^{S_3}        & = \frac{\delta_{aq}}{6} , &    
\end{align} 
where $\delta_{aq}$ and $\delta_{ad}$ are Kronecker symbols.  The comparison
with the EW mixing-induced contributions \refeq{eq:NLpsi4-EW-scaling} shows the
same dependence on $\Sigma^{ji}_{\rm LQ}$. Further they are enhanced by the
ratio $\alpha_s/\alpha_e \ln^{-1}(\muNP/\muEW)$. Numerically this amounts to
roughly $\approx 15 \ln^{-1}(\muNP/\muEW) \in [3, 7]$ for
$\muNP \in [1, 10]$~TeV and $\muEW = 100$~GeV, aside from the constant factors
$r_{4,a}^{\rm LQ}$ and corresponding ones in the SL-$\psi^4$ coefficients.
However, as will be shown in detail below and given in \refeq{eq:epe-seminum},
this numerical enhancement of QCD-penguin Wilson coefficients becomes outweighed
by another numerical enhancement of EW-penguin Wilson coefficients below $\muEW$
in the expression of $\epe$, leaving the EW mixing-induced contributions as the
dominant contributions in most LQ models.


\subsubsection{EW penguins
  \label{sec:1loop-LQ-dcpl-EW}
}

The one-loop contributions to $\epe$ from EW-penguin diagrams in
\reffig{fig:LQ-EW-peng} at the scale $\muNP$ are actually the next-to-leading
order (NLO) corrections to contributions from the tree-level decoupling in
\refsec{sec:tree-LQ-dcpl}, as will become evident once the RG evolution of SMEFT
in \refsec{sec:SMEFT-RGE} is taken into account. In fact, those diagrams in
\reffig{fig:LQ-EW-peng} that at low energies represent QED penguin diagrams,
contain infrared divergences that are cancelled in the matching on SMEFT by the
ultraviolet divergences of diagrams with SL-$\psi^4$ insertions when closing the
lepton lines to a loop and radiating off a $\SUtwoL$ or $\UoneY$ gauge boson
respectively. These are the very same diagrams that determine the anomalous
dimensions of SMEFT operators \cite{Alonso:2013hga}.  Parametrically such NLO
EW-penguin terms contribute to the NL-$\psi^4$ Wilson coefficients as in
\refeq{eq:NLpsi4-EW-scaling}, just without the logarithmic enhancement and
therefore we will not further consider them throughout.


\subsubsection{Box diagrams
  \label{sec:1loop-LQ-dcpl-boxes}
}

The most general quark transitions $Q_i \bar{Q}_k \to Q_j \bar{Q}_l$ from LQ
box diagrams with $Q = (q_L, d_R, u_R)$ generate in SMEFT 
\begin{align}
  {\cal L}_{\rm SMEFT} &
  \supset \sum_a \sum_{jikl} \wc[(o)]{a}{jikl} \, [{\cal P}_{a}^{(o)}]_{jikl}
\end{align}
with non-leptonic operators of colour-octet type
\begin{equation}
  \label{eq:box-SMEFT-operators}
\begin{aligned}
  \,\![{\cal P}_{qq}^{(o,1)}]_{jikl} &
  = [\bar{q}^{j,\alpha}_L \gamma_\mu q^{i,\beta}_L]
    [\bar{q}^{k,\beta}_L \gamma^\mu q^{l,\alpha}_L] , &
  & & S_1(LL), R_2(RR), S_3, U_1(LL), U_3
\\
  [{\cal P}_{qq}^{(o,3)}]_{jikl} &
  = [\bar{q}^{j,\alpha}_L \gamma_\mu \tau^I q^{i,\beta}_L]
    [\bar{q}^{k,\beta}_L \gamma^\mu \tau^I q^{l,\alpha}_L] , &
  & &  R_2(RR), S_{3}, V_2(RR), U_{3}
\\
  [{\cal P}_{dd}^{(o)}]_{jikl} &
  = [\bar{d}^{j,\alpha}_R \gamma_\mu d^{i,\beta}_R]
    [\bar{d}^{k,\beta}_R \gamma^\mu d^{l,\alpha}_R] , &
  & & \wTil{S}_1, \wTil{R}_2, U_1(RR), V_2(LL)
\\
  [{\cal P}_{qd}^{(o)}]_{jikl} &
  = [\bar{q}^{j,\alpha}_L \gamma_\mu q^{i,\beta}_L]
    [\bar{d}^{k,\beta}_R \gamma^\mu d^{l,\alpha}_R] , &
  & & U_1(LR), V_2(RL)
\\
  [{\cal P}_{qu}^{(o)}]_{jikl} &
  = [\bar{q}^{j,\alpha}_L \gamma_\mu q^{i,\beta}_L]
    [\bar{u}^{k,\beta}_R \gamma^\mu u^{l,\alpha}_R] , &
  & & S_1(LR), R_2 (RL)
\end{aligned}
\end{equation}
where $\alpha, \, \beta$ denote $\SUthreeC$ colour indices. Here we have
retained only those that contribute to down-type quark transitions and show
corresponding LQ models that give rise to each operator. Included are the
combinations of chirality $\chi\chi'$ of the couplings
$\Sigma_{\chi}^{ji} \Sigma_{\chi'}^{kl}$ for LQ models with two couplings
($U_1,V_2,S_1,R_2)$ that can be easily understood from \refeq{eq:Lag-LQ-S-AFW}
and \refeq{eq:Lag-LQ-V-AFW}. They are linear combinations of the NL-$\psi^4$
operators in \reftab{tab:4ferm}: $\Op[(1,8)]{qq,qd,qu},\, \Op{dd}$, which can be
seen upon using
\begin{align}
  \label{eq:SU3-TA-decomp}
  T^A_{ij} T^A_{kl} & 
  = \frac{1}{2} \left( \delta_{il} \delta_{jk} - \frac{1}{N_c} \delta_{ij} \delta_{kl} \right),
\end{align} 
or in the case of ${\cal P}_{qq,dd}^{(\ldots)}$ Fierz relations. 

The explicit matching results of the Wilson coefficients
$\wc[(o,\dots)]{a}{jikl}$ at $\muNP$ for scalar LQ models
$S_{1,3}, \wTil{S}_1, R_2$ and $\wTil{R}_2$ are provided in
\refapp{app:1loop-decoupl}. We will omit the RG evolution from $\muNP$ to
$\muEW$ as in the case of QCD penguin contributions. A main distinguishing
feature of boxes compared to QCD and EW penguins is that the gauge coupling is
replaced by an additional combination of LQ couplings
\begin{align}
  \wc[(o)]{a}{jikl} & \propto
  \frac{\Sigma^{kl}_{\chi'}}{(4\pi)^2} \frac{\Sigma^{ji}_\chi}{M^2_{\rm LQ}}.
\end{align}

%
%

\subsection{Renormalisation Group Equations
  \label{sec:SMEFT-RGE}
}

The RG equations have the general structure
\begin{align}
  \dotWc{a} & \equiv (4\pi)^2 \mu \frac{{\rm d} \Wc{a}}{{\rm d}\mu} 
  = \gamma_{ab}\, \Wc{b}
\end{align}
with $\gamma_{ab}$ being the entries of a very big anomalous dimension matrix
(ADM). The ADM is known for SMEFT at one-loop and the entries relevant here
have been presented in \cite{Alonso:2013hga}. For small
$\gamma_{ab}/(4\pi)^2 \ll 1$ the approximate solution retains only the first
leading logarithm (1stLLA)
\begin{align}
  \label{eq:SMEFT-RGE}
  \Wc{a} (\muEW) &
  =  \left[ \delta_{ab} 
     - \frac{\gamma_{ab}}{(4 \pi)^2} \ln \frac{\muNP}{\muEW} \right]
     \Wc{b} (\muNP) ,
\end{align}
which is sufficient as long as the logarithm is not too large, so that also
$\gamma_{ab}/(4\pi)^2\ln \frac{\muNP}{\muEW} \ll 1$ holds. We should stress that
RG effects due to top-quark Yukawa mixing considered recently in various
analyses \cite{Feruglio:2016gvd, Feruglio:2017rjo, Bobeth:2017xry} are absent
here \cite{Jenkins:2013wua}.

In what follows we list RG equations which govern the generation of NL-$\psi^4$
coefficients from SL-$\psi^4$ ones. For NL-$\psi^4$ $(\bar{L}L)(\bar{L}L)$ 
operators we find using \cite{Alonso:2013hga}
\begin{align}
  \label{eq:4-quark-RGE-qq1}
  \dotwc[(1)]{qq}{prst} & 
  = - \frac{1}{9} g^2_1 \left( 
      \wc[(1)]{\ell q}{wwst} \delta_{pr} + \wc[(1)]{\ell q}{wwpr} \delta_{st}
    + \wc{qe}{stww}          \delta_{pr} + \wc{qe}{prww}          \delta_{st} \right) ,
\\
  \label{eq:4-quark-RGE-qq3}
  \dotwc[(3)]{qq}{prst} &
  = + \frac{1}{3} g^2_2 \left( 
      \wc[(3)]{\ell q}{wwst} \delta_{pr} + \wc[(3)]{\ell q}{wwpr} \delta_{st} \right) .
\end{align}
For NL-$\psi^4$ $(\bar{R}R)(\bar{R}R)$ operators we find
\begin{align}
  \dotwc[]{uu}{prst} & 
  = - \frac{4}{9} g^2_1 \Big(
      \wc{eu}{wwst}     \delta_{pr} + \wc{eu}{wwpr}     \delta_{st}
    + \wc{\ell u}{wwst} \delta_{pr} + \wc{\ell u}{wwpr} \delta_{st} \Big) ,
\\ 
  \label{eq:4-quark-RGE-dd}
  \dotwc[]{dd}{prst} & 
  = + \frac{2}{9} g^2_1 \Big(
      \wc{ed}{wwst}     \delta_{pr} + \wc{ed}{wwpr}     \delta_{st}
    + \wc{\ell d}{wwst} \delta_{pr} + \wc{\ell d}{wwpr} \delta_{st} \Big) ,
\\
  \label{eq:4-quark-RGE-ud1}
  \dotwc[(1)]{ud}{prst} & 
  = + \frac{4}{9} g^2_1 \Big( 
        \wc{\ell u}{wwpr} \delta_{st} +   \wc{eu}{wwpr} \delta_{st}
    - 2 \wc{\ell d}{wwst} \delta_{pr} - 2 \wc{ed}{wwst} \delta_{pr} \Big) .
  & 
\end{align}
For NL-$\psi^4$ $(\bar{L}L)(\bar{R}R)$ operators we find 
\begin{align}
  \label{eq:4-quark-RGE-qu1}
  \dotwc[(1)]{qu}{prst} & 
  = - \frac{2}{9} g^2_1 \left(
       4 \wc[(1)]{\ell q}{wwpr} \delta_{st} + 4 \wc{qe}{prww} \delta_{st}
     +   \wc{\ell u}{wwst} \delta_{pr}      +   \wc{eu}{wwst} \delta_{pr} \right),
\\ 
  \label{eq:4-quark-RGE-qd1}
  \dotwc[(1)]{qd}{prst} & 
  = + \frac{2}{9} g^2_1 \left(
      2 \wc[(1)]{\ell q}{wwpr} \delta_{st} + 2 \wc{qe}{prww} \delta_{st}
    -   \wc{\ell d}{wwst}      \delta_{pr} -   \wc{ed}{wwst} \delta_{pr} \right) ,
\end{align}
and finally for all other NL-$\psi^4$ operators
\begin{equation}
\begin{aligned}
  \dotwc[(8)]{ud}{prst} & = 0 , &
  \dotwc[(8)]{qu}{prst} & = 0 , &
  \dotwc[(1)]{quqd}{prst} & = 0 , 
\\
  & &
  \dotwc[(8)]{qd}{prst} & = 0 , &
  \dotwc[(8)]{quqd}{prst} & = 0 .
\end{aligned}
\end{equation}

We observe that the $\SUtwoL \otimes \UoneY$ gauge-mixing of SL-$\psi^4$ into
NL-$\psi^4$ operators within SMEFT generates in 1stLLA only
$(\bar{L}L)(\bar{L}L)$, $(\bar{L}L)(\bar{R}R)$ and $(\bar{R}R)(\bar{R}R)$
NL-$\psi^4$ operators from the corresponding semi-leptonic classes.  The initial
Wilson coefficients of the semi-leptonic operators at the scale $\muNP$ enter
only summed over the lepton-flavour diagonal parts $\wc[(a)]{b}{ww\cdot\cdot}$
(and $\wc{qe}{\cdot\cdot ww}$), summation over the index $w = 1,2,3$ is implied,
because all leptons can run inside the loop.  In consequence the underlying
combination of LQ couplings is $\Sigma^{ji}_{\rm LQ}$, introduced in
\refeq{eq:def-Sigma_LQ}.  Further, the NL-$\psi^4$ Wilson coefficients at
$\muEW$ contain always one quark-flavour diagonal quark-bilinear since all ADMs
are $\propto \delta_{st}$ or $\delta_{pr}$ and as a consequence some terms will
not contribute to down-type $\Delta F = 1$ processes.

The $(\oL{L}R)(\oL{R}L)$ and $(\oL{L}R)(\oL{L}R)$ SL-$\psi^4$ operators
$\Op{\ell edq}$ and $\Op[(1,3)]{\ell equ}$ are only needed if they contribute to
semi-leptonic $K$ decays in order to derive constraints on the LQ couplings. On
the other hand, $(\oL{L}R)(\oL{L}R)$ NL-$\psi^4$ operators $\Op[(1,3)]{quqd}$
contribute to $\epe$ only in those LQ models that provide a direct one-loop
matching contribution at $\muNP$, i.e. ${\cal P}^{(o)}_{qd,\, qu}$ in
\refeq{eq:box-SMEFT-operators}.

The RG equations provide the Wilson coefficients of the SMEFT operators at the
electroweak scale $\muEW$, where electroweak symmetry breaking (EWSB) takes
place. At this point the transition from the weak to the mass eigenbasis for
gauge, quark and lepton fields can be done within SMEFT. The quark fields are
rotated by $3\times 3$ unitary rotations in flavour space
\begin{align}
  \psi_L & \to V^\psi_L \psi_L\,, & \psi_R & \to V^\psi_R \psi_R\,,
\end{align}
for $\psi = u,d$, such that
\begin{align}
  V^{\psi\dagger}_L m_\psi V^\psi_R & = m_\psi^{\rm diag}, &
  V & \equiv (V^{u}_L)^\dagger V^d_L \,,
\end{align}
with diagonal up- and down-quark mass matrices $m_\psi^{\rm diag}$. In general,
the non-diagonal mass matrices $m_\psi$ include the contributions of dim-6
operators. The quark-mixing matrix $V$ is unitary, similar to the CKM matrix of
the SM; however, in the presence of dim-6 contributions the numerical values are
different from those obtained in usual SM CKM-fits. Since we are interested in
down-type processes $\epe$ and rare Kaon processes, we will take the freedom to
choose the weak basis such that down-type quarks are already mass eigenstates,
which fixes $V_{L,R}^d = \One$, and assume without loss of generality
$V_R^u = \One$, yielding $q_L = (V^\dagger u_L, d_L)^T$.  Analogously, we choose
also the down-type lepton mass matrix to be diagonal and leave the neutrinos
\footnote{In SMEFT neutrinos receive masses from the dimension five Weinberg
  operator during EWSB.} in the flavour eigenbasis.  This defines the SMEFT
Wilson coefficients unambiguously and avoids the appearance of the PMNS
lepton-mixing matrix in interactions involving neutrinos.

%
%

\subsection{\boldmath 
  Non-leptonic operators: SMEFT on $\Delta F = 1$ EFT
  \label{sec:non-lept-match-SMEFT}
}

The tree-level matching of SMEFT on $\Delta F = 1$ low-energy EFT's at the scale
$\muEW$ is well-known for semi-leptonic processes \cite{D'Ambrosio:2002ex,
  Alonso:2014csa, Buras:2014fpa} and given for non-leptonic processes in
\cite{Aebischer:2015fzz}. We summarise the required parts in the following three
subsections. Starting with non-leptonic operators, we provide results relevant
for $\epe$ for the choice of the traditional basis of the QCD- and EW-penguin
operators \refeq{eq:QCD-peng-op} and \refeq{eq:QED-peng-op}, which differs from
\cite{Chetyrkin:1996vx}, and simplifies due to the particular flavour structure
\refeq{eq:4-quark-RGE-qq1} -- \refeq{eq:4-quark-RGE-qd1} of the EW gauge-mixing
of SL-$\psi^4$ into NL-$\psi^4$ SMEFT operators.  Further we summarise the
tree-level matching of SL-$\psi^4$ operators relevant for
$d_i \ell_a \to d_j \ell_b$, $d_i \nu_a \to d_j \nu_b$ and
$d_i \nu_a \to u_j \ell_b$.


\subsubsection{\boldmath 
  EW gauge-mixing
}

As already pointed out in \refsec{sec:SMEFT-RGE}, the EW gauge-mixing of
SL-$\psi^4$ into NL-$\psi^4$ operators leads to flavour-universal down-type and
up-type contributions that correspond almost exclusively to linear combinations
of QCD- and EW-penguin operators ($e_u = -2 e_d = + 2/3$)
\begin{equation}
  \label{eq:peng-decomposition}
\begin{aligned}
  (\bar s d)_{V-A}\,\sum_u (\bar u u)_{V-A} & 
  = \frac{1}{3}(Q_3 + 2 Q_9) , &
  (\bar s d)_{V-A}\,\sum_d (\bar d d)_{V-A} & 
  = \frac{2}{3}(Q_3 - Q_9) ,
\\
  (\bar s d)_{V-A}\,\sum_u (\bar uu)_{V+A} & 
  = \frac{1}{3}(Q_5 + 2 Q_7) , &
  (\bar s d)_{V-A}\,\sum_d (\bar d d)_{V+A} & 
  = \frac{2}{3}(Q_5 - Q_7) ,
\end{aligned}
\end{equation}
and analogously for chirality-flipped $Q'_{3,5,7,9}$ --- see definitions
\refeq{eq:QCD-peng-op} and \refeq{eq:QED-peng-op}, except for one contribution
from $\Op[(3)]{qq}$ as shown below.

Let us illustrate in some detail the matching for the $(\bar{L}L)(\bar{R}R)$
NL-$\psi^4$ operators $\Op[(1)]{qd}$ and $\Op[(1)]{qu}$. The ADM given in
\refeq{eq:4-quark-RGE-qd1} yields upon insertion into \refeq{eq:SMEFT-RGE} at
the scale $\muEW$
\begin{align}
  & \wc[(1)]{qd}{prst} \op[(1)]{qd}{prst} \nonumber 
  = - \frac{2}{9} \frac{g_1^2}{(4\pi)^2} \ln \frac{\muNP}{\muEW} \;
  \left( 2 \big[\Wc[(1)]{\ell q} + \Wc{qe} \big]_{wwpr} \delta_{st} 
         - \big[\Wc{\ell d} + \Wc{de} \big]_{wwst} \delta_{pr} \right)
\\ & \hskip 3.0cm
  \times \left[ (\bar{u}_L^p \gamma_\mu u_L^r) + (\bar{d}_L^p \gamma_\mu d_L^r) \right] 
                (\bar{d}_R^s \gamma^\mu d_R^t) .
\intertext{The $\delta_{pr, st}$-symbols give rise to the aforementioned 
flavour-diagonal quark-bilinears. In the transition to mass eigenstates
after EWSB, we keep only terms with $\bar{d}^p_{R(L)}\to \bar{s} P_{L(R)}$
and $d^r_{R(L)}\to P_{R(L)} d$ that contribute to $\bar{s} \to \bar{d}$ 
transitions ($P_{R,L} = (1 \pm \gamma_5)/2$) }
  &
  \simeq - \frac{1}{9} \frac{\alpha_e}{4\pi} \ln \frac{\muNP}{\muEW} \; 
    \frac{\big[\Wc[(1)]{\ell q} + \Wc{qe} \big]_{ww21}}{c_W^2} \;
    (\bar{s} d)_{V-A} \sum_{d} (\bar{d} d)_{V+A}
\\  \nonumber
  & \phantom{\to} +
    \frac{1}{18} \frac{\alpha_e}{4\pi} \ln \frac{\muNP}{\muEW} \; 
    \frac{\big[\Wc{\ell d} + \Wc{de} \big]_{ww21}}{c_W^2} \;
    (\bar{s} d)_{V+A} \left[ 
        \sum_{d} (\bar{d} d)_{V-A}
      + \sum_{k, i,j} V_{ik}^{} V_{jk}^\ast (\bar{u}^i u^j)_{V-A}
    \right] .
\intertext{Finally one finds with the unitarity of the mixing matrix 
$\sum_k V_{ik}^{} V_{jk}^* = \delta_{ij}$ and relations~\refeq{eq:peng-decomposition}}
   &
  \simeq \frac{\alpha_e}{4\pi} \ln \frac{\muNP}{\muEW} \left( - \frac{2}{27} 
    \frac{\big[\Wc[(1)]{\ell q} + \Wc{qe} \big]_{ww21}}{c_W^2} (Q_5 - Q_7)
  + \frac{1}{18}  
    \frac{\big[\Wc{\ell d} + \Wc{de} \big]_{ww21}}{c_W^2} \, Q'_5 
    \right) 
\end{align}
and similarly for the operator
\begin{align}
  & \wc[(1)]{qu}{prst} \op[(1)]{qu}{prst}
  \simeq \frac{\alpha_e}{4\pi} \ln \frac{\muNP}{\muEW} \frac{2}{27} 
    \frac{\big[\Wc[(1)]{\ell q} + \Wc{qe} \big]_{ww21}}{c_W^2} (Q_5 + 2 Q_7).
\end{align}
The total contribution of $(\bar{L}L)(\bar{R}R)$ operators is
\begin{equation}
\begin{aligned}
    \wc[(1)]{qu}{prst} & \op[(1)]{qu}{prst}
  + \wc[(1)]{qd}{prst} \op[(1)]{qd}{prst}
\\ &
  \simeq \frac{\alpha_e}{4\pi}  
    \frac{\big[\Wc[(1)]{\ell q} + \Wc{qe} \big]_{ww21}}{c_W^2}
    \frac{2}{9} \ln \frac{\muNP}{\muEW}  Q_7
  + \frac{\alpha_e}{4\pi}  
    \frac{\big[\Wc{\ell d} + \Wc{de} \big]_{ww21}}{c_W^2}
    \frac{1}{18} \ln \frac{\muNP}{\muEW}  Q'_5 
\end{aligned}
\end{equation}
free of $Q_5$ and all SL-$\psi^4$ Wilson coefficients are at the
scale $\muNP$.

The results for the other cases $(\bar{R}R)(\bar{R}R)$ and $(\bar{L}L)(\bar{L}L)$
are obtained analogously, 
\begin{align}
  \wc{dd}{prst} \op{dd}{prst} + \wc[(1)]{ud}{prst} \op[(1)]{ud}{prst}
  & \simeq
  \frac{\alpha_e}{4\pi} \; 
  \frac{\big[\Wc{\ell d} + \Wc{ed} \big]_{ww21}}{c_W^2} \;
  \frac{2}{9} \ln \frac{\muNP}{\muEW} Q'_9 ,
\\
  \wc[(1)]{qq}{prst} \op[(1)]{qq}{prst} 
  & \simeq 
  \frac{\alpha_e}{4\pi} \; 
  \frac{\big[\Wc[(1)]{\ell q} + \Wc{qe} \big]_{ww21}}{c_W^2} \;
  \frac{1}{18} \ln \frac{\muNP}{\muEW} Q_3 ,
\end{align}
where an additional term arises for $\Op[(3)]{qq}$
\begin{align}
  \wc[(3)]{qq}{prst} & \op[(3)]{qq}{prst} 
  \simeq \frac{\alpha_e}{4\pi} \; 
    \frac{\wc[(3)]{\ell q}{ww21}}{s_W^2} \;
    \frac{1}{18} \ln \frac{\muNP}{\muEW} \left(4\, Q_9 - Q_3 \right)
\\ & \nonumber \hskip 0.2cm
   - \frac{\alpha_e}{4\pi} \; \sum_{k,ij} 
     \left( V_{i1}^{} V_{jk}^\ast \frac{\wc[(3)]{\ell q}{ww2k}}{s_W^2}
          + V_{ik}^{} V_{j2}^\ast \frac{\wc[(3)]{\ell q}{wwk1}}{s_W^2} \right)
    \frac{1}{3} \ln \frac{\muNP}{\muEW} (\bar{s} d)_{V-A} (\bar{u}^i u^j)_{V-A} .
\end{align}
Although new physics can affect the quark-mixing matrix $V$ to deviate from the
SM CKM matrix, we assume that these effects do not lift the hierarchy in the
Cabibbo-angle $\lambda_C$ represented by the Wolfenstein parameterisation and
found in SM CKM fits. Assuming further that the Wilson coefficients
$\wc[(3)]{\ell q}{wwk'k}$ do not lift this hierarchy either, the additional term
in $\Op[(3)]{qq}$ becomes
\begin{equation}
  \label{eq:qq-3-intermediate}
\begin{aligned}
  \sum_{k,ij} &
     \left( V_{i1}^{} V_{jk}^\ast \frac{\wc[(3)]{\ell q}{ww2k}}{s_W^2}
          + V_{ik}^{} V_{j2}^\ast \frac{\wc[(3)]{\ell q}{wwk1}}{s_W^2} \right)
   (\bar{s} d)_{V-A} (\bar{u}^i u^j)_{V-A}
\\ &
   = \frac{\wc[(3)]{\ell q}{ww21}}{s_W^2} (\bar{s} d)_{V-A} 
   \Big[ |V_{11}|^2 (\bar{u} u)_{V-A} + |V_{22}|^2 (\bar{c} c)_{V-A} \Big]
\\ &
   + (\bar{s} d)_{V-A} (\bar{u} c)_{V-A} \frac{V_{11}^{} V_{22}^\ast}{s_W^2}
   \Big( \wc[(3)]{\ell q}{ww22} + \wc[(3)]{\ell q}{ww11} \Big)
   + {\cal O}(\lambda_C) .
\end{aligned}
\end{equation}
The $(\bar{u} c)_{V-A}$ part in the last line does not contribute to $\epe$,
whereas the $i = j=c$ part is loop-suppressed in principle. We still keep the
latter and use $|V_{11}|^2 = |V_{22}|^2 \approx 1 + {\cal O}(\lambda_C)$ as well
as \refeq{eq:peng-decomposition} to arrive at
\begin{align}
  \label{eq:qq3-approx}
  \mbox{eq.}~\refeq{eq:qq-3-intermediate} & 
  = \frac{\wc[(3)]{\ell q}{ww21}}{s_W^2} \frac{1}{3} (Q_3 + 2 Q_9) .
\end{align}

The matching conditions of $\Delta S = 1$ operators \refeq{eq:Heffeprime} at
$\muEW$ are given in terms of the SL-$\psi^4$ Wilson coefficients at $\muNP$
\begin{equation}
  \label{eq:DeltaS=1-matching}
\begin{aligned}
  C_3(\muEW) & 
  = - \frac{1}{9} \frac{\alpha_e}{4 \pi} \frac{v^2}{c_W^2}
    \frac{\big[\Wc[(1)]{\ell q} + \Wc{qe} \big]_{ww21}}{\lambda_u^{sd}}
    \ln \frac{\muNP}{\muEW}
  + \frac{1}{3} \frac{\alpha_e}{4 \pi} \frac{v^2}{s_W^2}
    \frac{\wc[(3)]{\ell q}{ww21}}{\lambda_u^{sd}} \ln \frac{\muNP}{\muEW},
\\
  C_7(\muEW) &
  = - \frac{4}{9} \frac{\alpha_e}{4 \pi} \frac{v^2}{c_W^2}
    \frac{\big[\Wc[(1)]{\ell q} + \Wc{qe} \big]_{ww21}}{\lambda_u^{sd}}
    \ln \frac{\muNP}{\muEW} ,
\\
  C'_9(\muEW) & = 4\, C'_5(\muEW)
  = - \frac{4}{9} \frac{\alpha_e}{4 \pi} \frac{v^2}{c_W^2}
    \frac{\big[\Wc{\ell d} + \Wc{ed} \big]_{ww21}}{\lambda_u^{sd}} 
    \ln \frac{\muNP}{\muEW} ,
\end{aligned}
\end{equation}
where $v^2 = (\sqrt{2} G_F)^{-1}$ and $c_W \equiv \cos \theta_W$. We used the
approximation \refeq{eq:qq3-approx}. These three expressions are fundamental for
EW-mixing effects in $\epe$ in LQ models.

There are three possible patterns of contributions to $\epe$ listed in
\reftab{tab:epe-LQ-contr}, showing also that LQ models $\wTil{U}_1$ and
$\wTil{V}_2$ do not contribute to $\epe$ via EW gauge-mixing.  In most models
$\epe$ is affected by $\Sigma^{ji}_{\chi,{\rm LQ}}$ with either $\chi = L$ or
$\chi = R$, but not both, the exceptions are vector LQ models $U_1$ and
$V_2$. For the first pattern involving $C_{3,7}$, the numerically largest impact
on $\epe$ will be due to the contribution from $C_7(\muEW)$ --- see
\refeq{eq:epe-seminum} and \reftab{tab:epe-seminum} --- either due to
$\Wc[(1)]{\ell q}$ or $\Wc{qe}$, such that $\Wc[(3)]{\ell q}$ is numerically
irrelevant for $\epe$. Let us note that in LQ models $\Wc[(3)]{\ell q}$ and
$\Wc[(1)]{\ell q}$ are not independent from each other but related through
$\Wc[(3)]{\ell q} \equiv r_{\rm LQ}\, \Wc[(1)]{\ell q}$ with
\begin{align}
  \label{eq:rel-lq3-lq1}
  r_{S_1} & = -1, & 
  r_{S_3} & =  \frac{1}{3}, &
  r_{U_1} & = 1 , &
  r_{U_3} & = -\frac{1}{3}, 
\end{align}
see \refapp{app:tree-decoupl}.  In the second pattern with $C'_{5,9}$ the
largest impact will be due to $C'_9 = 4\, C'_5$, where the $C'_5$ contributes
constructively. The third case of $C_{3,7}$ and $C'_{5,9}$ involves both
$\chi = L$ and $\chi = R$ LQ couplings, which can be in principle of different
size and prevent an apriori estimate of the relative numerical sizes of all
contributions, although $C_7$ is roughly enhanced by a factor of sixty compared
to $C'_9$, see \refeq{eq:epe-seminum} and \reftab{tab:epe-seminum}.  The latter
fact implies that models, which generate $\Wc[(1)]{\ell q}$ or $\Wc{qe}$ can
face easier the $\epe$ anomaly via the operator $Q_7$ than the other models.

\begin{table}[t]
\centering
\renewcommand{\arraystretch}{1.4}
\begin{tabular}{|c||c|c|}
\hline
  LQ model
& semi-leptonic SMEFT coeff.
& $\Delta S = 1$ coeff.
\\
\hline\hline
  $S_{1,3}$, $U_{3}$   
& $\Wc[(1,3)]{\ell q}$ ($L$)
& \multirow{2}{*}{$C_3$, $C_7$}
\\
  $R_2$
& $\Wc{qe}$  ($R$)
& 
\\
\hline
  $\wTil{S}_1$
& $\Wc{ed}$ ($R$)
& \multirow{2}{*}{$C'_5$, $C'_9$}
\\
  $\wTil{R}_2$
& $\Wc{\ell d}$ ($L$)
&
\\
\hline
  $U_1$
& $\Wc[(1,3)]{\ell q}$ ($L$), $\Wc{ed}$ ($R$)
& \multirow{2}{*}{$C_3$, $C_7$, $C'_5$, $C'_9$}
\\
  $V_2$
& $\Wc{qe}$ ($R$), $\Wc{\ell d}$ ($L$)
& 
\\
\hline
\end{tabular}
\renewcommand{\arraystretch}{1.0}
\caption{\label{tab:epe-LQ-contr}
  Classification of LQ models corresponding to their contribution to $\epe$
  via EW gauge-mixing and the involved semi-leptonic Wilson
  coefficient. The chirality of the LQ couplings entering the semi-leptonic 
  Wilson coefficients is shown in parenthesis, see \refapp{app:tree-decoupl}
  for details.
}
\end{table}


\subsubsection{\boldmath 
  QCD-penguins
}

Besides the EW mixing-induced contributions, the NL-$\psi^4$ coefficients
receive direct one-loop matching contributions at $\muNP$ from QCD- and
EW-penguin diagrams as well as box-type diagrams. As already discussed in
\refsec{sec:1loop-LQ-dcpl-QCD}, QCD-penguin contributions are parametrically
enhanced w.r.t. the mixing-induced contributions at $\muEW$. After EWSB, the
operators \refeq{eq:QCD-peng-SMEFT-op} are matched onto the $\Delta F = 1$
low-energy analogue yielding
\begin{equation}
  \label{eq:DF1-QCD-peng-matching}
\begin{aligned}
  -3\, C_3 = C_4 = -3\, C_5 = C_6 &
  = -\frac{v^2}{4} 
    \frac{\wc[(q)]{4}{21}}{\lambda_u^{sd}} 
  = -\frac{\alpha_s}{4\pi} \frac{v^2}{M_{\rm LQ}^2} \frac{r_{4,q}^{\rm LQ}}{4}
    \frac{\Sigma^{21}_{\rm LQ}}{\lambda_u^{sd}} ,
\\
  -3\, C_{3'} = C_{4'} = -3\, C_{5'} = C_{6'} &
  = -\frac{v^2}{4} 
    \frac{\wc[(d)]{4}{21}}{\lambda_u^{sd}}
  = -\frac{\alpha_s}{4\pi} \frac{v^2}{M_{\rm LQ}^2} \frac{r_{4,d}^{\rm LQ}}{4}
    \frac{\Sigma^{21}_{\rm LQ}}{\lambda_u^{sd}} ,
\end{aligned}
\end{equation}
at $\muEW$ using \refeq{eq:LQ-QCD-peng-matching}. This can be compared to the
contributions from EW gauge-mixing \refeq{eq:DeltaS=1-matching}, showing again
the enhancement factor $\alpha_s/\alpha_e \ln^{-1}(\muNP/\muEW)$. Yet, as we
will find in the next section at the end QCD penguin effects will be much
smaller than EW gauge-mixing and box-diagram contributions that we discuss next.


\subsubsection{\boldmath 
  Box diagrams
}

The LQ box-diagrams generate NL-$\psi^4$ operators
\refeq{eq:box-SMEFT-operators} of which the majority do not contribute directly
to $K\to\pi\pi$ transitions because the flavour indices do not involve the
required ones. Yet some of these operators can contribute indirectly due to RG
mixing into operators that contribute directly. We will first illustrate the
matching for the various operators ${\cal P}_{dd}^{(o)}$, since here the
transition from weak to mass eigenstates is trivial in the absence of $q_L$.
Note that due to equal Lorentz structure in both quark currents there is a
symmetry under simultaneous $i\leftrightarrow l$ and $j\leftrightarrow k$, such
that we might fix $j = 2$ since we are interested in $K\to \pi\pi$.  For the
time being we still use notational distinction $d^i_R \to P_R D_i$ between weak
and mass eigenstates by using capital $D_i = (d,s,b)_i$ for latter ones
\begin{align}
  {\cal L}_{\rm SMEFT} &  \nonumber
  \supset \sum_{ikl} \wc[(o)]{dd}{2ikl} 
    [\bar{d}^{2,\alpha}_R \gamma_\mu d^{i,\beta}_R]
    [\bar{d}^{k,\beta}_R \gamma^\mu d^{l,\alpha}_R]
\\ &
  = \sum_{ikl} \frac{\wc[(o)]{dd}{2ikl}}{4}
    (\bar{s}^\alpha D_i^\beta)_{V+A}
    (\bar{D}_k^\beta D_l^\alpha)_{V+A} .
\intertext{The operators with non-vanishing matrix elements to $K\to \pi\pi$ are
those that contain three $d$-quarks: $ikl=111$. For other operators to contribute
to the $\Delta S = 1$ transition $K\to\pi\pi$, at least one $d$ quark is required: 
$i=1$ or $l=1$, as well as the remaining two indices should be equal (either
$2$ or $3$, as $ikl=111$ is already covered above), because only then they contribute
via mixing into QCD- and EW-penguin operators when closing
the quark loop and radiating off either gluon or photon in the low-energy EFT
(same effects in SMEFT were neglected above). Thus the sum can be split
into} 
  & \nonumber
  = \frac{\wc[(o)]{dd}{2111}}{4} \;
    (\bar{s}^\alpha d^\beta)_{V+A} (\bar{d}^\beta d^\alpha)_{V+A}
\\ & \nonumber
  + \frac{1}{4} \sum_{k\neq 1} \left(
      \wc[(o)]{dd}{21kk} (\bar{s}^\alpha d^\beta)_{V+A} (\bar{D}_k^\beta D_k^\alpha)_{V+A}
    + \wc[(o)]{dd}{2kk1} (\bar{s}^\alpha D_k^\beta)_{V+A} (\bar{D}_k^\beta d^\alpha)_{V+A}
  \right)
\\ & 
  \label{eq:P8dd-1}
  + \frac{1}{4} \sum_{ikl} \wc[(o)]{dd}{2ikl}
    (\bar{s}^\alpha D_i^\beta)_{V+A}
    (\bar{D}_k^\beta D_l^\alpha)_{V+A}
\intertext{where the terms in the last line are such that they do not 
contribute to $K\to\pi\pi$ and are not part of the 1st and 2nd line. The
2nd term in the 2nd line contains actually only $k=3$, due to the aforementioned
symmetry. We rewrite the first term into a sum over $k$, yielding shifts
of the Wilson coefficients in the 2nd line}
  & \nonumber
  = \frac{\wc[(o)]{dd}{2111}}{4} \; 
    (\bar{s}^\alpha d^\beta)_{V+A} \sum_k (\bar{D}_k^\beta D_k^\alpha)_{V+A}
\\ &
  + \frac{1}{4} \sum_{k\neq 1}
      \left(\wc[(o)]{dd}{21kk} - \wc[(o)]{dd}{2111}\right) 
      (\bar{s}^\alpha d^\beta)_{V+A} (\bar{D}_k^\beta D_k^\alpha)_{V+A}
    + \ldots
\intertext{where the dots indicate the remaining terms in \refeq{eq:P8dd-1}
and make use of \refeq{eq:peng-decomposition}, taking into account the 
different colour structure,}
  & \nonumber
  = \frac{\wc[(o)]{dd}{2111}}{4} \; 
    \frac{2}{3} \left(Q_{4'} - Q_{10'} \right)
\\ &
  + \frac{1}{4} \sum_{k\neq 1}
      \left(\wc[(o)]{dd}{21kk} - \wc[(o)]{dd}{2111}\right) 
      (\bar{s}^\alpha d^\beta)_{V+A} (\bar{D}_k^\beta D_k^\alpha)_{V+A}
    + \ldots
\end{align}
In this way we have rewritten the operator $(\bar{s}d)(\bar{d}d)$ into QCD- and
EW-penguin operators and the operators $(\bar{s}d)(\bar{s}s)$ and
$(\bar{s}d)(\bar{b}b)$, which is a convenient choice of basis for
$K\to\pi\pi$. Taking into account normalisation factors \refeq{eq:Heffeprime},
it follows at $\muEW$
\begin{align}
  \label{eq:LQ-box-dd}
  C_{4'} = - C_{10'} &
  = - \frac{v^2}{3} \frac{\wc[(o)]{dd}{2111}}{\lambda_u^{sd}} .
\end{align}
Although operators $\sim (\bar{s}d)(\bar{s}s)$ and $\sim (\bar{s}d)(\bar{b}b)$
are loop suppressed in $\epe$ w.r.t. $(\bar{s}d)(\bar{d}d)$ since they enter via
RG mixing only, their Wilson coefficients might be numerically enhanced to
overcome the loop-suppression because they depend on different combinations of
LQ couplings. The mixing of $(\bar{s}d)(\bar{s}s)$ and $(\bar{s}d)(\bar{b}b)$
into QCD- and EW-penguins can be found in the literature as for example
\cite{Buras:2000if}, but we will neglect these effects here.

The operators ${\cal P}^{(o,1)}_{qq}$ and ${\cal P}^{(o,3)}_{qq}$ contribute
to $K\to \pi\pi$ as
\begin{equation}
    \label{eq:LQ-box-qq1}
\begin{aligned}
  C_{4} & 
  = -\frac{v^2}{6} \frac{\sum_{ji} (V_{1j}^{} V_{1i}^\ast + 2 \delta_{1j} \delta_{1i})
                         \wc[(o,1)]{qq}{21ji}}{\lambda_u^{sd}} , &
\\
  C_{10} & 
  = -\frac{v^2}{3} \frac{\sum_{ji} (V_{1j}^{} V_{1i}^\ast - \delta_{1j} \delta_{1i}) 
                         \wc[(o,1)]{qq}{21ji}}{\lambda_u^{sd}} ,
\end{aligned}
\end{equation}
and
\begin{equation}
  \label{eq:LQ-box-qq3}
\begin{aligned}
  C_9 = 2 \, C_3 & 
  = -\frac{2 v^2}{3} \frac{\sum_{ji} V_{1j}^{} V_{1i}^\ast \;
                         \wc[(o,3)]{qq}{2ji1}}{\lambda_u^{sd}}  ,
\\
  C_{4} & 
  = -\frac{v^2}{6} \frac{\sum_{ji} (-V_{1j}^{} V_{1i}^\ast + 2 \delta_{1j} \delta_{1i})
                         \wc[(o,3)]{qq}{21ji}}{\lambda_u^{sd}} , &
\\
  C_{10} & 
  = -\frac{v^2}{3} \frac{\sum_{ji} (-V_{1j}^{} V_{1i}^\ast - \delta_{1j} \delta_{1i}) 
                         \wc[(o,3)]{qq}{21ji}}{\lambda_u^{sd}} .
\end{aligned}
\end{equation}
The presence of $u_L$ in these operators leads to additional factors of
the quark-mixing matrix $V$ with summation over $\Sigma^{ji}_\chi$.

The contribution to $K\to \pi\pi$ from ${\cal P}^{(o)}_{qu}$ is found 
analogously to be
\begin{align}
  \label{eq:LQ-box-qu}
  C_8 = 2\, C_6 &
  = - \frac{v^2}{3} \frac{\wc[(o)]{qu}{2111}}{\lambda_u^{sd}} .
\end{align}
By comparison with \refeq{eq:box-SMEFT-operators}, this shows that in models
$S_1$ and $R_2$ the boxes give rise to the EW-penguin operators $Q_6$ and $Q_8$,
where $Q_8$ is strongly enhanced in $\epe$. The matching contributions given in
\refapp{app:1loop-decoupl} with
$\wc[(o)]{qu}{2111} \propto \Sigma^{11}_R \Sigma^{21}_L$ and
$\wc[(o)]{qu}{2111} \propto \Sigma^{11}_L \Sigma^{12}_R$ for $S_1$ and $R_2$
respectively, show that these contributions depend on both chirality couplings
$\chi = L,R$.  This goes hand in hand with the $\Delta F=2$ operator for
$D\oL{D}$-mixing analogous to $Q_{{\rm LR},2}^{ji}$ in \refeq{eq:DF2-operators}
that is strongly enhanced by QCD RG evolution, and which depends on the
combinations $\Sigma^{21}_R \Sigma^{21}_L$ and $\Sigma^{12}_L \Sigma^{12}_R$,
respectively.

With similar considerations, the contribution to $K\to \pi\pi$ from
${\cal P}^{(o)}_{qd}$ is found to be
\begin{equation}
  \label{eq:LQ-box-qd} 
\begin{aligned}
  C_6 = - C_8 &
  = - \frac{v^2}{3} \frac{\wc[(o)]{qd}{2111}}{\lambda_u^{sd}} , & \qquad
  C_{6'} & 
  = -\frac{v^2}{6} \frac{\sum_{ji} (V_{1j}^{} V_{1i}^\ast + 2 \delta_{1j} \delta_{1i}) 
                         \wc[(o)]{qd}{ji21}}{\lambda_u^{sd}} ,
\\
  && 
  C_{8'} & 
  = -\frac{v^2}{3} \frac{\sum_{ji} (V_{1j}^{} V_{1i}^\ast - \delta_{1j} \delta_{1i}) 
                         \wc[(o)]{qd}{ji21}}{\lambda_u^{sd}} .
\end{aligned}
\end{equation}
Note that $C_{8'}$ is Cabibbo-suppressed w.r.t. $C_{6'}$ and $C_{6,8}$, if one
were to use $|V_{ud}|^2 \approx 1$. Again operators $Q_{8,8'}$ are strongly
enhanced in $\epe$ such that for the corresponding models $U_1$ and $V_2$, see
\refeq{eq:box-SMEFT-operators}, these box-contributions could become important
depending on the size of the $\Sigma^{ji}_\chi$. Although for vector LQs we are
not able to calculate the coefficients $\Wc[(o)]{qd}$ without introducing
cut-offs, still we can give their dependence on the $\Sigma^{ji}_\chi$.

In summary the three main contributions from LQ decoupling are due to 1) EW
gauge-mixing of SL-$\psi^4$ into NL-$\psi^4$ operators, 2) QCD-penguins and 3)
box diagrams. As a result the corresponding Wilson coefficients of QCD- and
EW-penguin operators $C_i(\muEW)$ ($i = 3, \ldots, 10$) scale parametrically as
\begin{align}
  \label{eq:scaling-EW-mix-QCDpeng-box}
  \frac{e^2}{(4\pi)^2} \ln \frac{\muNP}{\muEW} \Sigma^{ji}_\chi 
  \qquad \leftrightarrow \qquad
  \frac{g_s^2}{(4\pi)^2} \Sigma^{ji}_\chi              
  \qquad \leftrightarrow \qquad
  \frac{\Sigma^{11}_{\chi'}}{(4\pi)^2} \Sigma^{ji}_\chi .
\end{align} 
Their relative sizes are thus fixed by
$e^2 \ln\muNP/\muEW \approx 0.1 \ln\muNP/\muEW \approx 0.2 \ldots 0.5$ for
$\muNP \in [1,\, 20]$~TeV, and $g_s^2 \approx 1.5$, whereas the yet-allowed size
of the complex-valued $\Sigma^{11}_{\chi'}$ is constrained by mostly tree-level
processes, depending strongly on the LQ model.  At the level of observables
different suppression/enhancement factors for each of the $C_i(\muEW)$ can
appear such that at this point no definite conclusions can be drawn about which
contribution is most important. We point out that concerning $\epe$, large
enhancement of the EW-penguin coefficients $C_{7,8}(\muEW)$ appear as can be
seen from \refeq{eq:epe-seminum}, which easily overcome the numerical
enhancement of LQ-QCD-penguins discussed here and leads to the dominance of
contributions due to EW gauge-mixing and/or LQ-boxes, depending on the LQ model.

%
%

\subsection{\boldmath 
  Semi-leptonic operators: SMEFT on $\Delta F = 1$ EFT
  \label{sec:semi-lept-SMEFT-matching}
}

The $\Delta F=1$ semi-leptonic FCNC processes $d_i \ell_a \to d_j \ell_b$ and
$d_i \nu_a\to d_j \nu_b$ are affected at tree-level by LQ exchange and provide
strong constraints on LQ couplings. For practical purposes we neglect the
running from $\muNP$ to $\muEW$ in SMEFT for the semi-leptonic operators if
self-mixing is present in \refeq{eq:SMEFT-RGE}.  The only exceptions are the
models $S_1$ and $U_1$ because they predict at the scale $\muNP$ the relation
$\Wc[(1)]{\ell q} = \mp \Wc[(3)]{\ell q}$, see \refeq{eq:rel-lq3-lq1}. As can be
seen from \refeq{eq:SMEFT-matching-ddllvv} below, as a consequence at tree level
their contribution to $d_i \ell_a\to d_j \ell_b$ or $d_i \nu_a\to d_j \nu_b$
vanishes, respectively.  Still, in this case a non-vanishing contribution at
$\muEW$ arises then due to gauge mixing of both operators
\cite{Feruglio:2016gvd}.  This mixing is given by \cite{Alonso:2013hga}
\begin{align}
  \label{eq:RGE-S1-lq1-lq3}
  \big[\Wc[(1)]{\ell q} +  \Wc[(3)]{\ell q}\big]_{prst}(\muEW) &
  = \frac{\alpha_e}{4 \pi} \frac{1}{s_W^2}
  \left( 2 \wc[(1)]{\ell q}{prww} \delta_{st}  
  \right) \ln \frac{\muNP}{\muEW} + \ldots
\\
  \label{eq:RGE-U1-lq1-lq3}
  \big[\Wc[(1)]{\ell q} - \Wc[(3)]{\ell q}\big]_{prst}(\muEW) &
  = \frac{\alpha_e}{4 \pi} \frac{1}{s_W^2}
  \left( 2 \wc[(1)]{\ell q}{prww} \delta_{st} - 12 \wc[(1)]{\ell q}{prst} 
  \right) \ln \frac{\muNP}{\muEW} + \ldots
\end{align}
where dots indicate neglected terms $\propto g_1$, which contribute only for
$s \neq t$ and constitute a correction of less than 4\%. From
\refeq{eq:RGE-S1-lq1-lq3} follows that even gauge-mixing does not induce
non-vanishing contributions to $d_i \ell_a \to d_j \ell_b$ in the $S_1$ model
for $i\neq j$. The dots indicate in principle also one-loop matching corrections
to $d_i \ell_a\to d_j \ell_b$ or $d_i \nu_a\to d_j \nu_b$ processes, which are
however not logarithmically enhanced. Once the data on this processes improve it
would be of interest to calculate them.

The new physics contribution to the Wilson coefficients of the $\Delta F=1$
semi-leptonic operators \refeq{eq:DF1-eff-H} at $\muEW$ in terms of the
semi-leptonic SMEFT Wilson coefficients at $\muEW$ is given as follows
\cite{Alonso:2014csa, Buras:2014fpa, Aebischer:2015fzz}
\begin{equation}
  \label{eq:SMEFT-matching-ddllvv}
\begin{aligned}
  C_{9,{\rm NP}}^{baji} &
  = \big[\Wc{qe} + \Wc[(1)]{\ell q} +\Wc[(3)]{\ell q} \big]_{baji} , &
 \quad
  C_{9',{\rm NP}}^{baji} &
  = \big[\Wc{ed} + \Wc{\ell d} \big]_{baji} ,
\\
  C_{10,{\rm NP}}^{baji} &
  = \big[\Wc{qe} - \Wc[(1)]{\ell q} - \Wc[(3)]{\ell q} \big]_{baji} , &
 \quad
  C_{10',{\rm NP}}^{baji} &
  = \big[\Wc{ed} - \Wc{\ell d} \big]_{baji} ,
\\
  C_{L,{\rm NP}}^{baji} &
  = \big[\Wc[(1)]{\ell q} - \Wc[(3)]{\ell q} \big]_{baji} , &
 \quad
  C_{R,{\rm NP}}^{baji} &
  = \big[\Wc{\ell d} \big]_{baji} ,
\\
  C_{S,{\rm NP}}^{baji}  &
   = - C_{P,{\rm NP}}^{baji} = \big[\Wc{\ell edq} \big]^\ast_{abij} ,  &
 \quad
   C_{S',{\rm NP}}^{baji}  &
  = C_{P',{\rm NP}}^{baji} = \big[\Wc{\ell edq} \big]_{baji} . 
\end{aligned}
\end{equation}
Here contributions from $Z$-mediating $\psi^2 H^2 D$--SMEFT operators
$\Op[(1,3)]{Hq}$ to $C_{9,10,L}$ and $\Op{Hd}$ to $C_{9',10',R}$, respectively,
have been omitted. In rare FCNC Kaon decays scalar and pseudo-scalar Wilson
coefficients are negligible and hence do not enter the phenomenological analysis
below.

For completeness we provide the low-energy effective Hamiltonian for
$d_i \nu_a \to u_j \ell_b$
\begin{align}
  {\cal H}_{d\to u\ell\nu} &
  = - \frac{4 G_F}{\sqrt{2}} V_{ji} \sum_k C_k^{baji} Q_k^{baji}
  + \mbox{h.c.}
\end{align}
that contains the operators
\begin{equation}
\begin{aligned}
  Q_{V_{L(R)}}^{baji} &
  = [\bar{u}_j \gamma_\mu P_{L(R)} d_i] [\bar{\ell}_b \gamma^\mu P_L \nu_a] ,
\\
  Q_{S_{L(R)}}^{baji} &
  = [\bar{u}_j P_{L(R)} d_i] [\bar{\ell}_b P_L \nu_a] ,
\\
  Q_{T}^{baji} &
  = [\bar{u}_j \sigma_{\mu\nu} P_{L} d_i] [\bar{\ell}_b \sigma^{\mu\nu} P_L \nu_a] .
\end{aligned}
\end{equation}
Their Wilson coefficients are \cite{Aebischer:2015fzz}
\begin{equation}
\begin{aligned}
  C_{V_L,{\rm NP}}^{baji} &
  = v^2 \frac{V_{jk} \wc[(3)]{\ell q}{baki}}{V_{ji}}, &
  C_{S_L,{\rm NP}}^{baji} &
  = \frac{v^2}{2} \frac{V_{jk} \wc[(1)]{\ell equ}{abik}^\ast}{V_{ji}} , & 
  C_{T,{\rm NP}}^{baji} &
  = \frac{v^2}{2} \frac{V_{jk} \wc[(3)]{\ell equ}{abik}^\ast}{V_{ji}} ,
\\
  C_{V_R,{\rm NP}}^{baji} &
  = 0, &
  C_{S_R,{\rm NP}}^{baji} &
  = \frac{v^2}{2} \frac{V_{jk} \wc{\ell edq}{abik}^\ast}{V_{ji}} , &
  &
\end{aligned}
\end{equation}
where summation over the index $k$ is implied. The SM contributes only to
$C_{V_L,{\rm SM}}^{baji} = - \delta^{ab}$.

From \refeq{eq:SMEFT-matching-ddllvv} and also \refeq{eq:RGE-U1-lq1-lq3} we
conclude that contributions to $\epe$ in all LQ models with non-vanishing
$\Wc[(1,3)]{\ell q}$ and/or $\Wc{\ell d}$ can be constrained by $\kpn$ and
$\klpn$ because of their dependence on imaginary parts of the relevant
semi-leptonic couplings. In the case of $U_1$ there are no NP contributions to
$\kpn$ and $\klpn$ at $\muNP$, but as seen in \refeq{eq:RGE-U1-lq1-lq3} they can
be generated through RG effects. However, as shown below, they appear to be too
small to provide a useful bound at present, although they could turn out to be
relevant when the data from NA62 and KOTO will be available.

As we only need the imaginary part of the relevant semi-leptonic couplings to
enhance $\epe$ the bound on $\klm$, being sensitive only to the real parts of
these couplings, does not play any role. On the other hand $\ksm$ and $\klpll$
are sensitive to imaginary parts and as we will see below already the
experimental upper bound on $\klpll$ in \refeq{KLee} and \refeq{KLmm} and the
new upper bound on $\ksm$ from LHCb \cite{Aaij:2017tia} in \refeq{ksmbound}
provide powerful constraints on the electronic and muonic LQ couplings in the
$U_1$ model. Similar comments apply to $R_2$ and $V_2$ where the contributions
to $\epe$ and $K \to \pi\nu\bar\nu$ are governed by different coefficients and
again the constraints on $\epe$ from $\klpll$ and $\ksm$ play important roles.

%
%

\subsection{\boldmath 
  $\Delta F = 2$ operators: SMEFT on $\Delta F = 2$ EFT
  \label{sec:DeltaF2-SMEFT-matching}
}

The matching equations of SMEFT on the low-energy effective theory 
\refeq{eq:DF2-hamiltonian} for down-type $\Delta F=2$ reads 
\cite{Aebischer:2015fzz}
\begin{equation}
  \label{eq:GSM-DF2-matching}
\begin{aligned}
  C_{\rm VLL}^{ji} & =
     - {\cal N}_{ji}^{-1} \left( \wc[(1)]{qq}{jiji} + \wc[(3)]{qq}{jiji} \right) , &
  C_{\rm VRR}^{ji} & = - {\cal N}_{ji}^{-1} \wc{dd}{jiji} ,
\\
  C_{{\rm LR}, 1}^{ji} & =
   -{\cal N}_{ji}^{-1} \left( \wc[(1)]{qd}{jiji} 
            - \frac{\wc[(8)]{qd}{jiji}}{2 N_c} \right) , &
  C_{{\rm LR}, 2}^{ji} & = {\cal N}_{ji}^{-1} \wc[(8)]{qd}{jiji} ,
\end{aligned}
\end{equation}
where ${\cal N}_{ji}$ is defined in \refeq{eq:DF2-hamiltonian} and all Wilson
coefficients are evaluated at the scale $\muEW$.  The
corresponding results for up-type $\Delta F=2$ processes can be obtained by
replacing the Wilson coefficients $\Wc{dd}\to \Wc{uu}$ and
$\Wc[(8)]{qd} \to \Wc[(8)]{qu}$.

We point out that semi-leptonic Wilson coefficients at $\muNP$ do not contribute
to non-leptonic $\Delta F = 2$ Wilson coefficients of down-type processes at
$\muEW$ via EW gauge-mixing as is the case for $\epe$ and has been discussed in
detail in \refsec{sec:SMEFT-RGE}.  This can be seen for $\Wc[(1,3)]{qq}$ from
\refeq{eq:4-quark-RGE-qq1} and \refeq{eq:4-quark-RGE-qq3}, which are
$\propto \delta_{pr}$ or $\delta_{st}$ and the same holds for $\Wc{dd}$, compare
\refeq{eq:4-quark-RGE-dd}.  These Wilson coefficients receive non-vanishing
contributions at one-loop at the scale $\muNP$ from box-diagams involving as
internal particles LQs and leptons. We provide explicit one-loop matching
results for the SMEFT Wilson coefficients at the scale $\muNP$ in
\refapp{app:1loop-decoupl} for the scalar LQ models $S_{1,3}$, $\wTil{S}_1$,
$R_2$ and $\wTil{R}_2$. In the case of vector LQs loop calculations are
problematic in the absence of a full UV completion, but we will be able to make
some statements on the Dirac structure of contributing operators in
\refsec{DMKEPS} with interesting implications for LQ contributions to $\epe$ and
rare decays in the case of $U_1$ and $V_2$ models.

%
%
%

\section{\boldmath Implications for $\epe$
  \label{sec:4}
}

The results of the previous section allow to determine the impact of LQ
contributions from EW gauge-mixing on $\epe$ in models with scalar and vector
LQs, whereas QCD penguin and box contributions are available for models with
scalar LQs. In the following we will assume that they are the origin of the
discrepancy between the SM prediction \refeq{eq:epe-LBGJJ} and the experimental
value \refeq{eq:epe-EXP} of $\epe$, responsible for at least a value of
$\kepe = 0.5$.

In the case of EW gauge-mixing contributions, $\epe$ depends on the imaginary
parts of the combinations
\begin{align}
  \label{eq:def-SMEFT-comb-epe}
  \Wc{L}(\muNP) &
  \equiv \sum_a \big[\Wc[(1)]{\ell q} + \Wc{qe} \big]_{aa21} \ln \frac{\muNP}{\muEW} , &
  \Wc{R}(\muNP) &
  \equiv \sum_a \big[\Wc{\ell d} + \Wc{ed} \big]_{aa21} \ln \frac{\muNP}{\muEW} ,
\end{align}
that appear in \refeq{eq:DeltaS=1-matching}. For the three cases summarised
in \reftab{tab:epe-LQ-contr} the bound~\refeq{eq:epe-SM+NP} on
$\kepe$ with \refeq{eq:epe-seminum} and \refeq{eq:DeltaS=1-matching} implies
\begin{equation}
  \label{eq:epeNP-SMEFT}
\begin{aligned}
  \mbox{I} &:& 
  \frac{0.5 \times 10^{-3}}{{\cal N}_{\epe}} & 
  \leq P_7 \; \mbox{Im} \left[ \Wc{L}(\muNP) \right]
\\
  \mbox{II} &:& 
  \frac{0.5 \times 10^{-3}}{{\cal N}_{\epe}} & 
  \leq - \left( \frac{P_5}{4} + P_9 \right) \mbox{Im} \left[ \Wc{R}(\muNP) \right]
\\
  \mbox{III} &:& 
  \frac{0.5 \times 10^{-3}}{{\cal N}_{\epe}} & 
  \leq P_7 \; \mbox{Im} \left[ \Wc{L}(\muNP) \right]
  - \left( \frac{P_5}{4} + P_9 \right) \mbox{Im} \left[ \Wc{R}(\muNP) \right]
\end{aligned}
\end{equation}
where we have neglected $P_3 \ll P_7$ (see \reftab{tab:epe-seminum}) and used
that $\lambda_u^{sd}$ is real.\footnote{Note that throughout the quark-mixing
  matrix $V$ is unitary, but in the presence of LQ contributions, the numerical
  values can differ from those obtained in SM fits for the CKM matrix.  We
  assume that LQ contributions do not lift the hierarchy in the Cabibbo angle.}
The numerical factor ${\cal N}_{\epe}$ is
\begin{align}
  {\cal N}_{\epe} &
  \equiv - \frac{4}{9} \frac{\alpha_e}{4\pi} \frac{v^2}{c_W^2} \frac{1}{\lambda_u^{sd}}
  \approx - 100 \,\mbox{GeV}^2.
\end{align}

The contributions of QCD-penguins \refeq{eq:DF1-QCD-peng-matching} and
box-diagrams \refeq{eq:LQ-box-dd}-\refeq{eq:LQ-box-qd} can be taken into account
for models with scalar LQs. According to \refeq{eq:scaling-EW-mix-QCDpeng-box},
they can be parametrically enhanced compared to the EW gauge-mixing, but the
strong hierarchy of $P_{i\neq7,8} \ll P_{7,8}$ in \refeq{eq:epe-seminum} can
lift this enhancements for models that generate $C_7$ via EW gauge-mixing.

The numerical analysis of $\epe$ in models $\wTil{S}_1$ and $\wTil{R}_2$ without
the enhanced contributions $\sim C_{7}$ from EW gauge-mixing nor $\sim C_8$ from
box-diagrams shows indeed
\begin{align}
  \label{eq:epe-wTil-S1,R2}
  (\epe)_{\rm NP} &
  = \frac{ (35  \, \mbox{GeV})^2}{M_{\rm LQ}^2} \left\{
    \begin{array}{rlcc}
      +\mbox{Im}\Big[ \frac{\Sigma^{21}_R}{\lambda_u^{sd}} & 
      \Big( \ln\frac{M_{\rm LQ}}{\muEW} + 0.55 + 2.50\, \Sigma^{11}_R \Big) \Big] 
      & \quad \mbox{for} & \quad \wTil{S}_1 \\[0.3cm]
      -\mbox{Im}\Big[ \frac{\Sigma^{12}_L}{\lambda_u^{sd}} & 
      \Big( \ln\frac{M_{\rm LQ}}{\muEW} - 1.10 - 5.00\, \Sigma^{11}_L \Big) \Big]
      & \quad \mbox{for} & \quad \wTil{R}_2 
    \end{array} \right.
\end{align}
with similar size of coefficients in parentheses for the 1st term
$\propto \ln(M_{\rm LQ}/\muEW)$ from EW gauge-mixing, the 2nd term from QCD
penguins and the 3rd term $\propto \Sigma^{11}_{\chi}$ from box diagrams. The
QCD penguins amount to a contribution of $24\,(12)$\% and $48\,(24)$\% of the EW
gauge-mixing term in both models respectively, for $\muEW = 100$~GeV and
$M_{\rm LQ} = 1 (10)$~TeV. The contribution of box-diagrams to $(\epe)_{\rm NP}$
depends strongly on the magnitude of $\Sigma^{11}_\chi$.  We note that
$\Sigma^{11}_\chi$ is by definition \refeq{eq:def-Sigma_LQ} real-valued and
strictly positive. This leads to the fixed constructive and destructive
interference behavior between EW gauge-mixing and box-diagram terms in both
models $\wTil{S}_1$ and $\wTil{R}_2$, respectively.

In the model $S_3$ the EW gauge-mixing dominates $(\epe)_{\rm NP}$ because it is
$\sim C_7$, whereas QCD-penguin and box-diagrams generate only $C_{i\neq 7,8}$
such that
\begin{equation}
  \label{eq:epe-S3} 
\begin{aligned}
  (\epe)_{\rm NP} &
  = \frac{ -(328 \, \mbox{GeV})^2}{M_{\rm LQ}^2} 
    \mbox{Im} \bigg[ \frac{\Sigma^{21}_L}{\lambda_u^{sd}} \;
\\ & \times 
      \Big( \ln\frac{M_{\rm LQ}}{\muEW} + 0.02 + 0.14\, \Sigma^{11}_L
            - 0.13 \, \frac{\sum_{ij} V_{1j}^{} V_{1i}^\ast \; \Sigma^{2j}_L \Sigma^{i1}_L }
                           {\Sigma^{21}_L} + \ldots\Big) \bigg] .
\end{aligned}
\end{equation}
Even stronger suppressed terms are indicated by the dots. Note the numerical
cancellation of the box contribution $\propto \Sigma^{11}_L$ with the one from
the sum for $ij= 11$ since $|V_{11}|^2 \approx 1$. In this model
$(\epe)_{\rm NP}$ is dominated by EW gauge-mixing, which leads to very strong
correlations with other rare Kaon processes.

In the models $S_1$ and $R_2$ the EW gauge-mixing is also enhanced in
$(\epe)_{\rm NP}$ by the large coefficient $P_7$, but here also box
contributions are enhanced by $P_8$ due to \refeq{eq:LQ-box-qu}, whereas QCD
penguins are negligible. In particular for $S_1$
\begin{equation}
  \label{eq:epe-S1} 
\begin{aligned}
  (\epe)_{\rm NP} &
  = \frac{ -(190 \, \mbox{GeV})^2}{M_{\rm LQ}^2} 
    \mbox{Im} \bigg[ \frac{\Sigma^{21}_L}{\lambda_u^{sd}} \;
\\ & \times 
      \Big( \ln\frac{M_{\rm LQ}}{\muEW} + 0.02 + 0.08\, \Sigma^{11}_L
            + 0.02 \, \sum_{ij} V_{1j}^{} V_{1i}^\ast \; \Sigma^{ji}_L 
            - 19.5 \, \Sigma^{11}_R \Big) \bigg]
\end{aligned}
\end{equation}
and $R_2$
\begin{equation}
  \label{eq:epe-R2} 
\begin{aligned}
  (\epe)_{\rm NP} &
  = \frac{ +(268 \, \mbox{GeV})^2}{M_{\rm LQ}^2} 
    \mbox{Im} \bigg[ \frac{\Sigma^{12}_R}{\lambda_u^{sd}} \;
\\ & \times 
      \Big( \ln\frac{M_{\rm LQ}}{\muEW} - 0.01 - 0.04\, \Sigma^{11}_R
            + 0.05 \, \frac{\sum_{ij} V_{1j}^{} V_{1i}^\ast \; 
                      \Sigma^{j2}_R \Sigma^{1i}_R }{\Sigma^{12}_R} 
            + 9.8 \, \Sigma^{11}_L \Big) \bigg] ,
\end{aligned}
\end{equation}
only the last terms from box diagrams $\sim P_8$ are sizeable in addition to the
EW gauge-mixing contributions. Again a fixed interference behavior arises in
both models due to the positive definite $\Sigma^{11}_\chi$.  Note that these
box terms involve both chiralities $\chi \neq \chi'$:
$(\epe)_{\rm NP} \propto \Sigma^{12}_\chi \, \Sigma^{11}_{\chi'}$.
  
The models with vector LQs yield for the EW gauge-mixing part only
\begin{align}
  \label{eq:epe-vector-LQs}
  (\epe)_{\rm NP} &
  = \frac{\ln(M_{\rm LQ}/\muEW)}{\lambda^{sd}_u \, M_{\rm LQ}^2} \times \left\{
    \begin{array}{llcc}
      + (268 \, \mbox{GeV})^2\, \mbox{Im} \big[ \Sigma^{21}_L - 0.0168\, \Sigma^{21}_R \big]
      & & \quad U_{1}       
    \\[0.2cm]
      - (380 \, \mbox{GeV})^2\, \mbox{Im} \big[ \Sigma^{21}_R - 0.0168\, \Sigma^{21}_L \big]
      &  \quad \mbox{for}  & \quad V_2
    \\[0.2cm]
      + (465 \, \mbox{GeV})^2\, \mbox{Im}\, \Sigma^{21}_L 
      & & \quad U_3      
    \end{array} \right.
\end{align}
where the couplings with chirality $\chi = L, R, L$ are enhanced in
$(\epe)_{\rm NP}$ by $C_7$ for models $U_1$, $V_2$ and $U_3$, respectively (see
\reftab{tab:epe-LQ-contr}).  It is evident that for models $U_1$, $V_2$ the
sub-scenarios $U_{1,L}$ and $V_{2,R}$ with only $\Sigma^{21}_{L,R}$
respectively, can accommodate large $(\epe)_{\rm NP}$ easier than the
sub-scenarios $U_{1,R}$ and $V_{2,L}$ with only $\Sigma^{21}_{R,L}$ couplings,
because the latter are suppressed by a factor 60.  The QCD-penguin and
box-diagram contributions do not receive additional enhancement in the model
$U_3$, such that in analogy to the scalar model $S_3$ in \refeq{eq:epe-S3},
where we could calculate analytic results for loop contributions, we believe
that EW gauge-mixing provides the numerically leading contribution.

From the above semi-numerical results for $(\epe)_{\rm NP}$ it is evident that
in the absence or small box-diagram contributions, the requirement of a specific
value of $\kepe$ would fix $\Sigma^{21}_\chi$ for a given value of $\muNP$. This
is indeed the case for the model $S_3$ and we can assume the same for the vector
LQ $U_3$, where QCD-penguin and box-diagram contributions would give rise to the
same $(V-A)\otimes(V-A)$ structures that are suppressed w.r.t. $C_7$ in
$(\epe)_{\rm NP}$. Indeed $|\mbox{Im}\,\Sigma^{21}_\chi|< 0.5$ for
$\muNP \lesssim 20$~TeV in both models when requiring $\kepe = 0.5$, such that
perturbativity issues with LQ couplings arise only for very large LQ masses, as
can be seen in \reffig{fig:Sigma-muNP}.

\newcommand{\myfigfraction}{0.40}
\newcommand{\myfigsepfraction}{0.08}

\begin{figure}
\centering
  \includegraphics[width=\myfigfraction \textwidth]{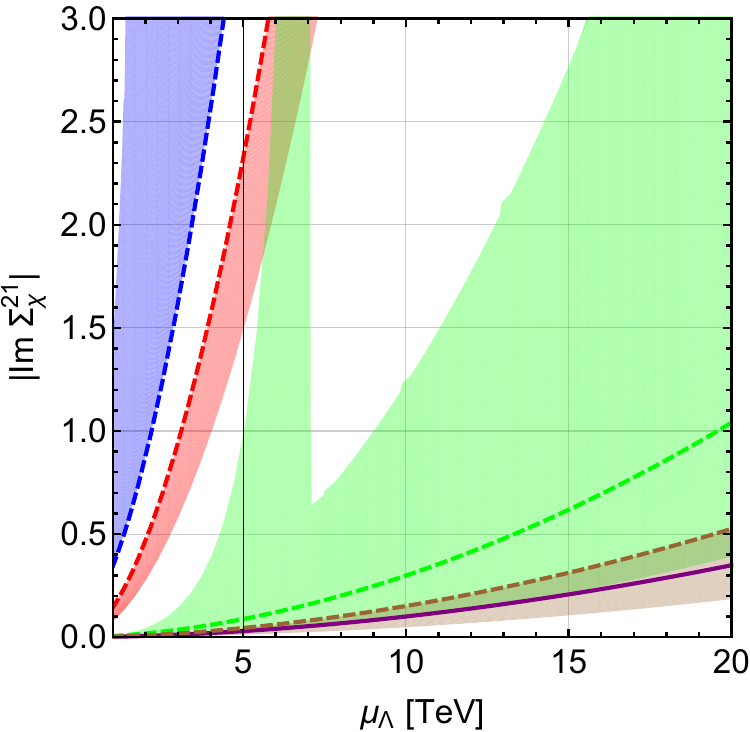}
  \hskip \myfigsepfraction \textwidth
  \includegraphics[width=\myfigfraction \textwidth]{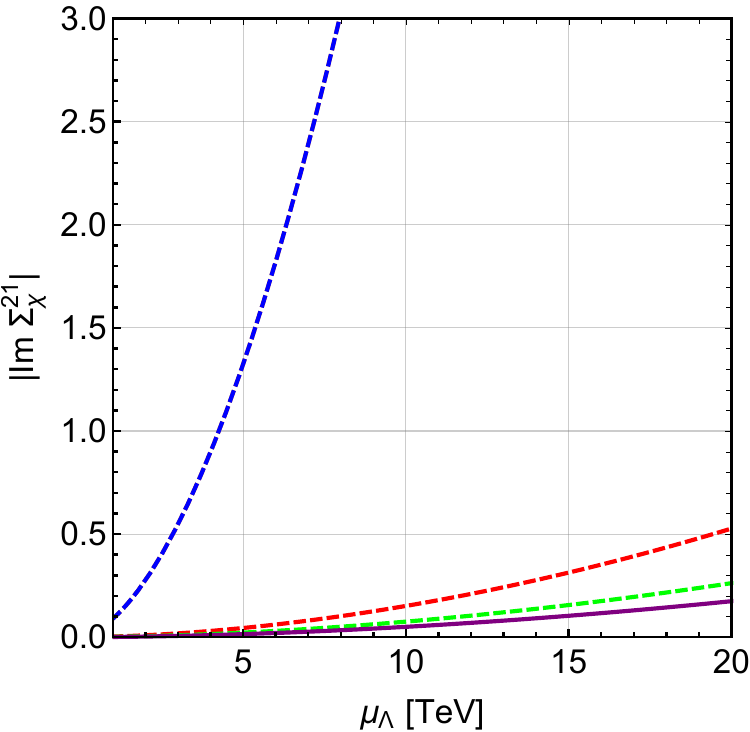}
  \caption{\small The $|{\rm Im}\, \Sigma^{21}_\chi|$ versus the LQ mass
    $\muNP \sim M_{\rm LQ}$ for fixed $\kepe = 0.5$. [Left] for scalar models
    $\wTil{S}_1$ [red], $\wTil{R}_2$ [blue], $S_3$ [purple], $S_1$ [green] and
    $R_2$ [brown]. The bands indicate the variation of
    $\Sigma^{11}_{L,R} \in [0.0,\, 1.0]$ and dashed lines are
    $\Sigma^{11}_{L,R} = 0.0$. [Right] for vector models assuming only EW
    gauge-mixing: $U_{1,R}$ and $V_{2,L}$ [blue, dashed],
    $U_{1,L}$ [red, dashed] and $V_{2,R}$ [green, dashed] and $U_3$ [purple].  }
\label{fig:Sigma-muNP}
\end{figure}

\begin{figure}
\centering
  \includegraphics[width=\myfigfraction \textwidth]{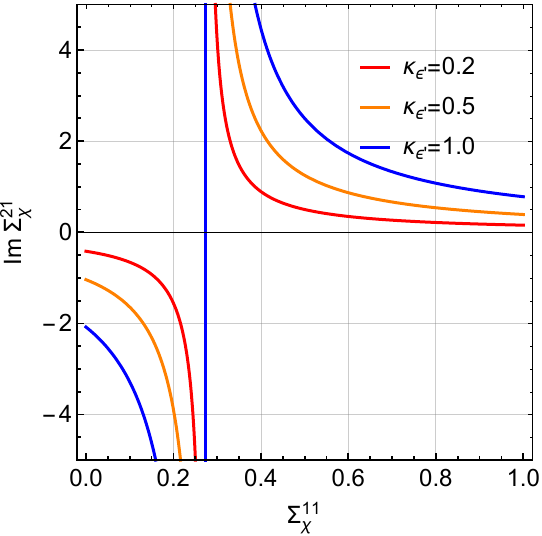}
  \hskip \myfigsepfraction \textwidth
  \includegraphics[width=\myfigfraction \textwidth]{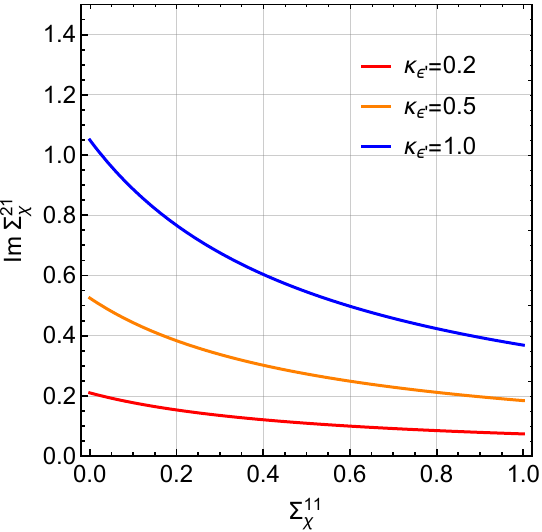}
  \caption{\small The dependence of the Im$\Sigma^{21}_L$ from box contribution
    $\Sigma^{11}_R$ in model $S_1$ [left] and $R_2$~[right] for different values
    of $\kepe$ and $M_{\rm LQ} = 20$~TeV.  }
\label{fig:crr-Sigma21_11-S1}
\end{figure}

As pointed out above, in other models the box diagrams have a fixed interference
behaviour with the EW gauge-mixing term. Therefore a fine-tuned cancellation of
the numerically leading contribution from box-diagrams and the EW gauge-mixing
\footnote{For models with scalar LQs the QCD-penguin contribution is included in
  the numerical analysis.} term can occur only in the models $\wTil{R}_2$ and
$S_1$ with destructive interference, rendering the subleading terms
important. The effect of destructive versus constructive interference on
$\mbox{Im}\, \Sigma^{21}_\chi$ is depicted in \reffig{fig:crr-Sigma21_11-S1} for
the two models $S_1$ and $R_2$, respectively, when varying
$\Sigma^{11}_{\chi'} \in [0.0,\, 1.0]$ for fixed values of $\kepe \neq 0$. In
the model $R_2$ the constructive interference allows to decrease
$\mbox{Im}\, \Sigma^{21}_\chi$ with increasing $\Sigma^{11}_{\chi'}$, which in
turn will lead to smaller effects in other rare Kaon processes that depend only
on $\mbox{Im}\, \Sigma^{21}_\chi$. On the other hand the destructive
interference in model $S_1$ leads for intermediate values of
$\Sigma^{11}_{\chi'}$ to a strong enhancement and sign flip of
$\mbox{Im}\, \Sigma^{21}_\chi$ in order to maintain a fixed value of $\kepe$
when the expression in parentheses in \refeq{eq:epe-S1} vanishes. In this case
rare Kaon processes would receive large contributions.

For the two models $\wTil{S}_1$ and $\wTil{R}_2$ the dependence of
$|\mbox{Im}\,\Sigma^{21}_\chi|$ on $\muNP$ is shown in \reffig{fig:Sigma-muNP}
when requiring $\kepe = 0.5$ and varying in the box-contribution
$\Sigma^{11}_{\chi'} \in [0.0,\, 1.0]$. The $|\mbox{Im}\, \Sigma^{21}_\chi|$
reaches fast a nonperturbative magnitude $> 3.0$ to be able to accommodate
$\kepe = 0.5$, preferring light LQ masses below $8$~TeV as a consequence of the
rather small scale in \refeq{eq:epe-wTil-S1,R2}.  Allowing for even larger
$\Sigma^{11}_{\chi'}$ cannot really ameliorate this situation. In consequence
there will be large enhancements of other rare Kaon processes. A similarly low
scale is present for sub-scenarios $U_{1,R}$ and $V_{2,L}$ in
\refeq{eq:epe-vector-LQs}.  The destructive interference can always lead to a
reduction of the effective scale, such that $|\mbox{Im}\, \Sigma^{21}_\chi|$ has
to become nonperturbative to explain $\kepe = 0.5$ for rather low LQ
masses. Thus it might be more appropriate to focus on either
\begin{enumerate}
\item negligible box contributions,
\item or constructive interference thereby restricting to $\Sigma^{11}_{\chi'} < 1.0$.
\end{enumerate}
These assumptions should increase the viability of the corresponding scenarios.
The results for models $S_1$ and $R_2$ in \reffig{fig:Sigma-muNP} show that
perturbativity of the couplings is guaranteed even at larger LQ masses $>20$~TeV
for suitable choices of $\Sigma^{11}_{\chi'}$.  Moreover at such high LQ masses,
even the constructive interference of the box contributions will reduce the
coupling only by a factor of about two compared to the case when they vanish,
showing that in these models the consideration of only EW gauge-mixing
contributions gives a representative picture for the impact of LQ effects on
$(\epe)_{\rm NP}$.

In the case of vector LQ models $U_{1,3}$ and $V_2$ we will use only the EW
gauge-mixing contribution in our numerical analysis of the perturbativity of
$|\mbox{Im}\, \Sigma^{21}_\chi|$ for $\kepe = 0.5$.  The case of $U_1$ and $V_2$
is at first sight more involved as having both left-handed and right-handed
couplings box contributions to $\epe$ could be important. In our analysis we
will first consider the sub-scenarios with left-handed or right-handed couplings
only.  In this way potentially large left-right contributions to $\Delta M_K$
are absent. The discussion of possible large box contributions to $\epe$ in
these models due to the simultaneous presence of left-handed and right-handed
couplings is postponed to \refsec{sec:5}.  The results in
\reffig{fig:Sigma-muNP} show that nonperturbativity of the couplings is only an
issue for $U_{1,R}$ and $V_{2,L}$.

In our numerical analysis we use analytical formulae and numerical
input as given in \cite{Bobeth:2016llm, Bobeth:2017xry} and described in
\refsec{sec:2}. 

%
%

\subsection{\boldmath
  Constraints from $K \to \pi \nu\bar\nu$
  \label{sec:crr-Kpivv-epe}
}

The decays $\klpn$ and $\kpn$ provide the most efficient constraints on the
combinations \refeq{eq:def-SMEFT-comb-epe} entering $\epe$ and apply to the LQ
models $S_{1,3}$, $\wTil{R}_2$ and $V_2$, $U_{1,3}$. We point out that for the
model $U_1$ with $\Wc[(1)]{\ell q} = \Wc[(3)]{\ell q}$ at $\muNP$ the first
non-vanishing contribution to $K\to \pi \nu\bar\nu$ at the scale $\muEW$ via
$C_L(\muEW) \propto [\Wc[(1)]{\ell q} - \Wc[(3)]{\ell q}] (\muEW) \propto
\alpha_e \ln(\muNP/\muEW) \Wc[(1)]{\ell q}(\muNP)$
is due to leading logarithms from gauge mixing and hence loop-suppressed
\cite{Feruglio:2016gvd}.  Still, below we will find that for $\kepe = 1.0$ this
effect enhances significantly branching ratios for $\klpn$. As explained in
\refsec{sec:Kpivv-basics} the branching fractions involve a sum over all lepton
flavours of the neutrinos in the final state. The LQ contribution in terms of
the SMEFT Wilson coefficients at $\muEW$ \refeq{eq:SMEFT-matching-ddllvv} enter
as
\begin{align}
  X_{\rm LQ}^{ab} &
  = - s_W^2 v^2 \frac{\pi}{\alpha_e}
    \frac{\big[ \Wc[(1)]{\ell q} - \Wc[(3)]{\ell q} + \Wc{\ell d} \big]_{ba21}}{\lambda_t^{sd}} ,
\end{align}
where we will make use of the model-specific relations \refeq{eq:rel-lq3-lq1}
to eliminate $\Wc[(3)]{\ell q}$.
Further, in LQ models the SM$\times$NP term
\begin{align}
  \sum_{a} \mbox{Im} (\lambda_t^{sd}X_{\rm LQ}^{aa}) &
  = - s_W^2 v^2 \frac{\pi}{\alpha_e}
    \sum_a \mbox{Im} \big[ \Wc[(1)]{\ell q} (1 - r_{\rm LQ}) + \Wc{\ell d} \big]_{aa21}\,,
\end{align}
with $r_{\rm LQ}$ given in \refeq{eq:rel-lq3-lq1}, is related to
\refeq{eq:def-SMEFT-comb-epe} entering $\epe$ since in a particular LQ model
only either $\Wc[(1,3)]{\ell q}$ or $\Wc{\ell d}$ are non-vanishing.  Note that
here the $\Wc[(n)]{m}$ are at the scale $\muEW$ whereas in
\refeq{eq:def-SMEFT-comb-epe} at $\muNP$. But for our purpose the self-mixing
can be neglected since it is loop-suppressed, such that we equate the Wilson
coefficients at both scales.

It is without much loss of generality to assume a hierarchy of the LQ couplings
such that a single $\wc[(n)]{m}{aa21} \propto g_{1a}^{k\chi} g_{2a}^{k\chi\ast}$
or $\propto h_{2a}^{k\chi} h_{1a}^{k\chi\ast}$ for specific $a = a'$ dominates
$(\epe)_{\rm NP}$, in particular one might expect weakest constraints on
third-generation lepton couplings of LQs. In consequence, also the SM$\times$NP
contribution to $\BR(\klpn)$ will be dominated by this specific coupling and
allow for a simple analytic correlation of $\epe$ and $\BR(\klpn)$.  Concerning
the NP$\times$NP term, the omission of terms $a\neq b$ in the sum in
\refeq{eq:Br-klpn-NP} will result always in a lower prediction compared to the
true value of $\BR(\klpn)$, i.e. a lower bound on the impact of LQ
contributions.

\begin{figure}[ht]
\centering
  \includegraphics[width=\myfigfraction \textwidth]{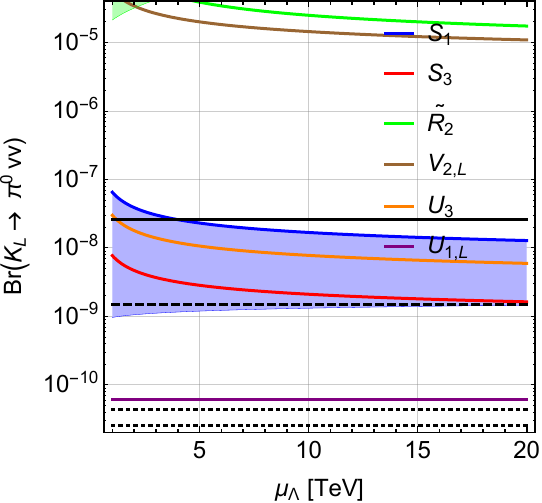}
  \hskip \myfigsepfraction \textwidth
  \includegraphics[width=\myfigfraction \textwidth]{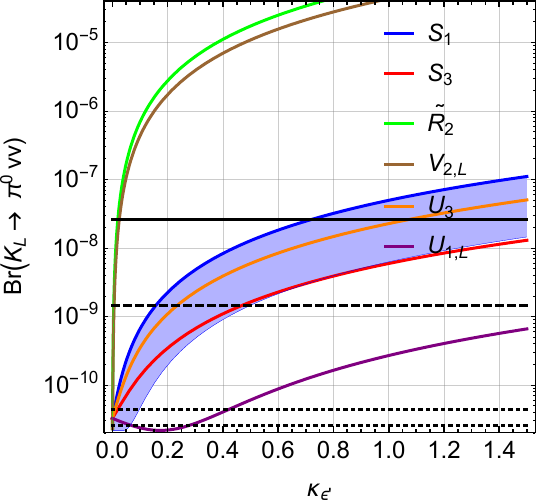}
\\[0.3cm]
  \includegraphics[width=\myfigfraction \textwidth]{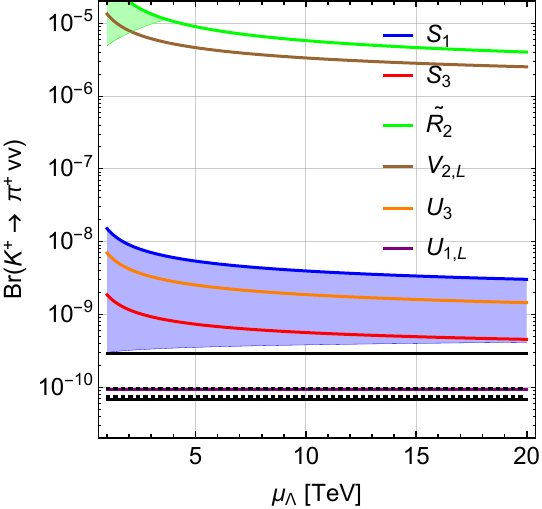}
  \hskip \myfigsepfraction \textwidth
  \includegraphics[width=\myfigfraction \textwidth]{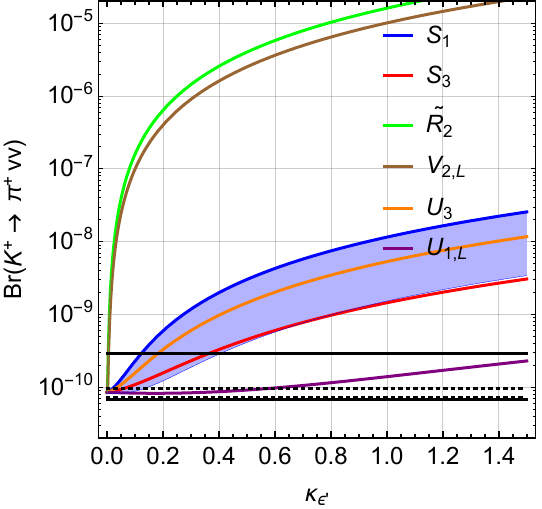}  
  \caption{\small The dependence of $\BR(\klpn)$ [upper] and $\BR(\kpn)$ [lower]
    on the new physics scale $\muNP \sim M_{\rm LQ}$ for $\kepe = 0.5$ [left]
    and on $\kepe$ for $\muNP = 20$~TeV [right], assuming the dominance of a
    single lepton-flavour coupling that is purely imaginary. For $U_{1,L}$ and
    $V_{2,L}$, it is assumed that only $L$-LQ couplings are present and saturate
    $\kepe$. The blue (green) band for $S_1$ ($\wTil{R}_2$) is due to variation
    of the box-diagrams via $\Sigma_{R(L)}^{11} \in [0.0,\, 1.0]$.  Further
    shown are the current upper experimental bound on $\BR(\klpn)$ and the
    measurement of $\BR(\kpn)$ \cite{Ahn:2009gb} [black solid], the Grossman-Nir
    bound from the current measurement of $\BR(\kpn)$ [black dashed] and the SM
    prediction \cite{Bobeth:2016llm} [black dotted].  }
\label{fig:crr-KLpivv-muNP}
\end{figure}

\begin{figure}
\centering
  \includegraphics[width=\myfigfraction \textwidth]{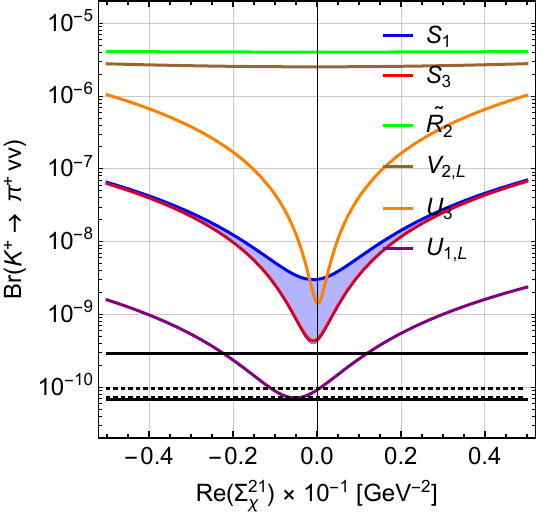}
  \hskip \myfigsepfraction \textwidth
  \includegraphics[width=\myfigfraction \textwidth]{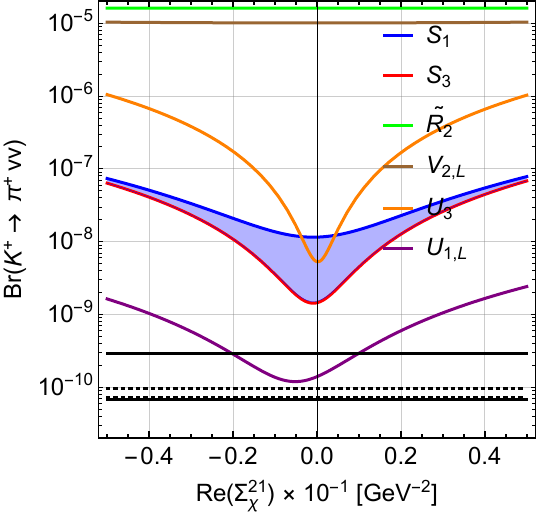}
  \caption{\small The dependence of $\BR(\kpn)$ on a non-zero real part of the
    dominant lepton-flavour coupling, for $\kepe = 0.5$ [left] and $\kepe = 1.0$
    [right], with $\muNP = 20$~TeV.  Further shown are the current measurement
    of $\BR(\kpn)$ \cite{Ahn:2009gb} [black solid] and the current SM
    predictions \cite{Bobeth:2016llm} [black dotted].  }
\label{fig:crr-Kplpivv-realP}
\end{figure}

With the assumption of the dominance of a single coupling and the requirement
that it induces at least a value of $\kepe = 0.5$, we can plot $\BR(\klpn)$ vs.
$\muNP \in [1, 20]$~TeV shown in \reffig{fig:crr-KLpivv-muNP}. We set
$\muEW = 100$~GeV and assume for the moment that box-diagram contributions to
$\epe$ discussed for scalar LQ models in
\refeq{eq:epe-wTil-S1,R2}--\refeq{eq:epe-R2} are vanishing.  The correlation
between $\BR(\klpn)$ and $\epe$ is due to their common dependence on
$\Wc[(1)]{\ell q}$ for $S_{1,3}$, $U_{3}$ or $\Wc{\ell d}$ for $\wTil{R}_2$. We
show also the correlations in $V_{2,L}$ and $U_{1,L}$ under the assumption that
only the $\chi = L$ couplings saturate $\kepe$ via $\Wc{\ell d}$ and
$\Wc[(1)]{\ell q}$, respectively.  This assumption is justified for these models
as in the presence of both $L$ and $R$ couplings a very strong enhancement of
$\Delta M_K$ through left-right operators would be possible placing strong
constraints on these couplings, despite that box diagrams cannot be calculated
reliably without a UV completion.

In general the enhancement of $\BR(\klpn)$ is smaller the larger $\muNP$. The
plot shows that for $\wTil{R}_2$ and $V_{2,L}$ the $\BR(\klpn)$ will be above
the current experimental bound $2.6 \times 10^{-8}$ \cite{Ahn:2009gb} and orders
of magnitude above the SM prediction, which excludes these models as an
explanation of $\kepe = 0.5$.  Furthermore the models $S_{1,3}$ and $U_3$ give
predictions above the Grossman-Nir bound (see \refsec{sec:Kpivv-basics}) and
they are almost two orders of magnitude above the SM prediction, thus being also
excluded for all practical purposes.  We note that it is expected that the final
analysis of the 2015 data collected with the KOTO experiment will approach the
sensitivity to the Grossman-Nir bound \cite{Beckford:2017gsf}.  We plot also
$\BR(\klpn)$ versus $\kepe$ for fixed $\muNP = 20$~TeV, showing that for larger
values of $\kepe$ the enhancement of $\BR(\klpn)$ becomes even more severe. The
couplings of the $U_{1,L}$ model enter only via RG effects described in
\refeq{eq:RGE-U1-lq1-lq3} and in this case the dependence on $\muNP$ cancels.
The enhancement of $\BR(\klpn) \sim 6\, (27) \times 10^{-11}$ is a factor 2 (9)
above the SM prediction for $\kepe = 0.5\, (1.0)$ and might be tested in the
long run of the KOTO experiment.

So far our numerical analysis neglected box-diagram contributions to $\epe$
presented for scalar LQ models in \refeq{eq:epe-wTil-S1,R2}--\refeq{eq:epe-R2}.
As pointed out there, for the model $S_3$ box contributions are suppressed and
we expect the same for $U_3$. We find for the model $\tilde{R}_2$ only
enhancement of $\BR(\klpn)$ when varying $\Sigma^{11}_L \in [0.0,\, 1.0]$,
except for small $M_{\rm LQ} \lesssim 4$~TeV, but of negligible
size. Box-diagrams in $\epe$ are more important in model $S_1$ as can be seen by
the band in \reffig{fig:crr-KLpivv-muNP} that is due to the variation of
$\Sigma^{11}_R \in [0.0,\, 1.0]$. This band shows only how box-diagrams lead to
a lowering of $\BR(\klpn)$, but for some values of
$\Sigma^{11}_R \in [0.0,\, 1.0]$ there is also enhancement w.r.t. to the
prediction at $\Sigma^{11}_R = 0$, which is not shown. Going beyond
$1.0 < \Sigma^{11}_R < 2.0$ will allow even lower $\BR(\klpn)$, but still
$\BR(\klpn) > 2 \times 10^{-10}$ is about one order of magnitude larger than the
SM prediction \refeq{eq:SM-Br-klpn} for $\kepe = 0.5$.

Despite the additional dependence on real parts of couplings the $\BR(\kpn)$
leads to similar conclusions, which can be expected from the qualitative
discussion above. We show the correlation of $\BR(\kpn)$ vs.
$\muNP \in [1, 20]$~TeV in \reffig{fig:crr-KLpivv-muNP} setting real parts of
couplings to zero. The effect of the real couplings is illustrated in
\reffig{fig:crr-Kplpivv-realP} for fixed $\muNP = 20$~TeV. The models
$\wTil{R}_2$ (and $V_{2,L}$) and $S_1$, $U_3$ require $\BR(\kpn) > 10^{-9}$,
which is at least a factor four above the central value of the current
measurement \cite{Ahn:2009gb}, whereas $S_3$ is close to the one sigma
region. For $\kepe = 1.0$, the branching ratios for models in question are all
above $10^{-9}$ and excluded, except for $U_{1,L}$, where the enhancement of
$\BR(\kpn)$ is very small: a factor 1.1 (1.6) for $\kepe = 0.5 \, (1.0)$. In the
near future the NA62 experiment at CERN will be able to measure $\BR(\kpn)$ with
10\% uncertainty at the level of the SM prediction, thus being able to
investigate these scenarios further. Also the improved value of $\kepe$ from
lattice QCD will be very important here.

%
%

\subsection{\boldmath
  Constraints from $\ksm$
}

\begin{figure}
\centering
  \includegraphics[width=\myfigfraction \textwidth]{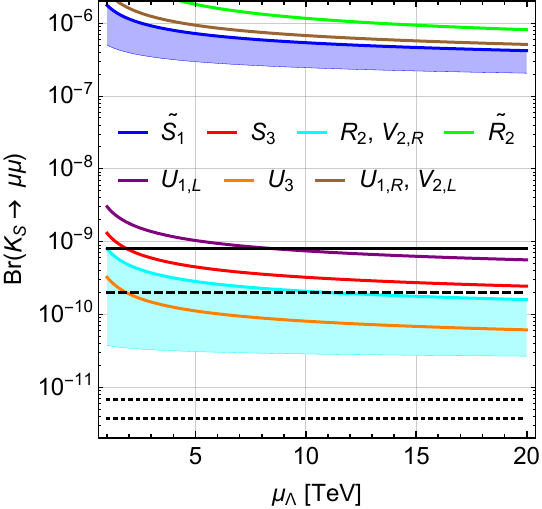}
  \hskip \myfigsepfraction \textwidth
  \includegraphics[width=\myfigfraction \textwidth]{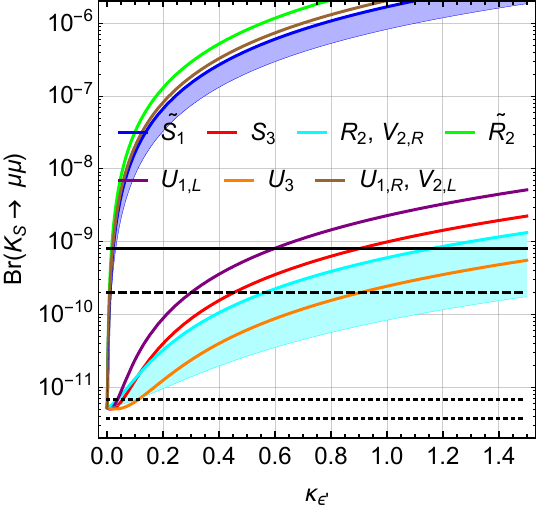}
  \caption{\small The dependence of $\BR(\ksm)$ on the new physics scale
    $\muNP \sim M_{\rm LQ}$ for $\kepe = 0.5$~[left] and on $\kepe$ for
    $\muNP = 20$~TeV [right]. For $U_1$ and $V_2$, it is assumed that either
    only $L$-LQ or only $R$-LQ couplings are present and saturate
    $\kepe$. Further shown are the current upper experimental bound on
    $\BR(\ksm)$ from LHCb [black solid], a conservative future prospect for LHCb
    with 23~fb$^{-1}$ [black dashed] and the SM prediction
    \cite{DAmbrosio:2017klp} [black dotted].  }
\label{fig:crr-KS2mu}
\end{figure}

The branching fraction of $\ksm$ provides constraints on the muonic LQ couplings
in models that generate $\Wc[(1,3)]{\ell q}$, $\Wc{qe}$, $\Wc{\ell d}$ and
$\Wc{ed}$, which are $\wTil{S}_1$, $R_2$, $\wTil{R}_2$, $S_3$, $U_{1,3}$ and
$V_2$. In the LQ model $S_1$ no contribution is generated due to EW gauge-mixing
\refeq{eq:RGE-S1-lq1-lq3}.

Contrary to $K\to\pi\nu\bar\nu$, the decay $\ksm$ depends only on the muonic LQ
couplings, such that a correlation between $\epe$ and $\ksm$ exists only if the
muonic LQ couplings were the origin of large $\kepe$. In such a case large NP
contributions to \refeq{eq:ksm-br-SD}
\begin{align}
  \label{eq:BrKS}
 \frac{ \sqrt{\BR(\ksm)_{\rm SD, NP}}}{ (180 \, {\rm GeV})^2} &
  = \left| {\rm Im} \big[ \Wc{qe} - \Wc[(1)]{\ell q} (1 + r_{\rm LQ})
                   + \Wc{\ell d} - \Wc{ed} \big]_{\mu\mu 21} \right|
\end{align}
are correlated with $\epe$ as can be seen from \refeq{eq:epeNP-SMEFT}.
For the convenience of the reader we provide here also the constraints
on the SMEFT SL-$\psi^4$ Wilson coefficients at $\muEW$ that enter \refeq{eq:BrKS}
when using the experimental bound from LHCb on $\BR(\ksm)$ \refeq{ksmbound}
at 90\% C.L.
\begin{align}
  \label{eq:SLpsi4-bound-Ksmm}
  \left| {\rm Im} \big[ \Wc{qe} - \Wc[(1)]{\ell q} - \Wc[(3)]{\ell q}
                   + \Wc{\ell d} - \Wc{ed} \big]_{\mu\mu21} \right|
  \leq (34\; \mbox{TeV})^{-2}.
\end{align}
Following the spirit of \cite{Carpentier:2010ue}, it allows easily to set bounds
on the imaginary parts when considering one Wilson coefficient
at a time.

Considerable simplifications take place in a given LQ model because not all
Wilson coefficients are present simultaneously. For example for $S_{3}$, $U_3$,
$U_{1,L}$ and $R_2$ ($r_{R_2} = 0$) the dominant LQ contribution to $\epe$
enters via $C_7$ as shown in \reftab{tab:epe-LQ-contr}, such that
\begin{align}
  (\epe)_{\rm NP} & 
  \le \frac{1.3 \times 10^{-4}}{(1 + r_{\rm LQ})} 
  \sqrt{\frac{\BR(\ksm)}{0.8\times 10^{-9}}} \ln\frac{\muNP}{\muEW}.  
\end{align}
While for $\muNP > 10 \tev$, values close to $(\epe)_{\rm NP} \sim 10^{-3}$ are
still allowed, the future improved upper bound on $\BR(\ksm)$ is likely to lower
the upper bound in question below $10^{-4}$.

The dependence of $\BR(\ksm)$ on $\muNP$ for $\kepe = 0.5$ and on $\kepe$ for
$\muNP = 20$~TeV is shown in \reffig{fig:crr-KS2mu} assuming the dominance of
muonic couplings. Under the latter assumption and requiring $\kepe \geq 0.5$,
the current bound on $\BR(\ksm)$ excludes models $\wTil{S}_1$, $\wTil{R}_2$,
$V_{2,L}$ and $U_{1,R}$ and in part also $U_{1,L}$. The models $S_3$, $R_2$,
$V_{2,R}$ will be all probed with higher statistics at LHCb and one can hope
that also $U_3$ will be testable \cite{Dettori:2017}. For the models
$\wTil{S}_1$ and $R_2$ the bands show the weakened constraint once allowing for
box-diagram contributions to $\epe$ due to the variation of
$\Sigma^{11}_{R,L} \in [0.0,\, 1.0]$, respectively, whereas in the model
$\wTil{R}_2$ they do not weaken the constraint.

Concerning $U_1$ and $V_2$ models, the bound given above could be in principle
eliminated through very high fine-tunning with the help of $\Wc{e d}$ and
$\Wc{\ell d}$, respectively. Although they contribute to $\BR(\ksm)$ without
important impact on $\epe$ where they modify only the coefficients $C_9^\prime$
and $C_5^\prime$ the presence of $\chi = L$ and $\chi = R$ couplings of same
size are strongly constrained by the bound from $\Delta M_K$.

%
%

\subsection{\boldmath Constraints from $K_L \to \pi^0 \ell\bar\ell$}

\begin{figure}[ht]
\centering
  \includegraphics[width=\myfigfraction \textwidth]{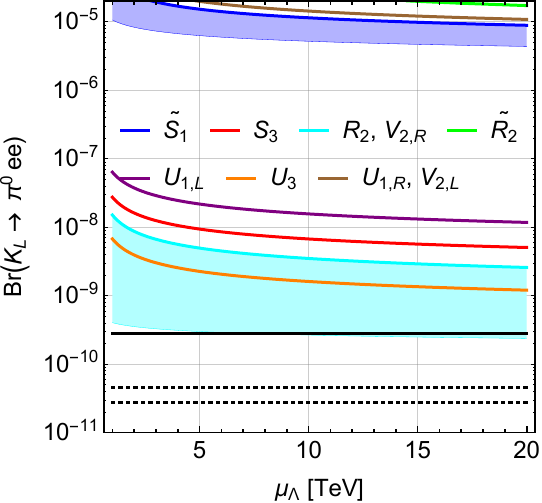}
  \hskip \myfigsepfraction \textwidth
  \includegraphics[width=\myfigfraction \textwidth]{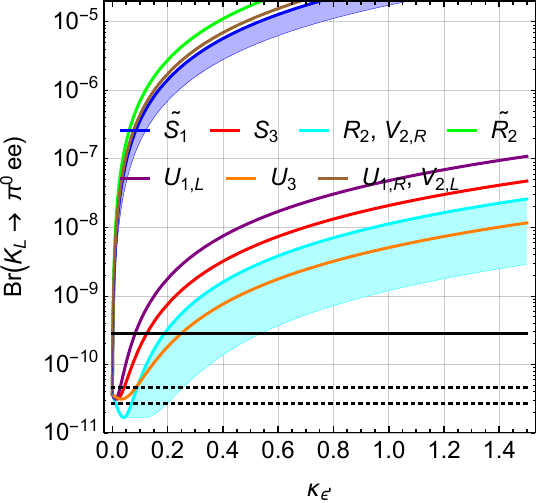}
\\[0.3cm]
  \includegraphics[width=\myfigfraction \textwidth]{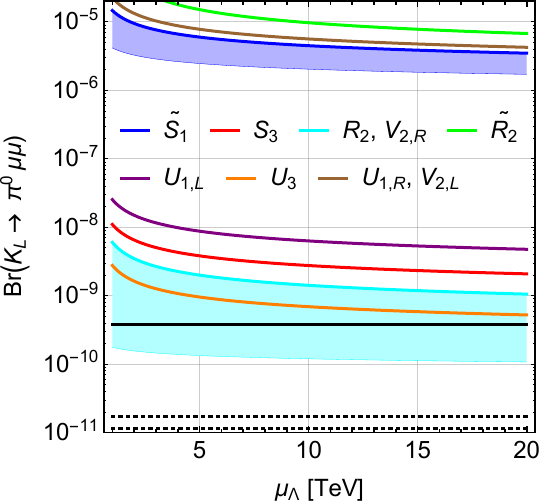}
  \hskip \myfigsepfraction \textwidth
  \includegraphics[width=\myfigfraction \textwidth]{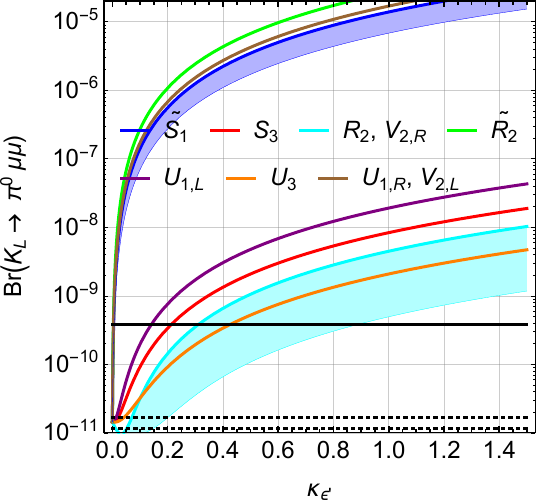}  
  \caption{\small The dependence of $\BR(\klpee)$ [upper] and $\BR(\klpmm)$
    [lower] on the new physics scale $\muNP \sim M_{\rm LQ}$ for $\kepe = 0.5$
    [left] and on $\kepe$ for $\muNP = 20$~TeV [right], assuming the dominance
    of the single electronic or muonic lepton-flavour coupling, respectively,
    that is purely imaginary. For $U_1$ and $V_2$, it is assumed that either
    only $L$-LQ or only $R$-LQ couplings are present and saturate $\kepe$.
    Further shown are the current upper experimental bounds on $\BR(\klpll)$
    [black solid] and the SM predictions [black dotted].  }
\label{fig:crr-KLpi0ll}
\end{figure}

The branching fractions of $\klpee$ and $\klpmm$ constrain the electronic and
muonic LQ couplings in models that generate $\Wc[(1,3)]{\ell q}$, $\Wc{qe}$,
$\Wc{\ell d}$ and $\Wc{ed}$ at tree-level, which are all models that contribute
to $\epe$, except for $S_1$. In contrast to \refeq{eq:SLpsi4-bound-Ksmm}, 
no such simple relation can be given here, but allowing one Wilson coefficient
to contribute at a time, we find similar bounds for the imaginary parts of all electronic 
and muonic SL-$\psi^4$ Wilson coefficients $a = \{qe,\; \ell q^{(1,3)},\; ed,\; 
\ell d\}$
\begin{align}
  \left| {\rm Im} \wc{a}{ee21} \right| \leq (58\; \mbox{TeV})^{-2} ,
\\
  \left| {\rm Im} \wc{a}{\mu\mu21} \right| \leq (50\; \mbox{TeV})^{-2} .
\end{align}
For muonic Wilson coefficients this bound is stronger than the bound
\refeq{eq:SLpsi4-bound-Ksmm} from $\ksm$, which is compatible with our
analysis that shows that the present constraint from $\ksm$ is weaker
than from $\klpll$.

The dependence of $\BR(\klpee)$ and $\BR(\klpmm)$ on $\muNP$ for $\kepe = 0.5$
and on $\kepe$ for $\muNP = 20$~TeV is shown in \reffig{fig:crr-KLpi0ll}
assuming the dominance of electronic and muonic couplings, respectively. These
plots are qualitatively analogous to $\BR(\ksm)$ in \reffig{fig:crr-KS2mu}, but
much more stringent due to the stronger experimental bounds on $\BR(\klpee)$ and
$\BR(\klpmm)$ and in addition also electronic LQ couplings are constrained.  All
LQ models predict enhancements of $\BR(\klpee)$ and $\BR(\klpmm)$ that violate
the current bounds once $\kepe \gtrsim 0.5$ for both $\ell = e, \mu$. This
demonstrates the importance of both observables in connection with LQ
contributions that predict NP to $\epe$. The only way to avoid these bounds but
still to enhance $\epe$ would be via non-vanishing tauonic LQ couplings, which
is ruled out for some LQ models by $K\to \pi\nu\bar\nu$ ($S_1$, $S_3$, $U_3$,
$\wTil{R}_2$ and $V_{2,L}$) and $\Delta M_K$ as will be discussed below
($\wTil{S}_1$, $\wTil{R}_2$ and with increasing size of $M_{\rm LQ}$ also $S_1$,
$S_3$ and $R_2$).

%

\subsection{\boldmath Constraints from $\Delta M_K$ and $\eps_K$}\label{DMKEPS}

As we have seen the strong correlations between $\epe$ and $K\to \pi\nu\bar\nu$
decays originated in the following features. First in both $\epe$ and
$K\to \pi\nu\bar\nu$ a summation over the lepton flavour indices of LQ couplings
appears. Second the mutual dependence on the imaginary parts of the
couplings. Although the latter applies also to $\klpll$ and $\ksm$ decays, the
former is absent such that only the electronic and muonic LQ couplings can lead
to correlations, whereas tauonic LQ couplings can lift them. In this respect,
the off-diagonal elements of the mass-mixing matrix $M_{12}^{sd}$ offer another
set of observables that are sensitive to a summation over lepton-flavour
indices, as can be seen from the expressions in \refapp{app:1loop-decoupl}.  The
relevant observables $\Delta M_K$ and $\eps_K$ were reviewed in
\refsec{sec:DeltaF2-obs-basics}.

As already pointed out in \refsec{sec:DeltaF2-SMEFT-matching}, the LQ
contributions to down-type $\Delta F = 2$ non-leptonic operators are of
different origin then those in $\epe$. They are actually generated at one-loop
at the scale $\muNP$ and provide a loop-suppressed matching contribution with
results given in \refapp{app:1loop-decoupl} for scalar LQ models. But these
results involve a summation of products of LQ couplings over the lepton-flavour
index, very much as appearing in the sum over the semi-leptonic SMEFT Wilson
coefficients entering $\epe$ in \refeq{eq:def-SMEFT-comb-epe}. Exploiting these
model-specific matching results one arrives at
\begin{align}
  C_{\rm VRR}^{sd} (\muEW) &
  = \frac{({\cal N}_{sd})^{-1}}{(4 \pi)^2}
    \left\{ \begin{array}{cc} {\displaystyle
      \frac{M_{\wTil{S}_1}^2}{2}} \Big( \sum_a \wc{ed}{aa21}(\muNP) \Big)^2 & 
      \qquad \mbox{for} \quad \wTil{S}_1 
    \\[0.4cm]
      M_{\wTil{R}_2}^2 \Big( \sum_a \wc{\ell d}{aa21}(\muNP) \Big)^2 & 
      \qquad \mbox{for} \quad \wTil{R}_2
    \end{array} \right.
\end{align}
where the running from $\muNP$ to $\muEW$ due to self-mixing of $\Wc{dd}$ has
been neglected for simplicity. The normalisation factor ${\cal N}_{sd}$ is
defined in \refeq{eq:DF2-hamiltonian}. Whereas $\epe$ is linear in the sum over
semi-leptonic Wilson coefficients, $\Delta M_K$ and $\eps_K$ depend
quadratically on it.  For LQ models $S_{1,3}$ and $R_2$ analogously
\begin{align}
  C_{\rm VLL}^{sd} (\muEW) &
  = \frac{({\cal N}_{sd})^{-1}}{(4 \pi)^2}
    \left\{ \begin{array}{cc}
      2 M_{S_1}^2 \Big( \sum_a \wc[(1)]{\ell q}{aa21}(\muNP) \Big)^2 & 
      \qquad \mbox{for} \quad S_1 
    \\[0.4cm] {\displaystyle
      \frac{M_{R_2}^2}{2}} \Big( \sum_a \wc{qe}{aa21}(\muNP) \Big)^2 & 
      \qquad \mbox{for} \quad R_2
    \\[0.5cm] {\displaystyle
      \frac{10}{9}} M_{S_3}^2 \Big( \sum_a \wc[(1)]{\ell q}{aa21}(\muNP) \Big)^2 & 
      \qquad \mbox{for} \quad S_3
    \end{array} \right.
\end{align}

\begin{figure}
\centering
  \includegraphics[width=\myfigfraction \textwidth]{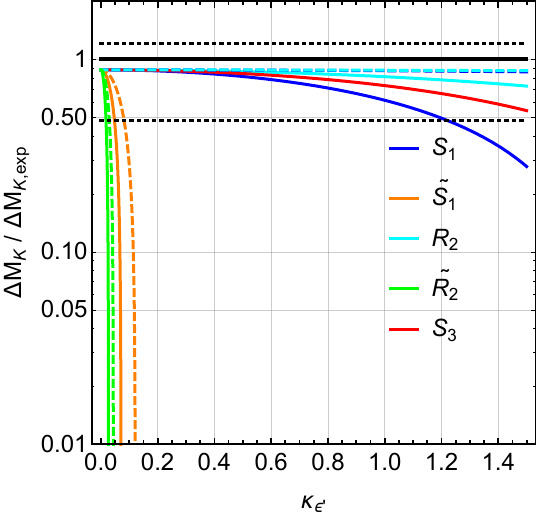}
  \hskip \myfigsepfraction \textwidth
  \includegraphics[width=\myfigfraction \textwidth]{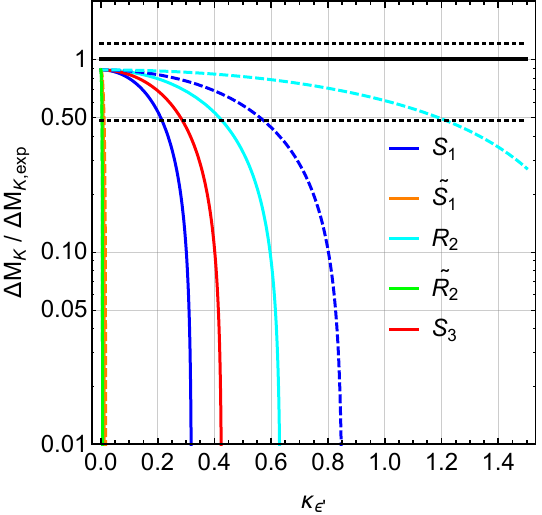}
  \caption{\small The dependence of the short-distance part of
    $\Delta M_K/\Delta M_{K,{\rm exp}}$ on $\kepe$ for
    $M_{\rm LQ} = 2$~TeV~[left] and $M_{\rm LQ} = 20$~TeV~[right].  The dashed
    lines show the effect of box-diagram contributions in $\epe$ for
    $\Sigma^{11}_{L,R} \in [0.0,\, 1.0]$.  Further shown are the experimental
    measurement [black solid] and the SM prediction [black dotted].  }
\label{fig:crr-DeltaMK-kepe}
\end{figure}

For example the correlation between $\kepe$ and $\Delta M_K$ takes the form
\begin{align}
  \label{DeltaMKR}
  \frac{(\Delta M_K)^{\rm LQ}}{(\Delta M_K)^{\rm exp}} &
  = -1.3\,\left( \frac{\kepe}{\ln(\muNP/\muEW)}\,\frac{M_{\rm LQ}}{2\tev}\right)^2 
\end{align}
in the $S_3$ model in which $C_7$ dominates the NP contribution to $\epe$.  It
should be noted that suppression of $\Delta M_K$ by the LQ contribution
increases with increasing $M_{\rm LQ}$, a feature found already in
\cite{Buras:2015jaq} in the context of $Z^\prime$ models. This shows that
$\Delta M_K$ can provide powerful constraints on scalar LQ models, even though
numerically enhanced left-right operators do not contribute.

The strong correlation of $\epe$ and the short-distance part of $\Delta M_K$ is
specific for each LQ model and shown in \reffig{fig:crr-DeltaMK-kepe} for
$M_{\rm LQ} = 2,\, 20$~TeV. The LQ models $\wTil{S}_1$ and $\wTil{R}_2$ lead
even for very small $\kepe \lesssim 0.05$ to a strong decrease by orders of
magnitude independently of $M_{\rm LQ}$. The correlation of $\epe$ and
$\Delta M_K$ can be dampened for low $M_{\rm LQ}$ of a few TeV in models
$S_{1,3}$ and $R_2$, but at large scales $M_{\rm LQ} \gtrsim 10$~TeV again
strong suppression of $\Delta M_K$ for $\kepe \gtrsim 0.3$ sets in disfavouring
them as a explanation of $\epe$, with weakest constraints on the model $R_2$. At
this point strictly imaginary couplings were assumed. In
\reffig{fig:crr-DeltaMK-realP} we show the impact of small real contributions to
the couplings on $\Delta M_K$ and $\eps_K$ for $\kepe = 0.5$ and
$M_{\rm LQ} = 2,\, 20$~TeV. Although it seems that fine-tuning between real and
imaginary parts can bring $\Delta M_K$ in agreement with data, actually $\eps_K$
becomes changed by orders of magnitude, even for $M_{\rm LQ}$ of a few TeV. On
the other hand the presence of box-diagram contributions in $\epe$ can further
weaken the constraints from $\Delta M_K$, but at the same time are subject to
constraints from $D^0-\oL{D}^0$ mixing, which are sensitive to left-right
operators.

\begin{figure}[ht]
\centering
  \includegraphics[width=\myfigfraction \textwidth]{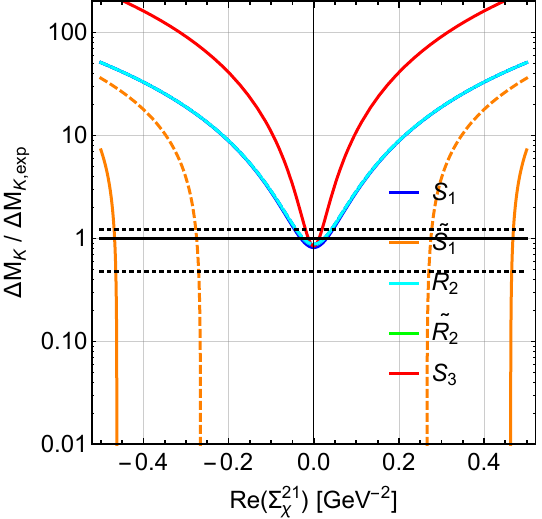}
  \hskip \myfigsepfraction \textwidth
  \includegraphics[width=\myfigfraction \textwidth]{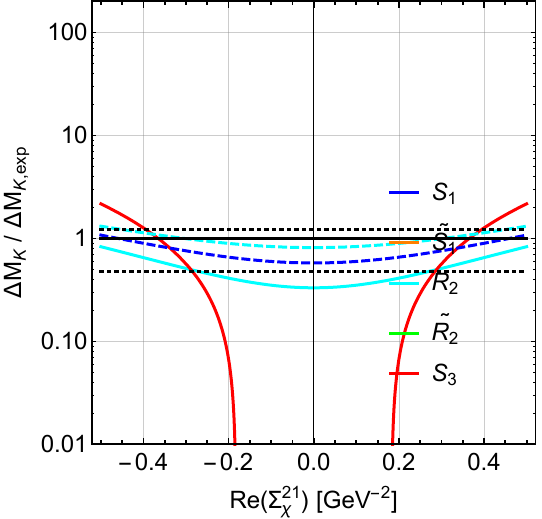}
\\
  \includegraphics[width=\myfigfraction \textwidth]{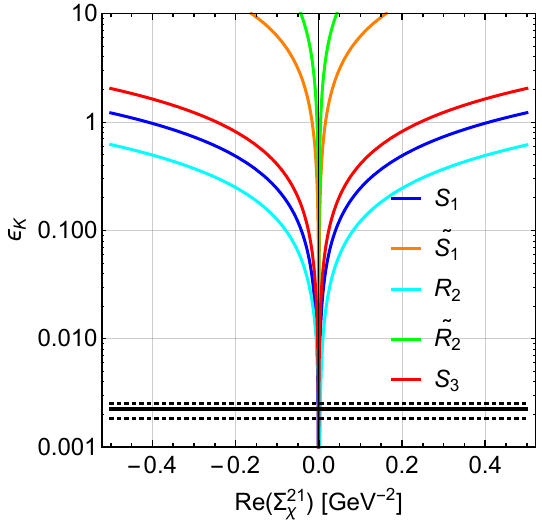}
  \hskip \myfigsepfraction \textwidth
  \includegraphics[width=\myfigfraction \textwidth]{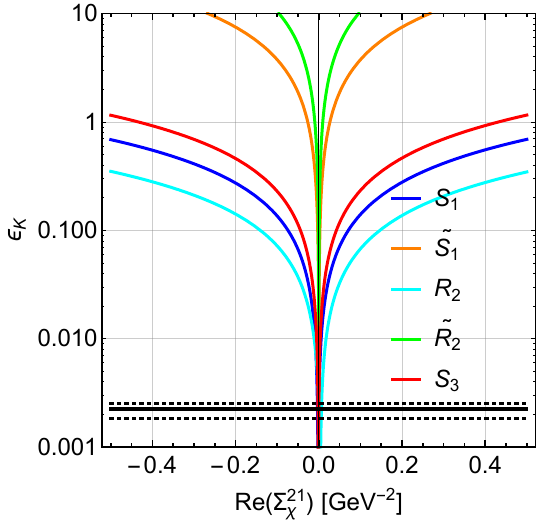}
\caption{\small The dependence of the short-distance part of
    $\Delta M_K/\Delta M_{K,{\rm exp}}$~[upper] and $\eps_K$~[lower] on real
    parts of couplings for $M_{\rm LQ} = 2$~TeV~[left] and
    $M_{\rm LQ} = 20$~TeV~[right] for $\kepe = 0.5$. Further shown are the
    experimental measurement [black solid] and the SM prediction [black dotted].
  }
\label{fig:crr-DeltaMK-realP}
\end{figure}

While no reliable calculations of $\Delta M_K$ and $\eps_K$ can be performed in
models with vector LQs without invoking a UV completion we would like to make an
observation on the models $U_1$ and $V_2$, which will turn out to be relevant
soon. As seen in \refeq{eq:Lag-LQ-V-AFW} in these models two couplings with
$\chi=L$ and $\chi'=R$ are present. If both are non-zero, strongly enhanced
left-right operators contributing to $\Delta M_K$ and $\eps_K$ will be
present. In fact the chiral enhancement of the hadronic matrix elements of such
operators combined with RG evolution brings in an enhancement of these
contributions by two orders of magnitude relative to VLL and VRR cases
\cite{Bobeth:2017xry} constraining strongly this model in the presence of large
imaginary couplings. Thus one might set one of the two couplings to zero, what
we have done while presenting the numerical results above.  We expect therefore
strong constraints from $\Delta M_K$ and $\eps_K$ on the couplings of these
models even without this approximation.
 
%
%
%

\section{\boldmath
  Summary, Conclusions and Outlook
  \label{sec:5}
}

In this paper we have presented for the first time the analysis of $\epe$ in LQ
models and provided general formulae in the framework of SMEFT for models in
which non-leptonic operators governing $\epe$ are generated from semi-leptonic
operators through electroweak (EW) renormalisation group (RG) effects. We have
also performed the one-loop decoupling for scalar LQ models.

Our analysis showed the strong correlation of rare Kaon processes with $\epe$.
They imply strong constraints on LQ models from the rare Kaon sector in the case
of a future confirmation of the $\epe$ anomaly by lattice QCD. They can be
further strengthened with improved measurements of $\kpn$ by NA62, $\klpn$ by
KOTO and $\ksm$ by LHCb, as well as a improved lattice result for $\Delta M_K$
in the Standard Model (SM). Hopefully also $\klpll$ decays will one day help in this
context.  Within our approximations, we were able to consider most relevant
contributions to $\epe$ from EW gauge-mixing for both scalar and vector LQ
models, and from one-loop decoupling in a complete manner for scalar LQ models.
On the one hand, the EW gauge-mixing generates numerically enhanced EW-penguin
operator $Q_7$ in models $S_{1,3}$, $R_2$, $U_{1,L}$, $V_{2,R}$ and $U_3$. On
the other hand, the box-diagram contributions exhibit in LQ models with
left-handed and right-handed couplings ($S_1$, $R_2$, $U_1$ and $V_2$) the
remarkable feature that they can generate EW-penguin operators $Q_{8,8'}$
already at the LQ scale. In turn they are numerically strongly enhanced in
$\epe$ through RG effects and their hadronic matrix elements.  Notably, the
latter contributions involve both LQ couplings of the corresponding models and
would vanish if either of them were zero. The main results of our analysis might
be summarised as follows
\begin{itemize}
\item The models with only one LQ coupling $\wTil{S}_1$, $\wTil{R}_2$, $S_3$ and
  $U_3$ lead to large enhancement of the branching fractions of $\klpn$ and
  $\kpn$ if $(\epe)_{\rm NP}$ is non-vanishing, such that for moderate
  enhancements $\kepe = 0.5$ the current bounds on both decays exclude these
  models as a possible explanation of the $\epe$ anomaly. Here the box-diagram
  contributions to $\epe$ have been included, thereby assuming the involved
  couplings to stay perturbative. We expect that even going beyond this
  assumption will be in general insufficient to explain the $\epe$ anomaly, even
  for the vector LQ model $U_3$, where we are not able to calculate box diagrams
  in a cut-off independent manner without specifying a UV completion.
\item The model $S_1$ shows also strong enhancement of $\klpn$ and $\kpn$ above
  the current Grossman-Nir bound even when box-diagram contributions are
  included as long as $\Sigma^{11}_R < 1.0$, but for larger values of
  $\Sigma^{11}_R$ a stronger bound on $\klpn$ and $\kpn$ is required to
  conclusively exclude it as an explanation of the $\epe$ anomaly.
\item The sub-scenario of vector LQ model $V_{2,L}$ predicts also huge
  enhancements of $\klpn$ and $\kpn$ for $\kepe \gtrsim 0.1$ if only EW
  gauge-mixing is included in $\epe$. From our experience with $\tilde{R}_2$ we
  expect that the inclusion of box-diagrams to $\epe$ will not be able to avoid
  this strong enhancement, leaving $V_{2,L}$ as a very unlikely explanation of
  the $\epe$ anomaly.
\item For the other models $R_2$, $U_{1,L}$, $U_{1,R}$ and $V_{2,R}$ we see
  large enhancements in $\ksm$ and $\klpll$, which put stringent constraints on
  electron and muon couplings and hence on the LQ parameter space.  But tauonic
  LQ couplings remain unconstrained and can serve as an explanation of
  $\epe$. In the scalar LQ model $R_2$ this statement includes box-diagram
  contributions to $\epe$ with $\Sigma^{11}_R < 1.0$.
\item For the scalar LQ models $S_1$ and $R_2$ the box-diagrams for
  $\Delta F = 2$ processes are calculable and here $\Delta M_K$ provides
  complementary bounds also on tauonic LQ couplings, but their effectivity
  becomes weak with smaller LQ masses for the case of purely imaginary
  $\Sigma^{21}_{\chi}$. If $\Sigma^{21}_\chi$ has also a small real part then
  $\Delta M_K$ and also $\eps_K$ become powerful constraints on large deviations
  in $\epe$ from the SM.
\end{itemize} 

The LQ models which have the best chance to explain the $\epe$ anomaly are then
the scalar LQ models $R_2$ and in part $S_1$ and two vector LQ models $U_1$ and
$V_2$. Among these four models only the model $U_1$ \footnote{In most of the
  literature actually only the sub-scenario $U_{1,L}$ is considered.} has a
chance to explain the $B$ physics anomalies if only one LQ representation is
considered. But as suggestions have been made to explain $B$ physics anomalies
by considering simultaneously two LQ representations \cite{Crivellin:2017zlb,
  Buttazzo:2017ixm, Dorsner:2017ufx} and $B$ physics anomalies could disappear
one day we analysed all these models.

We have pointed out that in models $R_2$ and $S_1$ the large contribution to
$\epe$ via box-diagrams \refeq{eq:LQ-box-qu} involves both couplings
$\sim \Sigma^{21}_\chi \Sigma^{11}_{\chi'}$ with $\chi \neq \chi'$. A similar
combination $\sim V_{m2}^{} V_{n1}^\ast\, \Sigma^{mn}_\chi \Sigma^{21}_{\chi'}$
can be bounded in principle by $D^0-\oL{D}^0$ mixing, providing also constraints
on the tauonic LQ couplings. Indeed it is well known that LR operators have
enhanced matrix elements not only for $K^0-\oL{K}^0$ mixing, but also for
$D^0-\oL{D}^0$ mixing \cite{Gedalia:2009kh, Isidori:2010kg}. However, due to the
presence of the quark-mixing matrix $V$ only a detailed global analysis can
provide a conclusive answer how strong $D^0-\oL{D}^0$ mixing can constrain
box-diagram contributions to $\epe$. Such an analysis is beyond the scope of our
paper.

In vector LQ models $U_1$ and $V_2$ the $\Delta F = 2$ constraints will be even
stronger since here LR operators contribute to $\Delta M_K$ and $\eps_K$, where
they are strongly enhanced, see \cite{Bobeth:2017xry} for recent updates. The
box-diagrams contribute to $\epe$ via the EW-penguin operator $Q_8$ (and also
$Q_{8'}$, but with Cabibbo suppression) \refeq{eq:LQ-box-qd} involving the
combinations $\sim \Sigma^{21}_\chi \Sigma^{11}_{\chi'}$ with $\chi \neq \chi'$,
whereas $\Delta F=2$ observables depend on
$\sim \Sigma^{21}_\chi \Sigma^{21}_{\chi'}$.  Again tauonic couplings are in
principle also subject to constraints, but similar to scalar LQ models, also
here only a global analysis on the basis of a UV completion can provide a
conclusive answer on the effectivity of these constraints.

Whether the LQ models where the $\epe$ anomaly seems to be still compatible with
present constraints from rare Kaon processes are challenged by other existing
constraints goes beyond the scope of our work. This would require dedicated
global analysis of each model. In the case of vector LQ models a UV completion
should be considered, as for example proposed in \cite{Diaz:2017lit,
  DiLuzio:2017vat, Calibbi:2017qbu, Bordone:2017bld}. These UV completions
contain usually new gauge and scalar sectors, subject to additional constraints
beyond flavour physics. On the other hand, UV completions based on
models with partial compositeness \cite{Gripaios:2014tna} or composite Higgs
models \cite{Barbieri:2016las, Barbieri:2017tuq} also lack full predictability 
due to the strongly interacting dynamics in these models, requiring nonperturbative
methods. We conclude therefore that the inclusion of box-diagram 
contributions to $\epe$ with both left-handed and right-handed couplings can improve
the situation in models $R_2$, $S_1$, $V_2$ and $U_1$ but this improvement might
be insufficient to explain the $\epe$ anomaly in LQ models, in particular if
$\kepe$ will turn out to be close to unity. 

It should also be emphasized that the presence of significant right-handed
couplings goes against the present wisdom based on $B$ physics anomalies that
new physics is dominated by left-handed currents, see in particular 
\cite{Buttazzo:2017ixm}. But as the $U_1$ model is favoured by $B$ physics
anomalies our analysis challenges model builders to find a UV completion for
this model that includes also right-handed couplings and couplings to the first
generation and while explaining the $\epe$ anomaly, satisfies
all existing constraints, in particular describes $B$ physics anomalies and 
is consistent with the bounds on $\Delta M_K$ and $\varepsilon_K$ that are 
very strong in the presence of left-handed and right-handed couplings.

While the vector LQ $U_1$ performs best as a single representation in the case
of $B$-physics anomalies, models with two or more LQ representations have been
considered in the literature in the context of these anomalies. The question
then arises whether with two LQ representations the results for $\epe$ would
improve.  We comment here briefly on two such models with scalar LQs, one
involving $S_1$ and $S_3$ representations \cite{Crivellin:2017zlb,
  Buttazzo:2017ixm} and the second $S_3$ and $\tilde R_2$ representations
\cite{Dorsner:2017ufx}.

\begin{table}[t]
\centering
\renewcommand{\arraystretch}{1.4}
\resizebox{\textwidth}{!}{
\begin{tabular}{|c||ccccc||ccc|ccc|c|}
\hline
  
& $S_1$
& $\wTil{S}_1$
& $R_2$
& $\wTil{R}_2$
& $S_3$
& $U_{1,L}$
& $U_{1,R}$
& $U_1$
& $V_{2,L}$
& $V_{2,R}$
& $V_2$
& $U_3$
\\
\hline\hline
  $\klpn$
& $\dagger$
& ---
& ---
& $\dagger$
& $\dagger$
&
& ---
& 
& $\dagger$
& ---
&
& $\dagger$
\\
\hline
  $\kpn$
& $\dagger$
& ---
& ---
& $\dagger$
& $\dagger$
&
& ---
& 
& $\dagger$
& ---
&
& $\dagger$
\\
\hline
  $\klpll$
& ---
& $\dagger_{e,\mu}$
& $\dagger_{e,\mu}$
& $\dagger_{e,\mu}$
& $\dagger_{e,\mu}$
& $\dagger_{e,\mu}$
& $\dagger_{e,\mu}$
&
& $\dagger_{e,\mu}$
& $\dagger_{e,\mu}$
&
& $\dagger_{e,\mu}$
\\
\hline
  $\ksm$
& ---
& $\dagger_{\mu}$
& 
& $\dagger_{\mu}$
& 
& 
& $\dagger_{\mu}$
&
& $\dagger_{\mu}$
& 
&
& 
\\
\hline
  $\Delta M_K$, $\eps_K$
& $\dagger_\muNP$
& $\dagger$
& 
& $\dagger$
& $\dagger_\muNP$
& loop
& loop
& loop
& loop
& loop
& loop
& loop
\\
\hline\hline
  $\sum$
& $\dagger$
& $\dagger$
& $\dagger_{e,\mu}$
& $\dagger$
& $\dagger$
& $\dagger_{e,\mu}$
& $\dagger_{e,\mu}$
&
& $\dagger$
& $\dagger_{e,\mu}$
&
& $\dagger$
\\
\hline
\end{tabular}
}
\renewcommand{\arraystretch}{1.0}
\caption{\label{tab:LQ-models}
  Overview of strong conflicts with current bounds/measurements on rare
  Kaon processes when requiring $(\epe)_{\rm NP} = \kepe \times 10^{-3}$ with 
  $\kepe \gtrsim 0.5$ for scalar and vector LQ models. In vector LQ models only the 
  EW gauge-mixing is included in $\epe$, see main text for details. The symbols
  in the table correspond to: 
  ``---'' = no contributions in this LQ model; 
  ``loop'' = mediated by loop corrections in vector LQ models; 
  ``$\dagger$'' = ruled out for all lepton flavours $\ell = e,\mu,\tau$; 
  ``$\dagger_{\ell}$'' = ruled out for lepton flavour $\ell$; 
  ``$\dagger_\muNP$'' = ruled out for all lepton flavour if LQ mass is large enough.
}
\end{table}

Looking at \refeq{eq:epe-S3} and \refeq{eq:epe-S1} we observe that in the case
of a model with $S_1$ and $S_3$ the value of the coupling
$\Sigma_{\chi, \text{LQ}}^{sd}$ can be decreased for a given $\kepe$. Assuming that the
couplings in these two representations are equal the coupling in question can be
decreased by a factor 1.33 implying the reduction of the branching ratio for
$\klpn$ in the case of $S_3$ model first by a factor 1.8 with a smaller effect
in $\kpn$. But as in both cases now also $S_1$ contributes the change is
smaller. While this still improves the situation the model is still predicting
values of $\mathcal{B}(\klpn)$ close to the Grossman-Nir bound and similar
comments apply to $\kpn$ where the change is smaller. In the case of
$K_L \to \pi^0 \ell\bar\ell$ the representation $S_1$ does not contribute and
one can see by inspecting \reffig{fig:crr-KLpi0ll} that this reduction of the
coupling and of the branching ratio does not really solve the problem.

As far as combination of $S_3$ and $\tilde R_2$ is concerned the great disparity
in the effectiveness of these two representations to enhance $\epe$ seen in
\refeq{eq:epe-wTil-S1,R2} and \refeq{eq:epe-S3} tells us that the results of the
$S_3$ model remain practically unchanged. These two examples indicate that even
invoking more representations it will be difficult to enhance sufficiently
$\epe$, in particular if $\kepe$ close to unity will be required.

The goal of our paper was to demonstrate on the basis of Kaon physics alone
that the explanation of a possible $\epe$ anomaly within the context of LQ models
was very unlikely. Any additional constraint on the couplings of LQs would 
further strengthen this conclusion. Such constraints could come in particular 
from $B$ physics anomalies but this would require the imposition of 
flavour symmetries that would relate $K$ and $B$ decays. In connection with 
the latter it has been demonstrated in \cite{Aloni:2017ixa} that the imposition 
of minimal flavour violation (MFV) on LQ models excludes the explanation of $B$ 
physics anomalies within these models. We would like to emphasize that in the
case of $\epe$ MFV is broken from the beginning as only significant new
CP-violating phases have a chance to explain the anomaly in question.

Another possible constraint could come from the simultaneous considerations 
of flavour symmetries responsible from the observed spectrum of fermion masses.
In the context of $B$ physics anomalies this issue has been addressed in 
\cite{Gripaios:2014tna} for the scalar LQ model $S_3$ in the framework of 
partial compositeness and for $\wTil{R}_2$ and $S_3$ models imposing various
flavour symmetries in \cite{Varzielas:2015iva}. Our analysis shows that in these
frameworks the bounds on $\klpn$ and $\kpn$ are violated when requiring 
$\kepe \geq 0.05$ for $\wTil{R}_2$ and  $\kepe \geq 0.4$ for $S_3$.  
It will be interesting to generalize such studies to include $\epe$ and in 
particular $\kpn$ after the result from NA62 will be known. 
In case $\epe$ and $B$ physics anomalies would persist and the measurement 
of the $\kpn$ branching ratio would significantly deviate from the rather
precise SM prediction, a valuable information on family structure of BSM models could
be obtained. We hope to return to this issue once the $\epe$ anomaly will
be confirmed and the experimental status of $B$ physics anomalies and of $\kpn$
will improve.

Our findings for the ten LQ models listed in \reftab{tab:LQ-q-numbers} as far as
the $\epe$ anomaly in correlation with rare Kaon processes is concerned are
summarised in \reftab{tab:LQ-models} \footnote{The models $\tilde U_1$ and
  $\tilde V_2$ are absent in this table because they do not provide new
  contributions to $\epe$.} The different symbols appearing in this table are
explained in the caption of this table.

Finally, in most papers analysing LQ and other models in the context of $B$
physics anomalies it is a common practice to set NP couplings to Kaon and other
light physics sectors to zero.  If the $\epe$ anomaly will be confirmed by
future lattice results all these analyses have to be reconsidered.

The main messages of our analysis to take home are the following ones. If the
future improved lattice calculation will confirm the $\epe$ anomaly at the
level $(\epe)_{\rm NP}\ge 5\times 10^{-4}$ LQs are likely not responsible for it.
But if the $\epe$ anomaly will disappear one day, large NP effects in rare $K$ decays
that are still consistent with present bounds will be allowed.

%
%
%

\section*{Acknowledgements}
This research was supported by the DFG cluster of excellence ``Origin and Structure
of the Universe''. We thank Svjetlana Fajfer and David Straub for useful
discussions.

%
%
%

\appendix

\section{\boldmath LQ Lagrangian}
\label{app:LQ-Lag}

Here we summarise our conventions for the LQ Lagrangian, which follows
\cite{Buchmuller:1986zs}. The transformation properties of the spin $S=0$
(scalar) and $S=1$ (vector) LQ's under the SM group
$\GSM \equiv \SUthreeC \otimes \SUtwoL \otimes \UoneY$ are summarised in
\reftab{tab:LQ-q-numbers}. We have modified the definition of the $\SUtwoL$
doublet LQ's $R_2$, $\wTil{R}_2$ and $V_2$, $\wTil{V}_2$ to follow the standard
conventions for quark, lepton and Higgs doublets in which upper and lower
$\SUtwoL$ components correspond to isospin $+1/2$ and $-1/2$, respectively,
opposite to the original convention in \cite{Buchmuller:1986zs}.

\def\oh{\frac{1}{2}}
\begin{table}
\centering
\renewcommand{\arraystretch}{1.3}
\setlength{\tabcolsep}{7pt}
\begin{tabular}{|c||ccccc|ccccc|}
\hline
  LQ 
& $S_1$    & $\wTil{S}_1$ & $R_2$   & $\wTil{R}_2$ & $S_3$    & $U_1$   & $\wTil{U}_1$ & $V_2$    & $\wTil{V}_2$ & $U_3$
\\
\hline \hline
  $\SUthreeC$ 
& $3^\ast$ & $3^\ast$     & $3\;\;$ & $3\;\;$      & $3^\ast$ & $3\;\;$ & $3\;\;$      & $3^\ast$ & $3^\ast$     & $3$
\\
  $\SUtwoL$
& $1\;$    & $1\;$        & $2\;\;$ & $2\;\;$      & $3\;$  & $1\;\;$ & $1\;\;$        & $2\;\,$  & $2\;\,$      & $3$  
\\
  $\UoneY$
& $2/3$    & $8/3$        & $7/3$ & $1/3$          & $2/3$    & $4/3$   & $10/3$       & $5/3$    & $-1/3$       & $4/3$     
\\
\hline
  $T_3$
& $0$      & $0$          
& $\begin{array}{cc} +\oh \\ -\oh \end{array}$  
& $\begin{array}{cc} +\oh \\ -\oh \end{array}$ 
& $\begin{array}{cc} +1 \\ 0 \\-1 \end{array}$         
& $0$      & $0$ 
& $\begin{array}{cc} +\oh \\ -\oh \end{array}$  
& $\begin{array}{cc} +\oh \\ -\oh \end{array}$ 
& $\begin{array}{cc} +1 \\ 0 \\-1 \end{array}$         
\\
  $Q_{\rm em}$
& $1/3$ & $4/3$
& $\begin{array}{cc}  5/3 \\  2/3 \end{array}$
& $\begin{array}{cc} +2/3 \\ -1/3 \end{array}$
& $\begin{array}{cc} +4/3 \\ +1/3 \\ -2/3 \end{array}$
& $2/3$ & $5/3$
& $\begin{array}{cc}  4/3 \\  1/3 \end{array}$
& $\begin{array}{cc} +1/3 \\ -2/3 \end{array}$
& $\begin{array}{cc} +5/3 \\ +2/3 \\ -1/3 \end{array}$
\\
\hline
\end{tabular}
\renewcommand{\arraystretch}{1.0}
\setlength{\tabcolsep}{2pt}
\caption{
  \label{tab:LQ-q-numbers}
  Quantum numbers of LQ's under the gauge groups of the SM, and
  the isospin $T_3$ and electric charge $Q_{\rm em}$ of their 
  $\SUtwoL$ components, $Q_{\rm em} = T_3 + Y/2$.
}
\end{table}

The couplings of scalar LQs to quarks ($q_L, u_R, d_R$) and leptons 
($\ell_L, e_R$) in the unbroken $\SUtwoL \otimes \UoneY$ phase are
\begin{equation}
  \label{eq:Lag-LQ-S-AFW}
\begin{aligned}
  {\cal L}_S & 
  = \left( g^{1L} \left[\oL{q^c_L} i\tau^2 \ell_L\right] 
         + g^{1R} \left[\oL{u^c_R} e_R \right] \right) S_1
  + \widetilde{g}^{1R} \left[\oL{d^c_R} e_R \right]\wTil{S}_1
\\ & 
  + h^{2L} \left[\oL{u}_R R_2^T i \tau^2 \ell_L\right] 
  + h^{2R} \left[\oL{q}_L R_2 e_R \right]
  + \widetilde{h}^{2L} \left[\oL{d}_R \wTil{R}_2^T i \tau^2 \ell_L \right]
\\ & 
  + g^{3L} \left[\oL{q^c_L} i\tau^2 \vec{\boldsymbol{\tau}} \ell_L \right]
    \boldsymbol{\cdot \, \vec{S}_{3}} 
  + \, \mbox{h.c.} ,
\end{aligned}
\end{equation}
and for vector LQs
\begin{equation}
  \label{eq:Lag-LQ-V-AFW}
\begin{aligned}
  {\cal L}_V &
  = \left( h^{1L} \left[\oL{q}_L \gamma_\mu \ell_L\right] 
         + h^{1R} \left[\oL{d}_R \gamma_\mu e_R \right] \right) U_1^{\mu}
  + \widetilde{h}^{1R} \left[\oL{u}_R \gamma_\mu e_R \right]\wTil{U}_1^{\mu}
\\ & 
  + g^{2L} \left[\oL{d^c_R} \gamma_\mu  (V_2^{\mu})^T i\tau^2 \ell_L\right] 
  + g^{2R} \left[\oL{q^c_L} \gamma_\mu i\tau^2 V_2^{\mu} e_R \right]
  + \widetilde{g}^{2L} \left[\oL{u^c_R} \gamma_\mu (\wTil{V}_2^{\mu})^T i\tau^2 \ell_L \right] 
\\ & 
  + h^{3L} \left[\oL{q}_L \gamma_\mu \vec{\boldsymbol{\tau}} \ell_L \right]
    \boldsymbol{\cdot \, \vec{U}_{3}^{\mu}} 
  + \, \mbox{h.c.} 
\end{aligned}
\end{equation}
where the generation indices on quark and lepton fields have been suppressed.
The LQ couplings $g^{a\chi}$, $\wTil{g}^{a\chi}$ and $h^{a\chi}$,
$\wTil{h}^{a\chi}$ ($a = 1,2,3$ and $\chi = L,R$) are $3 \times 3$
complex-valued matrices in the generation space of the quarks and leptons.
Above $(i\tau^2)_{ab} = \varepsilon_{ab}$ is the second Pauli matrix,
$\varepsilon_{12} = - \varepsilon_{21} = + 1$. Charge-conjugated fields are
denoted as $\psi^c \equiv C \oL{\psi}^T$ with the charge-conjugation matrix $C$.

%
%
%

\section{\boldmath LQ tree-level decoupling}
\label{app:tree-decoupl}

The tree-level decoupling of LQs gives rise to semi-leptonic operators in SMEFT.
Here we summarise the results of their Wilson coefficients at the LQ scale
$\muNP$.  The semi-leptonic operators are listed in \reftab{tab:4ferm}.  There
are two classes of diagrams to consider, depending on whether charge-conjugated
fields are involved or not. We follow \cite{Denner:1992vza} for Feynman rules
and consider the tree-level matching for $Q_i + L_a \to Q_j + L_b$.

\begin{align}
  S_1 & : &
  \wc[(1)]{\ell q}{baji} = - \wc[(3)]{\ell q}{baji} & 
  = \frac{g^{1L}_{ia} g^{1L\ast}_{jb}}{4 M^2} ,
\\ &&
  \wc{eu}{baji} & 
  = \frac{g^{1R}_{ia} g^{1R\ast}_{jb}}{2 M^2} ,
\\ &&
  \wc[(1)]{\ell equ}{abij}^\ast = - 4 \wc[(3)]{\ell equ}{abij}^\ast & 
  = \frac{g^{1L}_{ia} g^{1R\ast}_{jb}}{2 M^2} ,
\\[0.2cm]
  \label{eq:LQ-SL-match:tilS_1-ed}
  \wTil{S}_1 & : &
  \wc{ed}{baji} & 
  = \frac{\wTil{g}^{1R}_{ia} \wTil{g}^{1R\ast}_{jb}}{2 M^2} ,
\\[0.2cm]
  R_2 & : &
  \wc{\ell u}{baji} & 
  = - \frac{h^{2L}_{ja} h^{2L\ast}_{ib}}{2 M^2} ,
\\ &&
  \wc{qe}{baji} & 
  = - \frac{h^{2R}_{ja} h^{2R\ast}_{ib}}{2 M^2} ,
\\ &&
  \wc[(1)]{\ell equ}{abij}^\ast = 4 \wc[(3)]{\ell equ}{abij}^\ast & 
  = \frac{h^{2L}_{ja} h^{2R\ast}_{ib}}{2 M^2} ,
\\[0.2cm]
  \wTil{R}_2 & : &
  \wc{\ell d}{baji} & 
  = - \frac{\wTil{h}^{2L}_{ja} \wTil{h}^{2L\ast}_{ib}}{2 M^2} ,
\\[0.2cm]
  S_3 & : &
  \wc[(1)]{\ell q}{baji} = 3 \wc[(3)]{\ell q}{baji} & 
  = \frac{3}{4} \frac{g^{3L}_{ia} g^{3L\ast}_{jb}}{M^2} ,
\end{align}

\begin{align}
  U_1 & : &
  \wc[(1)]{\ell q}{baji} = \wc[(3)]{\ell q}{baji} & 
  = - \frac{h^{1L}_{ja} h^{1L\ast}_{ib}}{2 M^2} ,
\\ &&
  \wc{ed}{baji} & 
  = - \frac{h^{1R}_{ja} h^{1R\ast}_{ib}}{M^2} ,
\\ &&
  \wc{\ell edq}{abij}^\ast & 
  = 2 \frac{h^{1L}_{ja} h^{1R\ast}_{ib}}{M^2} ,
\\[0.2cm]
  \wTil{U}_1 & : &
  \wc{eu}{baji} & 
  = - \frac{\wTil{h}^{1R}_{ja} \wTil{h}^{1R\ast}_{ib}}{M^2} ,
\\[0.2cm]
  V_2 & : &
  \wc{\ell d}{baji} & 
  = \frac{g^{2L}_{ia} g^{2L\ast}_{jb}}{M^2} ,
\\ &&
  \wc{qe}{baji} & 
  = \frac{g^{2R}_{ia} g^{2R\ast}_{jb}}{M^2} ,
\\ &&
  \wc{\ell edq}{abij}^\ast & 
  = -2 \frac{g^{2L}_{ia} g^{2R\ast}_{jb}}{M^2} ,
\\[0.2cm]
  \wTil{V}_2 & : &
  \wc{\ell u}{baji} & 
  = \frac{\wTil{g}^{2L}_{ia} \wTil{g}^{2L\ast}_{jb}}{M^2} ,
\\[0.2cm]
  U_3 & : &
  \wc[(1)]{\ell q}{baji} = -3 \wc[(3)]{\ell q}{baji} & 
  = -\frac{3}{2} \frac{h^{3L}_{ja} h^{3L\ast}_{ib}}{M^2} .
\end{align}

%
%
%

\section{\boldmath LQ one-loop decoupling}
\label{app:1loop-decoupl}

For scalar LQs it is possible to calculate one-loop decoupling for non-leptonic
processes, which contribute directly to the non-leptonic operators in SMEFT that
mediate $\Delta F=1$ processes like $\epe$, but also to $\Delta F=2$ processes
$\Delta M_K$ and $\eps_K$.

Here we provide the general results for $Q_i \bar{Q}_k \to Q_j \bar{Q}_l$ for
the choice of operators \refeq{eq:box-SMEFT-operators} that contribute to
down-type quark transitions ($ji \neq kl$) and are valid for $\Delta F = 1$:
\begin{align}
  S_1 : & &
  \wc[(o,1)]{qq}{jikl} & 
  = - \frac{\Sigma^{kl}_L}{(4\pi)^2} \frac{\Sigma^{ji}_L}{4 M^2} ,
\\
  & &
  \wc[(o)]{qu}{jikl} & 
  = - \frac{\Sigma^{kl}_R}{(4\pi)^2} \frac{\Sigma^{ji}_L}{4 M^2} ,
\\
  \wTil{S}_1 : & &
  \wc[(o)]{dd}{jikl} & 
  = - \frac{\Sigma^{kl}_R}{(4\pi)^2} \frac{\Sigma^{ji}_R}{4 M^2} ,
\\
  R_2 : & &
  \wc[(o,1)]{qq}{jikl} = \wc[(o,3)]{qq}{jikl} &
  = - \frac{\Sigma^{lk}_R}{(4\pi)^2} \frac{\Sigma^{ij}_R}{8 M^2} ,
\\
  & &
  \wc[(o)]{qu}{jikl} &
  = - \frac{\Sigma^{lk}_L}{(4\pi)^2} \frac{\Sigma^{ij}_R}{4 M^2} ,
\\
  \wTil{R}_2 : & &
  \wc[(o)]{dd}{jikl} &
  = - \frac{\Sigma^{lk}_L}{(4\pi)^2} \frac{\Sigma^{ij}_L}{2 M^2} ,
\\
  S_3 : & & 
  \wc[(o,1)]{qq}{jikl} = \frac{3}{2} \wc[(o,3)]{qq}{jikl} & 
  = - \frac{\Sigma^{kl}_L}{(4\pi)^2} \frac{3 \,\Sigma^{ji}_L}{4 M^2} .
\end{align}

The results for $\Delta F =2$ matching onto SMEFT operators $\Op[(1,3)]{qq}$,
$\Op{dd, uu}$ and $\Op[(8)]{qu}$ is given here, where for completeness also the
operators $\Op[(1,8)]{qu}$ and $\Op{uu}$ are listed that mediate up-type
$\Delta F = 2$ processes\footnote{We thank the authors of 
\cite{Gherardi:2020det} to inform us about the mistakes in equations 
\refeq{eq:DF2-SMEFT-S1} and \refeq{eq:DF2-SMEFT-S3} that they noticed
in the context of their analysis.}
\begin{align}
  \label{eq:DF2-SMEFT-S1}
  S_1 & : &
  \wc[(1)]{qq}{jiji} = \wc[(3)]{qq}{jiji} &
  = -\frac{\Sigma_L^{ji}}{(4\pi)^2} \frac{\Sigma_L^{ji}}{16 M^2} , 
\\ & & 
  6 \wc[(1)]{qu}{jiji} = \wc[(8)]{qu}{jiji} &
  = -\frac{\Sigma_R^{ji}}{(4\pi)^2} \frac{\Sigma_L^{ji}}{2 M^2} ,    
\\ & & 
  \wc{uu}{jiji} &
  = -\frac{\Sigma_R^{ji}}{(4\pi)^2} \frac{\Sigma_R^{ji}}{8 M^2} ,    
\\[0.2cm]
  \wTil{S}_1 & : &
  \wc{dd}{jiji} &
  = -\frac{\Sigma^{ji}_R}{(4\pi)^2} \frac{\Sigma^{ji}_R}{8 M^2} ,
\\[0.2cm]
  R_2 & : & 
  \wc[(3)]{qq}{jiji} = 0 , \qquad \wc[(1)]{qq}{jiji} &
  = -\frac{\Sigma_R^{ij}}{(4\pi)^2} \frac{\Sigma_R^{ij}}{8 M^2} ,  
\\ & & 
  6 \wc[(1)]{qu}{jiji} = \wc[(8)]{qu}{jiji} &
  = -\frac{\Sigma_L^{ij}}{(4\pi)^2} \frac{\Sigma_R^{ij}}{2 M^2} ,
\\ & & 
  \wc{uu}{jiji} &
  = -\frac{\Sigma_L^{ij}}{(4\pi)^2} \frac{\Sigma_L^{ij}}{4 M^2} ,    
\\[0.2cm]
  \wTil{R}_2 & : &
  \wc{dd}{jiji} &
  = -\frac{\Sigma^{ij}_L}{(4\pi)^2} \frac{\Sigma^{ij}_L}{4 M^2} ,
\\[0.2cm]
  \label{eq:DF2-SMEFT-S3}
  S_3 & : &
  \wc[(1)]{qq}{jiji} = 9 \wc[(3)]{qq}{jiji} &
  = -\frac{\Sigma^{ji}_L}{(4\pi)^2} \frac{9\Sigma^{ji}_L}{16 M^2} .
\end{align}

%
%
%

\section{\boldmath $d_j\to d_i\, q\bar{q}$ and $\epe$}
\label{app:d->dqq}

The effective Lagrangian for $\bar{s} \to \bar{d} q\bar{q}$ ($i\neq j$) is
adopted from \cite{Buras:1993dy} with the definition of the operators given in
\refeq{eq:QCD-peng-op} and \refeq{eq:QED-peng-op}. At the scale $\muEW$
($N_f = 5$) it reads
\begin{equation}
  \label{eq:Heffeprime}
\begin{aligned}
  {\cal H}_{d\to dq\bar{q}} 
  = \frac{G_F}{\sqrt{2}} \, V_{ud}^{} V_{us}^\ast
    \Big\{ & (1-\tau) \big[z_1 (Q_1 - Q_1^c) + z_2 (Q_2 - Q_2^c)\big]
\\ & 
  + \sum_{a=3}^{10} (\tau v_a + v_a^{\rm NP}) Q_a
  + \sum_{a=3}^{10} v'_a Q'_a \Big\} + \mbox{h.c.} ,
\end{aligned}
\end{equation}
where $Q_{1,2}^{(c)}$ denote current-current operators. The sum over $a$ extends
over the QCD- and EW-penguin operators and we included their chirality-flipped
counterparts $Q'_a = Q_a [\gamma_5 \to -\gamma_5]$.  The Wilson coefficients are
denoted as $z_a$, $v_a^{(\rm NP)}$ and $v'_a$, taken at the scale $\muEW$. For
the SM-part, CKM unitarity was used,
\begin{align}
  \tau & \equiv -\frac{V_{td}^{} V_{ts}^\ast}{V_{ud}^{} V_{us}^\ast} ,
\end{align}
and we introduced a new physics contribution $v_a^{\rm NP}$ as shown above,
which is related to the LQ-contribution \refeq{eq:DeltaS=1-matching} as
\begin{align}
  v_a^{\rm NP} & = C_a , & 
  v'_a & = C'_a .
\end{align}

The RG evolution at NLO in QCD and QED leads to the effective Hamiltonian at a
scale $\mu\lesssim \mu_c \sim m_c$ ($N_f=3$)
\begin{align}
  \label{eq:EFT:ddqq:Nf3}
  {\cal H}_{d\to dq\bar{q}} &
  = \frac{G_F}{\sqrt{2}} \, V_{ud}^{} V_{us}^\ast
    \left\{ z_1 Q_1 + z_2 Q_2 + 
    \sum_{a=3}^{10} [z_a + \tau y_a + v_a^{\rm NP}] Q_a 
  + \sum_{a=3}^{10} v'_a Q'_a \right\} + \mbox{h.c.} ,
\end{align}
after decoupling of $b$- and $c$-quarks at scales $\mu_{b,c}$
\cite{Buras:1993dy}, where $y_a \equiv v_a - z_a$ and all Wilson coefficients
are at the scale $\mu$.

The contributions of new physics can then be accounted for in $\epe$ by the
replacement
\begin{align}
  \label{eq:epe-NP-contr}
  y_a(\mu) & \to 
  y_a(\mu) + \frac{v_a^{\rm NP} (\mu)- v'_a(\mu)}{\tau} ,
\end{align}
where the minus sign is due to
$\langle (\pi\pi)_I | Q_a | K \rangle = - \langle (\pi\pi)_I | Q'_a | K \rangle$
for the pseudo-scalar pions in the final state \cite{Kagan:2004ia}. For the
readers convenience we provide a semi-numerical formula for $\epe$ from
\cite{Bobeth:2016llm} with initial conditions of Wilson coefficients from new
physics in QCD- and EW-penguins $a = 3^{(\prime)}, \ldots, 10^{(\prime)}$ at
the electroweak scale $\muEW$:
\begin{equation}
  \label{eq:epe-seminum}
\begin{aligned}
  \frac{\varepsilon'}{\varepsilon} & =
  \left[-2.58 + 24.01 B_6^{(1/2)} - 12.70 B_8^{(3/2)}\right] \times 10^{-4}
  + \sum_a P_a \, \mbox{Im}(v_a^{\rm NP} - v'_a)[\muEW].
\end{aligned}
\end{equation}
The coefficients are 
\begin{align}
  P_a & 
  = p_a^{(0)} + p_a^{(6)} B_6^{(1/2)} + p_a^{(8)} B_8^{(3/2)}
\end{align}
with $p^{(n)}_a$ given in \reftab{tab:epe-seminum}, where the last column gives
$P_a$ for $B_6^{(1/2)}(\mu) = 0.57$ and $B_8^{(3/2)}(\mu) = 0.76$. For this
purpose $\muEW = M_W$, $\mu_b = m_b(m_b)$, $\mu_c = 1.3$~GeV and
$\mu = 1.53$~GeV have been used. The central value of the SM prediction is
$(\epe)_{\rm SM} = 1.5 \times 10^{-4}$ compared to $1.9 \times 10^{-4}$ in
\cite{Buras:2015yba} due to different numerical inputs.

\begin{table}[!tb]
\renewcommand{\arraystretch}{1.3}
\centering
\begin{tabular}{|c|rrr|r||c|rrr|r|}
\hline
  $a$
& $p^{(0)}_a$ & $p^{(6)}_a$ & $p^{(8)}_a$ & $P_a$
& $a$
& $p^{(0)}_a$ & $p^{(6)}_a$ & $p^{(8)}_a$ & $P_a$
\\
\hline\hline
  3
&    $7.45$ &  $-3.40$ &   $-3.50$ &    $2.85$
&  7
& $-102.02$ &  $-1.32$ & $2040.38$ & $1447.91$
\\
  4
&   $-15.3$ & $-15.59$ &    $9.39$ &  $-17.05$
& 8
& $-428.11$ &   $-6.9$ & $6908.01$ & $4818.04$
\\
  5
&    $1.70$ &  $30.62$ &  $-18.74$ &    $4.91$
& 9
&   $36.72$ &   $4.42$ &  $-21.28$ &   $23.06$
\\
  6
&    $8.63$ & $115.28$ &  $-47.69$ &   $38.10$
& 10
&    $9.57$ &  $-3.96$ &   $-4.80$ &    $3.66$
\\
\hline
\end{tabular}  
\renewcommand{\arraystretch}{1.0}
  \caption{
    Values of the coefficients entering the semi-numerical formula of $\epe$
    in Eq.~\refeq{eq:epe-seminum}. 
    The last column gives $P_a$ for $B_6^{(1/2)} = 0.57$ and $B_8^{(3/2)} = 0.76$,
    the central values of these parameters obtained in \cite{Buras:2015yba} from
    \cite{Bai:2015nea}.
  }
  \label{tab:epe-seminum}
\end{table}

%
%

\renewcommand{\refname}{R\lowercase{eferences}}

\addcontentsline{toc}{section}{References}

\bibliographystyle{JHEP}

\small

\bibliography{Bookallrefs}

\end{document}